\documentclass[aps,prd,twocolumn,showpacs,floatfix,preprintnumbers,amsmath,amssymb,nofootinbib,groupedaddress, 
superscriptaddress, notitlepage]{revtex4-2}
\usepackage{import}
\usepackage{graphicx}
\usepackage{diagbox}
\usepackage{chngcntr}
\usepackage{enumitem}
\usepackage{stmaryrd}
\usepackage{appendix}
\usepackage{cancel}
\usepackage{tikz}
\usetikzlibrary{positioning, shapes.geometric, arrows.meta}
\usepackage{dcolumn}
\usepackage{bm}
\usepackage[colorlinks=true,citecolor=Plum,linkcolor=Plum,urlcolor=Plum, backref=false,pdfborder={0 0 0}]{hyperref}
\usepackage{cancel}
\usepackage[dvipsnames]{xcolor}
\usepackage{mathtools}
\usepackage{mathbbol}
\usepackage{amsthm}
\usepackage{amsmath}
\usepackage{booktabs} 
\usepackage{mathtools}
\usepackage{physics}
\usepackage{xcolor}
\usepackage{adjustbox}
\usepackage{placeins}
\usepackage[T1]{fontenc}
\usepackage{lipsum}
\usepackage{csquotes}

\newcommand{\wscos}{\texttt{WxSCOS}}
\newcommand{\unwise}{\texttt{unWISE}}
\newcommand{\blue}{\texttt{`blue'}}
\newcommand{\green}{\texttt{`green'}}
\newcommand{\red}{\texttt{`red'}}
\newcommand{\planck}{\textit{Planck}}
\newcommand{\catwise}{\texttt{CatWISE}}
\newcommand{\lcdm}{\ensuremath{\Lambda}CDM}
\newcommand{\hvec}[1]{\hat{\bm{#1}}}
\newcommand{\Weff}[1]{\ensuremath{\mathcal{W}^{#1}}}
\newcommand{\WeffTH}{\Weff{{\rm TH}}}
\newcommand{\WeffG}{\Weff{{\rm G}}}
\newcommand{\Weffksz}[1]{\ensuremath{\mathcal W^{\rm kSZ}_{#1}}}
\newcommand{\kmpers}{{\rm km}\,s^{-1}}
\newcommand{\hinvMpc}{\ h^{-1}{\rm\,Mpc}}

\makeatletter
\renewcommand{\p@subsection}{}
\renewcommand{\p@subsubsection}{}
\makeatother
\newcommand{\mnras}{Monthly Notices of the Royal Astronomical Society}
\newcommand{\apjl}{ApJ Letters}
\newcommand{\aap}{Astronomy and Astrophysics}

\begin{document}
\title{Measuring cosmic bulk flow with kinetic Sunyaev-Zel'dovich \\ velocity reconstruction}
%

\author{Suroor \textsc{Seher Gandhi}}\email[Email: ]{ssehergandhi@pitp.ca}
\affiliation{Perimeter Institute for Theoretical Physics, 31 Caroline St N, Waterloo, ON N2L 2Y5, Canada} 

\author{Matthew C. \textsc{Johnson}}
\affiliation{Perimeter Institute for Theoretical Physics, 31 Caroline St N, Waterloo, ON N2L 2Y5, Canada} \affiliation{Department of Physics and Astronomy, York University, Toronto, ON M3J 1P3, Canada}
\author{Jordan \textsc{Krywonos}}
\affiliation{Perimeter Institute for Theoretical Physics, 31 Caroline St N, Waterloo, ON N2L 2Y5, Canada} \affiliation{Department of Physics and Astronomy, York University, Toronto, ON M3J 1P3, Canada}
\author{Michael J. \textsc{Hudson}}
\affiliation{Department of Physics and Astronomy, University of Waterloo, 200 University Avenue West, Waterloo, Ontario N2L 3G1, Canada} \affiliation{Waterloo Centre for Astrophysics, University of Waterloo, Waterloo, Ontario N2L 3G1, Canada} \affiliation{Perimeter Institute for Theoretical Physics, 31 Caroline St N, Waterloo, ON N2L 2Y5, Canada}
%
\begin{abstract}
Cosmic bulk flow---the volume-averaged peculiar velocity of matter---serves as a fundamental test of the Cosmological Principle when probed on gigaparsec (Gpc) scales. 
Historically, however, measurements of cosmic bulk flow have been limited to $R\lesssim 100\ h^{-1}{\rm Mpc}$.
We present an application of kinetic Sunyaev-Zel'dovich (kSZ) velocity reconstruction to constrain the bulk flow on cosmological scales, over a volume of effective radius $R\sim2000\ h^{-1} {\rm Mpc}$. 
We use the \texttt{WISE}$\times$\texttt{SuperCOSMOS}\ and \unwise\ galaxy catalogs, combined with CMB temperature maps from \planck\ to reconstruct large-scale velocities in six tomographic bins ranging $0.1\lesssim z \lesssim 1.5$. 
We place some of the tightest upper limits to date on bulk velocity at $200 \lesssim R\,[h^{-1}{\rm Mpc}]\lesssim 2000$, finding results fully consistent with the $\Lambda\rm CDM$ bulk flow expectation. 
Our \unwise\ constraints are in strong tension with the \catwise\ quasar number-count dipole measurement if that dipole is due to a coherent bulk flow $\sim 370\ \kmpers$ at $R\sim1000\ h^{-1} {\rm Mpc}$.
We also derive constraints on the matter power spectrum at low-$k$ ($k\lesssim10^{-3}\, {\rm Mpc}^{-1}$) with low-$z$ ($z\sim 1$) galaxy samples.
Alongside these cosmological constraints, we introduce a novel approach to map the optical depth bias---an inherent astrophysical degeneracy in kSZ velocity reconstruction---across different data combinations.
Our work bridges the theoretical gap between bulk flow and kSZ-reconstructed velocities, and expands the horizon of bulk velocity measurements out to Gpc scales.
\end{abstract}
\maketitle

\section{Introduction}\label{sec: intro}
The peculiar velocity field on large scales ($\gtrsim 100\, h^{-1}$ Mpc) is a valuable probe for constraining cosmological models. Large-scale matter and velocity perturbations are well-described by linear theory \cite[e.g.,][]{Sunyaev_Zeldovich_1980, Peebles_1980}, providing a regime where robust comparisons against observations are possible.
One of the fundamental quantities derived from the peculiar velocity field is its volume average, known as bulk flow or bulk velocity. In a homogeneous and isotropic universe, these bulk motions are expected to diminish as the averaging scale increases.
Homogeneity and isotropy on large scales are postulated by the cosmological principle---the tenet underpinning the standard model of cosmology, $\Lambda$CDM. Tracing bulk flow as a function of scale helps assess if the coherent motion of matter converges to the cosmic rest frame, and thus serves as a powerful test of the cosmological principle.

Many previous works have measured bulk flow at different scales (e.g., \cite{Scrimgeour_2015_6dfgs, Planck_2013, Lavaux_2013}; for a recent comprehensive list of references, see \cite{Watkins_2025}). Consistently over the past two decades, a number of such studies have found excess bulk velocities in tension with $\Lambda$CDM predictions at distances $\sim 100$--$300 \hinvMpc$ \cite[e.g.,][]{Kashlinsky_2008, Watkins_2015a, Peery_2018, Whitford_2023, Watkins_2023}. However, consensus has not yet been reached on what the origin of this excess might be---systematics, incomplete modeling, or new physics \cite[for recent discussions of the continued uncertainty, see][]{Whitford_2023, Turner_2024, Watkins_2025}. This persistent ambiguity and its critical implications for the cosmological principle motivate the use of independent probes that can map the velocity field out to cosmological distances $\sim 100$--$1000 \hinvMpc$.

Reaching these scales, however, presents a fundamental challenge. Traditional approaches to mapping out the peculiar velocity field (and hence measuring bulk flow) rely on distances, deriving peculiar velocity from observed redshift $z_{\rm obs}$ and distance $d$ via $v_{\rm pec} \approx c z_{\rm obs} - H_0d$. However, the uncertainties associated with standard distance indicators (e.g., Tully-Fisher \cite{Tully_1977} and Fundamental Plane \cite{Dressler_1987_FP} relatuons, Type Ia Supernovae \cite{Phillips_1993_SNeIa}, Cepheids \cite{Leavitt_1912_cepheids}, Tip of the Red Giant Branch \cite{Lee_1993_TRGB}) scale proportionately with distance, thereby limiting bulk flow measurements to $R\sim 100\ h^{-1}{\rm Mpc}$.

The kinetic Sunyaev-Zel'dovich (kSZ) distortion of the cosmic microwave background (CMB) \cite{Sunyaev_1980_ksz} is precisely the probe that can overcome this limitation. 
The kSZ effect is a \textit{direct}, distance-independent measure of peculiar velocities, scaling only with the electron density and peculiar velocity ($\Theta^{\rm kSZ}\propto n_e\, v_e$). This is a significant advantage that allows the extraction of velocity information at distances far beyond traditional~methods. 

Physically, the kSZ effect arises from the Doppler shifting of CMB photons scattering off the free electron plasma in large-scale structure (LSS), thereby imprinting the remote peculiar velocity field (i.e., the coherent motion of remote plasma relative to the cosmic rest frame) onto the CMB. In this work, we extract the aforementioned remote velocity signal using a technique known as kSZ tomography or kSZ velocity reconstruction \cite[e.g., ][]{Shao_2011, Munshi_2016, Terrana_2017, Deutsch_2018, Smith_2018, Cayuso_2023, Bloch_2024}. This approach leverages the cross-correlation between the CMB and LSS tracers across multiple tomographic redshift bins. 
We implement the velocity reconstruction via the quadratic estimator formalism (developed in Ref. \cite{Deutsch_2018} and refined in Ref. \cite{Bloch_2024}), utilizing galaxies as tracers of the free electron distribution. 
As demonstrated by multiple previous works, kSZ-based velocities are highly capable of probing the homogeneity and isotropy of the universe~\cite[e.g.,][]{Zhang_2010, Zhang_2011, Cayuso_2020}.

\begin{figure}
    \centering
    \includegraphics[width=\linewidth]{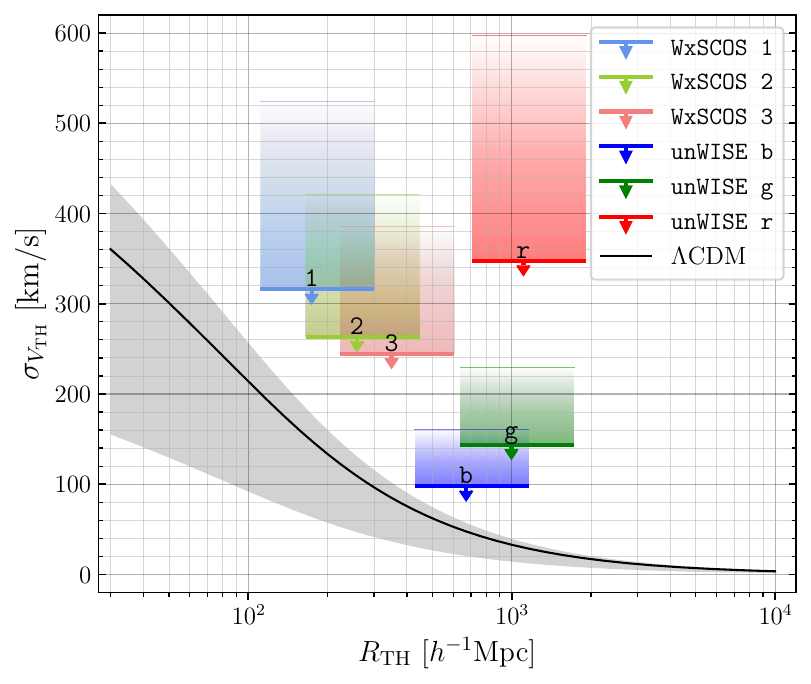}
    \caption{The main results of this work: 1$\sigma$ upper limits (marked by downward arrows) on the variance of bulk velocities as a function of a tophat (`TH') distance scale, derived using kSZ velocity reconstruction across six galaxy subsamples.
    We use two galaxy catalogs, \texttt{WISE}$\times$\texttt{SuperCOSMOS}\ (\texttt{WxSCOS}) \cite{Bilicki_2016wscos} and \unwise\ \cite{Schlafly_2019unwise, Krolewski_2020unwise}, split into three bins each (labeled by $\{\texttt{`1'},\texttt{`2'},\texttt{`3'},\texttt{`b'},\texttt{`g'},\texttt{`r'}\}$ respectively, described in \S\ref{sec: data}).
    The shaded vertical colored regions go up to the 2$\sigma$ upper limit.
    The horizontal extent of the error-bars is indicative of the width of the galaxy subsample window functions.    
    The black curve depicts the theoretical bulk velocity variance expectation in $\Lambda$CDM (Eq.~\eqref{eq: sigmaV} with a tophat window function). The gray shaded region represents the central 68\% probability interval around the theoretical peak velocity in $\Lambda$CDM [Eqs. (\ref{eq: Delta V interval integral}--\ref{eq: gray shaded range})].
    We provide a detailed discussion in \S\ref{sec: bulk flow upper limits}, and a comparison to other bulk flow measurements in the literature in Figure~\ref{fig: comparison plot}.
    }
    \label{fig: vbulk}
\end{figure}

\citet{Krywonos_2024} recently showed the high constraining power of the kSZ velocity reconstruction pipeline of Ref. \cite{Bloch_2024} at low-$\ell$, using the data combination of \unwise\ \blue\ galaxies \cite{Schlafly_2019unwise, Krolewski_2020unwise} and \planck\ CMB temperature maps \cite{Planck18_2020}. 
This demonstration of high sensitivity on large scales with existing datasets motivates expanding the horizon of cosmological observables---like bulk flow---that have typically been distance-limited.
One of the main contributions of our work is to formalize the theoretical link between the dipole ($\ell=1$) of the kSZ-reconstructed velocity field and bulk flow. 
Furthermore, using the galaxy catalogs \texttt{WISE}$\times$\texttt{SuperCOSMOS} (\wscos) \cite{Bilicki_2016wscos} and \unwise\ \cite{Schlafly_2019unwise}, each divided into three subsamples, we constrain bulk flow out to $\sim$Gpc scales (Figure~\ref{fig: vbulk}). 
%
%

Measuring bulk flow with kSZ was initially proposed in Refs. \cite{Haehnelt_1996, Kashlinsky_2000} via the use of galaxy clusters, and was later implemented in multiple studies \cite[e.g.,][]{Kashlinsky_2008, Kashlinsky_2009, Kashlinsky_2010, Osborne_2011, Planck_2013}. There are two important distinctions between these works and the kSZ velocity reconstruction approach to bulk flow used in our work. 
First, while cluster catalogs are limited by sparse sky coverage and low number densities, we utilize galaxy catalogs with high number counts that trace the continuous large-scale structure across large fractions of the sky.
Second, cluster-based bulk flow measurements that do reach $\sim$Gpc scales (e.g., \cite{Planck_2013}) are cumulative: estimates of bulk flow at a given distance $R$ include all clusters in the sample within a sphere of that radius. Consequently, they are heavily weighted by the higher sample density of nearby clusters in the local universe. Whereas kSZ velocity reconstruction tomographically probes the remote bulk flow within a redshift shell, and our upper limits (shown in Figure \ref{fig: vbulk}) map these shell-based velocities to effectively equivalent spheres of radius~$R$. 

Another method of measuring bulk flow with kSZ is based on forward-modeled templates created from galaxy catalogs \cite[e.g.][]{Lavaux_2013}, thereby mitigating the shot noise inherent in cluster samples. 
While templates are adaptable to various velocity models, the kSZ velocity reconstruction approach lends itself to a highly flexible framework for recovering the detailed structure of the velocity field. 
%
Our data combination---\wscos\ and \unwise\ galaxies with \planck\ \cite{Planck18_2020} CMB maps---also represents a significant improvement in sample size and cosmological reach compared to the analysis in \cite{Lavaux_2013} based on 2M++ galaxies \cite{Lavaux_2011_2M++} and WMAP \cite{Jarosik_2011_WMAP7}.

However, utilizing galaxy catalogs for velocity reconstruction (as is done in this work) entails a trade-off between statistical power and astrophysical uncertainty. While the hot gas profiles of massive clusters are relatively well-constrained by X-ray and thermal-SZ observations, the distribution of ionized gas around galaxies is more uncertain and sensitive to baryonic feedback processes. Consequently, while galaxy-based velocity reconstruction dramatically reduces shot noise, it introduces a perfect degeneracy between the amplitude of the velocity field and the uncertain optical depth of the galaxy sample (also known as optical depth degeneracy, and often parameterized as a velocity bias, $b_v$) \cite[e.g., ][]{Battaglia_2016, Flender_2017, Smith_2018, Madhavacheril_2019}. We account for this degeneracy by translating the ACT$\times$DESI measurement of $b_v$ from Ref. \cite{McCarthy:2025brx} into $b_v$-priors applicable to our data combinations of \planck$\times$\wscos\ and \planck$\times$\unwise. We then marginalize over $b_v$ when deriving bounds on bulk flow.
This paper is organized as follows: we provide theoretical background in \S\ref{sec: theory}, including a discussion of idealized window functions and the kSZ velocity reconstruction window function. 
\S\ref{sec: data} provides details of the data products we use. 
In \S\ref{sec: methods}, we describe how we get constraints on bulk flow from kSZ velocity reconstruction. 
We describe our results in \S\ref{sec: results}, place them in context and discuss their implications in \S\ref{sec: discussion}, and conclude in \S\ref{sec: conclusion}.

Appendices \ref{app: vel recon} and \ref{app: galaxy distributions} review aspects of the kSZ velocity reconstruction procedure.
Appendix \ref{app: bv analysis} details how we incorporate data-informed constraints on the optical depth bias.
In Appendix \ref{app: vbulk comparison}, we explain how we translate other existing bulk flow measurements to values that can be consistently compared with our upper limits (summarized in Figure \ref{fig: comparison plot} and Table \ref{tab: bulk_flow_lit}).

\section{Theoretical Background}\label{sec: theory}
\subsection{Bulk Flow Theory}\label{sec: theory_signal}

The bulk velocity, $\bm V$, is a volume-average of the peculiar velocity, defined within some volume usually centered around the Local Group. To calculate this, we begin with the line-of-sight integral over the peculiar velocity field, $\bm v_{\rm pec}(\bm x, \eta)$, weighted by a given window function (writing comoving spatial coordinates as $\bm x = \chi \hvec n$):
\begin{eqnarray}\label{eq: Vbulk definition}
    \bm V(\hvec n) = \int d\chi\ W(\chi\hvec n)\, \bm v_{\rm pec}(\chi\hvec n, \eta(\chi)),
\end{eqnarray}
where $\hvec n$ is the unit vector along the line of sight, $\chi$ is the comoving radial distance, $\eta(\chi)$ is conformal time, and the range of the integral is determined by the window function $W(\chi\hvec n)$.
In theory, $W(\chi\hvec n)$ can be a simple Gaussian or tophat function, and in practice, it can encapsulate survey selection functions, geometric projections, or physically-motivated weights based on galaxy stellar mass, luminosity, etc \citep{Li_2012}.

We can express the peculiar velocity field in terms of its Fourier transform to get
\begin{eqnarray}\label{eq: Vbulk with vpec(k)}
    \bm V(\hvec n) = \int d\chi\ W(\chi\hvec n) \int \frac{d^3\bm k}{(2\pi)^3}\bm v_{\rm pec}(\bm k, \eta(\chi))\, e^{i\bm k\cdot \chi \hvec n}, 
\end{eqnarray}
where, assuming linear perturbation theory, 
\begin{eqnarray}\label{eq: vpec(k) vector}
    \bm v_{\rm pec}(\bm k, \eta(\chi)) = i f(\chi) H(\chi) a(\chi) \frac{\delta_m(\bm k, \eta(\chi))}{k}\hvec k.
\end{eqnarray}
In the above equation, $f(\chi) = d\ln D/d\ln a$ is the growth rate (with $D(a)$ being the linear growth factor of density fluctuations), $H(\chi)$ is the Hubble parameter, $a(\chi)$ is the scale factor, and $\delta_m$ is the matter overdensity field.

The variance of the amplitude of the bulk velocity can be estimated using the variance of Eq.~\eqref{eq: Vbulk with vpec(k)}:
\begin{eqnarray}
    \big\langle \left|\bm V\right|^2 \big\rangle &=& \int d^2\hvec n\int d\chi\, W(\chi\hvec n) \int d^2\hvec n' \int d\chi'\, W(\chi'\hvec n') \nonumber \\
    &&\times \int \frac{d^3\bm k}{(2\pi)^3} e^{i\bm k\cdot\chi\hvec n} \int \frac{d^3\bm k'}{(2\pi)^3} e^{-i\bm k'\cdot\chi'\hvec n'}  \nonumber \\
    &&\times \left\langle \bm v_{\rm pec}(\bm k,\eta(\chi))\cdot \bm v^*_{\rm pec}(\bm k',\eta(\chi')) \right\rangle \\
    %
    \label{eq: V auto-correlation template}
    &\equiv& \int_{-\infty}^{\infty}d\ln k\ \mathcal W^2(k)\ P_{mm}(k),
\end{eqnarray}
where we have used Eq.~\eqref{eq: vpec(k) vector} to write the last line in terms of $P_{mm}(k)$, the matter power spectrum at redshift $z=0$. 
We have introduced an effective window function $\mathcal W(k)$ (which incorporates the angular average over the 3D survey window function), whose explicit form will depend on the technique and survey window used to measure bulk flow. 
In this sense, Eq.~\eqref{eq: V auto-correlation template} acts as a template, which we will return to repeatedly to demonstrate the commonality between measuring bulk flow with standard methods and with kSZ velocity reconstruction. The explicit form of $\mathcal W$ is provided in later sections where we discuss the respective methods [e.g., Eqs.~\eqref{eq: tophat Wf}, \eqref{eq: Gaussian Wf}, \eqref{appeq: ksz Wf any ell}].

\subsection{Comparing measurements of bulk flow: differences in window functions}\label{sec: reconciling diffs in Wfs}
Differences in the specific form of $\mathcal W(k)$ arising from varying survey selections and/or weight-assignment methods can give rise to different estimates of the bulk flow. Importantly, comparison between measurements across surveys and methods (albeit non-trivial) requires matching the effective window functions in a self-consistent way. 
This is a longstanding challenge especially for tracking the evolution of bulk flow as a function of increasing redshift, as multiple surveys need to be patched together to achieve broad redshift coverage.
One way of standardizing comparison is to find a mapping between a survey-specific window function and a theoretical, idealized window function. This simplified approach is the one we adopt. 
More sophisticated approaches exist—such as the minimum variance (MV) estimator developed by \cite{Watkins_2009}, which more rigorously addresses observational mismatches across surveys; we defer such analyses to future work.

We begin this section by briefly discussing two standard theoretical forms of $\mathcal{W}$---spherical top-hat ($\mathcal{W}^{\rm TH}$) and Gaussian ($\mathcal{W}^{\rm G}$)--- that serve as the target (ideal-coverage) windows for many existing bulk flow measurements \cite[e.g.,][]{Watkins_2009, Feldman_2010_MV_ppr2, Li_2012, Scrimgeour_2015_6dfgs, Peery_2018, Watkins_2023}. We then introduce the procedure to measure bulk flow with kSZ velocity reconstruction, which uses a third window function, qualitatively distinct from $\mathcal{W}^{\rm TH}$ or $\mathcal{W}^{\rm G}$. 

We leverage Eq.~\eqref{eq: V auto-correlation template} as a template that fits each of these three cases, allowing us to clearly demonstrate that kSZ can extract the same physical quantity as the other methods of measuring the bulk flow that use $\mathcal W^{\rm TH}$ or $\mathcal W^{\rm G}$ as their standard. In \S\ref{sec: step 2}, we present a method to map between the kSZ effective window function and the idealized $\mathcal{W}^{\rm TH}$ or $\mathcal{W}^{\rm G}$.

\subsubsection{Bulk Flow with Idealized Window Functions}\label{sec: ideal Wf}
Spherical Top-hat (`TH') and Gaussian (`G') window functions [$\mathcal{W}^{\rm TH}(k;R)$ and $\mathcal{W}^{\rm G}(k;R)$] are two examples of idealized window functions that assume isotropic coverage over an associated characteristic scale, $R$. Peculiar velocities on scales smaller than $R$ get averaged over, and those on larger scales are suppressed, thereby effectively computing the bulk velocity within a given volume, coherent on scales $\sim R$.

Explicitly, these two effective window functions take the following forms in Fourier space:%
\footnote{Note we adopt slightly modified definitions of the top-hat (`TH') and Gaussian (`G') window functions by including the $H_0^2 f_0^2 k /(2\pi^2)$ factor, so as to fit the form of Eq.~\eqref{eq: V auto-correlation template}. In more standard notation, Eqs.~(\ref{eq: tophat Wf}, \ref{eq: Gaussian Wf}) can be re-expressed as $\left(\mathcal{W}^{\rm TH,G}(k;R)\right)^2 =  H_0^2 f_0^2 k /(2\pi^2) \left(W^{\rm TH,G}(k;R)\right)^2$. \label{ftnote: window_funcs}}
\begin{eqnarray}
    \label{eq: tophat Wf}
    \left(\mathcal{W}^{\rm TH}(k;R)\right)^2 &\equiv& \frac{H_0^2 f_0^2}{2\pi^2} \, k \left[\frac{3\left(\sin{kR} - kR\cos{kR}\right)}{(kR)^3}\right]^2 \\
    \label{eq: Gaussian Wf}
    \left(\mathcal{W}^{\rm G}(k;R)\right)^2 &\equiv& \frac{H_0^2 f_0^2}{2\pi^2} \, k \exp(-k^2R^2),
\end{eqnarray}
where $H_0$ and $f_0$ are the functions $H(\chi)$ and $f(\chi)$ evaluated at $\chi(z=0)=0$.
Eqs.~\eqref{eq: tophat Wf} or \eqref{eq: Gaussian Wf} can then be used in Eq.~\eqref{eq: V auto-correlation template} to estimate the amplitude of bulk velocity through the variance \cite{Scrimgeour_2015_6dfgs}, %
\begin{eqnarray}\label{eq: sigmaV}
    \sigma_V^2(R) &=& \left\langle |\bm V(R)|^2\right\rangle \nonumber\\
    &\equiv& \int_{-\infty}^\infty d\ln k\ \left(\Weff{\rm{TH,G}}(k;R)\right)^2 P_{mm}(k),
\end{eqnarray}
where $R$ sets the scale over which peculiar velocities are averaged to compute the bulk flow.
Galaxy and galaxy-cluster surveys (which generally have anisotropic and/or incomplete coverage) then use this theoretical estimate of bulk velocity as the benchmark against which to compare their measurements of bulk flow. Multiple methods exist to weight measured peculiar velocities in optimal ways to enable a comparison with theory \cite[e.g.,][]{Kaiser_1988_MLE, Watkins_2009}, the details of which are not directly relevant for this discussion. It is worth emphasizing, however, that such a comparison between theoretical and measured bulk flow (and hence, between idealized and realistic window functions) is a crucial aspect of any study aiming to use bulk flow to test cosmological models. 
\subsubsection{Bulk Flow with the kSZ Velocity Reconstruction Effective Window Function}\label{sec: BF from kszv theory}
The quantity encoded in kSZ-reconstructed velocity maps is the peculiar velocity field projected along the line of sight, averaged over a window function:
\begin{eqnarray}\label{eq: los v}
    v(\hvec n) \equiv \int d\chi\, W(\chi \hvec n)\left(\bm v_{\rm pec}(\chi\hvec n, \eta(\chi)) \cdot \hvec n\right).
\end{eqnarray}
But this is precisely the velocity projected along the line of sight, which can be rewritten as [using Eq.~\eqref{eq: Vbulk definition}],
\begin{eqnarray}\label{eq: Vbulk projected}
    \bm V(\hvec n)\cdot\hvec n = \left(\int d\chi\, W(\chi\hvec n) \bm v_{\rm pec}(\chi\hvec n, \eta(\chi))\right)\cdot \hvec n.
\end{eqnarray}

To then statistically estimate the amplitude of bulk flow, we can compute the variance of Eq.~\eqref{eq: Vbulk projected}, which is related  the velocity angular power spectrum, $C_\ell^{vv}$, as
\begin{eqnarray}\label{eq: sum 2l+1 over 4pi Clvv}
    \left\langle |\bm V(\hvec n)\cdot\hvec n|^2\right\rangle = \sum_\ell \frac{2\ell+1}{4\pi}C_\ell^{vv}. 
\end{eqnarray}

In Appendix~\ref{app: derivation}, we show that Eq.~\eqref{eq: sum 2l+1 over 4pi Clvv} can be written~as
\begin{eqnarray}\label{eq: kszv bulk flow variance}
    &&3\left\langle |\bm V(\hvec n)\cdot\hvec n|^2\right\rangle \nonumber \\
    &&~~~~~~~~~~~~~\equiv \int_{-\infty}^\infty d\ln k\ \sum_\ell\left(\mathcal W^{\rm kSZ}_\ell(k)\right)^2 P_{mm}(k),\ \ \ \ \ \\
    %
    \label{eq: 3(2l+1) over 4pi Clvv}
    &&~~~~~~~~~~~~~= 3 \sum_\ell\frac{2\ell+1}{4\pi}C_\ell^{vv},
\end{eqnarray}
where $\mathcal{W}_\ell^{\rm kSZ}(k)$ is the $\ell$-component of the kSZ effective window function, the explicit form of which is given in Eq.~\eqref{appeq: ksz Wf any ell}. 
We underscore here that Eq.~\eqref{eq: kszv bulk flow variance} is the harmonic-space kSZ velocity reconstruction counterpart of Eq.~\eqref{eq: V auto-correlation template}, and also has the same form as Eq.~\eqref{eq: sigmaV}. The factor of $3$ on the left-hand-side arises because $\langle|\bm V|^2\rangle$ [Eqs.~(\ref{eq: V auto-correlation template}, \ref{eq: sigmaV})] is the variance of the 3D bulk velocity vector, whereas $\langle|\bm V(\hvec n)\cdot\hvec n|^2\rangle$ is the variance of the line-of-sight component alone. The two are related~by 
$$\langle|\bm V|^2\rangle = 3\langle|\bm V(\hvec n)\cdot\hvec n|^2\rangle.$$

Throughout this work, we simplify Eqs.~(\ref{eq: kszv bulk flow variance}, \ref{eq: 3(2l+1) over 4pi Clvv}) by making the approximation that the sum on the right-hand-side is dominated by the $\ell=1$ component,
\begin{eqnarray}\label{eq: kszv bulk flow variance approx}
    3\left\langle |\bm V(\hvec n)\cdot\hvec n|^2\right\rangle &\approx& \int_{-\infty}^\infty d\ln k\ \left(\mathcal W^{\rm kSZ}_1(k)\right)^2 P_{mm}(k),\ \ \ \ \ \\
    \label{eq: 9 over 4pi C1vv}
    &\approx&\frac{9}{4\pi}C_1^{vv}.
\end{eqnarray}
This is a valid assumption in the regime where the window function is much wider than the coherence length of velocities ($\gtrsim 100h^{-1}$Mpc), which is true for the galaxy subsamples we use. 
Figure~\ref{fig: ClvvUL_Nrec} also demonstrates that the theoretical signal $C_\ell^{vv;\rm\lcdm}$ (black curves) is dominated by $\ell=1$; for all galaxy subsamples we use, the theory quadrupole is at least $2\times$ smaller than the dipole.
Although we do not measure a statistically significant detection of the dipole in the kSZ-reconstructed velocity sky-maps, we can obtain an upper limit on it (we discuss how this is done in \S\ref{sec: C1vv upper lim}).%

\section{Data}\label{sec: data}
The process of kSZ velocity reconstruction requires galaxy- and CMB temperature maps.
These two types of data products are then combined to construct velocity~maps. 
%

\subsection{Three \texttt{WISE}$\times$\texttt{SuperCOSMOS} (\wscos) galaxy subsamples}\label{sec: WxSCOS}
We split the $\sim 20 \times 10^6$ unmasked galaxies in the \texttt{WISE}$\times$\texttt{SuperCOSMOS}\ (\wscos) catalog \cite{Bilicki_2016wscos} into three bins labeled \texttt{WxSCOS}\ \texttt{`1'}, \texttt{`2'}, and \texttt{`3'}. Each bin has $\sim 6.8 \times 10^6$ galaxies before masking. After applying the survey mask, \wscos\ \texttt{`1'}, \texttt{`2'}, and \texttt{`3'} have $\sim 6\times 10^6,\ 6.2\times 10^6$, and $6.3\times 10^6$ galaxies, with median redshifts $z_{\rm med} \approx  0.11, 0.19$, and $0.27$, respectively. 
The redshift distributions and $1\sigma$ redshift errors for the three $z$-bins are shown in the top panel of Figure~\ref{fig: bdndz}. 
The sky fraction available with \texttt{WxSCOS} after masking is $f_{\rm sky}\approx 68\%$.

We create galaxy number-count maps, $N_g(\hvec n)$, for these subsamples at NSIDE of $2048$ using \texttt{healpy} \cite{healpy2005, Zonca2019_healpy}, where $N_g$ is the number count per pixel. To these raw number-count maps, we apply empirically-derived corrective weights, $W^{\rm corr}(\hvec n)$, to account for systematics as follows:
$$N^{\rm corr}_g(\hvec n) = W^{\rm corr}(\hvec n) N_g(\hvec n).$$
Then, we convert them to galaxy overdensity maps, 
$$\delta_g(\hvec n) = N^{\rm corr}_g(\hvec n)/\bar N^{\rm corr}_g - 1,$$
where $\bar N^{\rm corr}_g = N^{\rm corr}_{g\,\rm tot}/(f_{\rm sky}N_{\rm pix})$ is the average number of galaxies per unmasked pixel in the corrected number-count map. 
Finally, the galaxy overdensity maps are used in the kSZ velocity reconstruction pipeline [Eq.~\eqref{appeq: quadratic estimator}]. For more details on the corrective weights, see Appendix~\ref{app: vel recon},~\S\ref{sec: corrective wts}.
\subsection{Three \unwise\ galaxy subsamples}\label{sec: unWISE}
We use the three tomographic galaxy subsamples, \unwise\ \cite{Schlafly_2019unwise} \blue, \green, and \red\ created by \citet{Krolewski_2020unwise}, at median redshifts $z_{\rm med} \approx 0.6, 1.1$,  and $1.5$ respectively. After applying the catalog mask, the \blue\ subsample has $\sim 81\times10^6$ galaxies, \green\ has $\sim 44\times10^6$, and \red\ has $\sim 2.5\times 10^6$. 
The redshift distributions and $1\sigma$ redshift errors for the three $z$-bins are shown in the bottom panel of Figure~\ref{fig: bdndz}.
With \unwise, the unmasked sky fraction available is $f_{\rm sky}\approx 58\%$.

As mentioned above for \wscos, we similarly create systematics-corrected galaxy number-count maps, $N_g^{\rm corr}(\hvec n)$, for the \unwise\ subsamples as well. 
Then we generate galaxy overdensity maps [$\delta_g(\hvec n)$] used in the kSZ velocity reconstruction procedure. 
For more details on how the corrective weights are derived, see Appendix~\ref{app: vel recon},~\S\ref{sec: corrective wts}.

\subsection{\planck\ CMB temperature map}
For the CMB temperature map $\Theta(\hvec n)$ needed for kSZ velocity reconstruction, we use the Spectral Matching Independent Component Analysis (\texttt{SMICA}) map from \planck\ \cite{Planck18_2020, Planck_2020_maps}. This map was created using a technique from Ref. \cite{Tegmark_2003_compsepmap} that linearly weights each \planck\ frequency map in harmonic space. The weighting is such that the variance of a desired spectral component is minimized, which in this case is the blackbody spectrum containing the primary CMB and kSZ.

\subsection{kSZ-reconstructed velocity maps}
The derived data product that we use are kSZ-reconstructed line-of-sight velocity maps, $v(\hvec n)$, which we create by combining the corrected galaxy overdensity maps [$\delta_g(\hvec n)$] and the CMB \texttt{SMICA} map [$\Theta(\hvec n)$] mentioned above. 
We build a velocity map corresponding to each of the six galaxy subsamples (\wscos\ \texttt{1}, \texttt{2}, \texttt{3}, \unwise\ \blue, \green, and \red) following the kSZ velocity reconstruction pipeline of Ref. \cite{Bloch_2024}. 
We review their pipeline in Appendix \ref{app: vel recon}, and show the six reconstructed velocity maps in Figure \ref{fig: velocity maps}.
The procedure for deriving bulk flow constraints that we describe next in \S\ref{sec: methods} is applied to each of the six velocity~maps.

\section{Methods: Constraining Bulk Flow from kSZ Velocity Reconstruction}\label{sec: methods}
%
\begin{figure*}
\centering
\begin{tikzpicture}[
    node distance=3.0cm, 
    proc/.style={
        rectangle,
        rounded corners,
        draw=black, very thick,
        inner sep=10pt, 
        align=center, 
        font=\normalsize,
    },
    arrow/.style={
        -{Stealth[scale=1.5]}, 
        thick,
        shorten >=2pt,
        shorten <=2pt,
    },
    label/.style={
        font=\small,
        align=center,
        inner sep=4pt
    }
]


\node[proc] (map) {
    kSZ-reconstructed\\
    velocity map\\ 
    \small{(Figure \ref{fig: velocity maps})}
};

\node[proc, right=2.5cm of map] (c1) {
    $\mathcal C_1^{vv}$
};

\node[proc, right=2.1cm of c1] (amp) {
    $A$\\
    \small{[Eq. \eqref{eq: Aalpha}]}
};

\node[proc, right=2.1cm of amp] (sigma) {
    $\sigma_{V_{\rm TH}}^2(R_{\rm TH})$\\
    \small{[Eq. \eqref{eq: sigma_V sq upper lim line-1}]}
};


\draw[arrow] (map) -- node[label, above] {Constrain map\\ dipole power}
node[label, below] {\S\ref{sec: C1vv upper lim}} (c1);

\draw[arrow] (c1) -- 
    node[label, above] {Eq.~\eqref{eq: C1vv to Ak}} 
    node[label, below] {\S\ref{sec: step 1}}
    (amp);

\draw[arrow] (amp) -- 
    node[label, above] {Eq. \eqref{eq: sigmaV}}
    node[label, below] {\S\ref{sec: step 2}}
    (sigma);

\end{tikzpicture}
\caption{Graphical depiction of the methodology to derive tophat bulk flow variance [$\sigma_{V_{\rm TH}}(R_{\rm TH})$] upper limits from tomographic kSZ reconstructed velocity maps. We implement this procedure for each galaxy subsample. For a detailed description, see~\S\ref{sec: methods}.}
\label{fig: method flowchart}
\end{figure*}
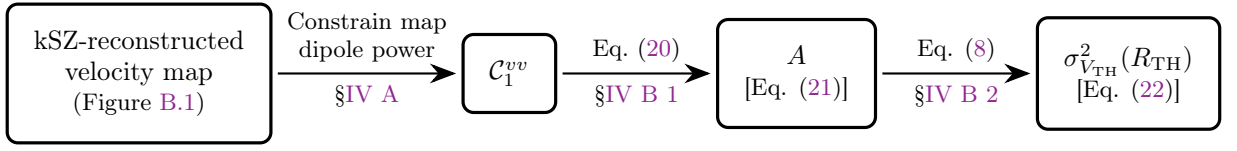

%
In \S\ref{sec: BF from kszv theory}, we explained that the dipole ($\ell=1$) component of a kSZ velocity sky-map captures bulk flow within a survey window. But to place kSZ-derived bulk flow measurements in a cosmological context, we need to compare them with theoretical predictions (e.g., from $\Lambda$CDM).
One way to do this, starting from a constraint on bulk velocity derived from kSZ (that uses \Weffksz{1}), is to place a constraint on the bulk velocity that \textit{would be} observed with an idealized survey window (e.g., \WeffTH or \WeffG).

First, we build velocity maps using the kSZ velocity reconstruction pipeline of Ref. \cite{Bloch_2024} (see Appendix \ref{app: vel recon} for an overview) for our six galaxy subsamples. In this section, we describe how we obtain a 1$\sigma$ upper limit on the full-sky dipole variance, $\mathcal C_1^{vv;1\sigma}$, from the kSZ-reconstructed velocity maps (\S\ref{sec: C1vv upper lim}), followed by our procedure for translating the velocity map dipole to an upper limit on bulk flow variance (\S\ref{sec: procedure}). Figure \ref{fig: method flowchart} gives a graphical summary of our method.

\subsection{Constraint on the full-sky $\mathcal C_1^{vv}$ from the cut-sky kSZ-reconstructed  $\hat C_1^{vv;\rm cut}$} \label{sec: C1vv upper lim}
Our analysis must account for the fact that we do not have access to data that covers the full sky, but rather, only a masked (or cut) fraction of it. We write down the following posterior for the full-sky dipole $\mathcal C_1^{vv}$, given the cut-sky kSZ-reconstructed velocity map dipole $\hat C_1^{vv;{\rm cut}}$ \cite[adapted from Eq. (31) of Ref.][]{Krywonos_2024}: 
\begin{eqnarray}\label{eq: C1vv posterior}
    && P(\mathcal C_1^{vv} | \hat C_1^{vv;\rm cut}) \propto \int db_v\, P(b_v) \int d\phi \int d\theta\, \sin\theta \nonumber \\
    && \times \frac{1}{\mathcal V^{3/2}} \exp\bigg[-\frac{3}{2 \mathcal V}\Big(\hat C_1^{vv;\rm cut} + \mathcal C_1^{vv;\rm cut}(b_v, \mathcal C_1^{vv}) \nonumber \\
    && \ \ \ \ \ \ \ \ \ \ \ \ \ \ \ \ \ \ \ \ \ \ \ \ \ - 2 \sqrt{\hat C_1^{vv;\rm cut} \mathcal C_1^{vv;\rm cut}(b_v, \mathcal C_1^{vv})} \cos\theta \Big) \bigg]. \nonumber \\
\end{eqnarray}
This is a generalized posterior for the full-sky dipole denoted by the placeholder $\mathcal C_1^{vv}$, where $\mathcal C_1^{vv}$ can have two interpretations; it~can be:\\
(\textit{i})~the underlying theoretical velocity variance, in which case we shall set the placeholder $\mathcal C_1^{vv}= C_1^{vv}$ in Eq.~\eqref{eq: C1vv posterior}, or,\\
(\textit{ii})~the velocity dipole realization in our specific universe, such that $\mathcal C_1^{vv}= \hat C_1^{vv}$.\\ 
We will discuss these in more detail below.
In either case, the cut-sky dipole model $\mathcal C_1^{vv;\rm cut}$ is related to the full-sky dipole via
\begin{eqnarray}\label{eq: cut sky dipole model}
    \mathcal C_1^{vv;\rm cut}(b_v, \mathcal C_1^{vv}) = b_v^2 M_{11} \mathcal C_1^{vv},
\end{eqnarray}
where $b_v$ is the optical depth bias parameter, and $M_{11}$ is the $\ell = \ell' = 1$ element of the mode-mode coupling matrix $M_{\ell \ell'}$ \cite{Hivon_2002_master}. This expression holds when the $\ell=1$ component dominates the signal, which is the case for each of the samples considered here (see Fig.~\ref{fig: ClvvUL_Nrec}).

When constraining the underlying theory (i.e., case~(\textit{i}) above), the variance $\mathcal V$ includes both cosmic variance and the cut-sky reconstruction noise, and thus $\mathcal{V} = \mathcal C_1^{vv;\rm cut}(b_v, C_1^{vv;\rm \lcdm}) + N_{\rm rec}^{\rm cut}$. 
The \lcdm\ angular power spectrum is computed as [using Eqs.~(\ref{eq: kszv bulk flow variance approx}, \ref{eq: 9 over 4pi C1vv})]
\begin{eqnarray}\label{eq: lcdm C1vv}
    C_\ell^{vv;\rm \lcdm} = \frac{4\pi}{9}\int_{-\infty}^\infty d\ln k\, \big[\mathcal W^{\rm kSZ}_\ell(k)\big]^2 P^{\rm\lcdm}_{mm}(k),
\end{eqnarray} 
where $\big(\mathcal W^{\rm kSZ}_\ell(k)\big)^2$ is defined in Eq.~\eqref{appeq: ksz Wf any ell}, and $P^{\rm\lcdm}_{mm}(k)$ is computed using \texttt{CAMB} \cite{CAMB_site}.
$N_{\rm rec}^{\rm cut} = f_{\rm sky}N_{\rm rec}$, where $f_{\rm sky}$ is the unmasked sky fraction. The explicit form of the full-sky reconstruction noise $N_{\rm rec}$ is given in Eq.~\eqref{appeq: Nrec}.
Case (\textit{i}) is applicable when we constrain the underlying theoretical bulk flow variance $\sigma_V(R)$ (Figure~\ref{fig: vbulk}, \S\ref{sec: bulk flow upper limits}), and the matter power spectrum $P_{mm}(k)$ (Figure~\ref{fig: Pmm upper limits}, \S\ref{sec: Pmm upper limits}).

On the other hand, to get bulk velocity constraints on the realization of our universe (case~(\textit{ii}) above), the uncertainty comes solely from the reconstruction noise, and thus we set $\mathcal V = N_{\rm rec}^{\rm cut}$ in Eq.~\eqref{eq: C1vv posterior}. This case is applicable when we compare our constraints to other measurements of bulk velocity in our Galactic neighborhood (Figure~\ref{fig: comparison plot},~\S\ref{sec: comparison with other measurements}).

The other components of Eq.~\eqref{eq: C1vv posterior} are as follows:
the dipole of the masked kSZ-reconstructed velocity map, $\hat C_1^{vv;\rm cut}$, is obtained with \texttt{healpy}'s \cite{healpy2005, Zonca2019_healpy} \texttt{anafast} function. 
$P(b_v)$ is a prior on $b_v$ (shown in Figure~\ref{fig: Pbv_results}), informed by the high signal-to-noise $b_v$ measurement of Ref.~\cite{McCarthy:2025brx} that used DESI galaxies and ACT CMB data. 
We derive $P(b_v)$ by translating the DESI$\times$ACT $b_v$ measurement to apply to our galaxy- and CMB-data combinations, i.e., \unwise\ and \texttt{WxSCOS}$\times$\planck. More details on how we implement this translation are in \S\ref{sec: bv disc} and Appendix~\ref{app: bv analysis}.

With the posterior in Eq.~\eqref{eq: C1vv posterior}, we can then find the 1$\sigma$ upper limit on the full-sky dipole, $\mathcal C_1^{vv;1\sigma}$, such that:
\begin{eqnarray}\label{eq: C1vv_ul definition}
    \frac{\int_0^{\mathcal C_1^{vv;1\sigma}} P(\mathcal C_1^{vv} | \hat C_1^{vv;{\rm cut}}) \, d\mathcal C_1^{vv}}{\int_0^\infty P(\mathcal C_1^{vv} | \hat C_1^{vv;{\rm cut}}) \, d\mathcal C_1^{vv}} = 0.68.
\end{eqnarray}
\subsection{Tophat bulk flow variance $\sigma_{V_{\rm TH}}^2(R_{\rm TH})$ from the full-sky dipole $\mathcal C_1^{vv}$}\label{sec: procedure}
In this section, we present our methodology for deriving a constraint on the tophat (`TH') bulk flow variance, $\sigma^2_{V_{\rm TH}}(R_{\rm TH})$ from the full-sky dipole $\mathcal C_1^{vv}$ (which we constrain in \S\ref{sec: C1vv upper lim}).
Leveraging the fact that the $k$-range probed by our kSZ velocity maps is in the regime where the matter power spectrum is linear in $k$, we parametrize $P_{mm}(k) = A k$ (the Harrison-Zel'dovich power spectrum). With this parametrization, we implement the following two-step procedure (for each galaxy survey subsample): 
\begin{enumerate}
    \item Having constrained $\mathcal C_1^{vv}$,
    we infer the power spectrum amplitude, $A$,
    by setting $P_{mm}(k) = A\,k$
    in Eq.~\eqref{eq: kszv bulk flow variance approx}.
    \item With the amplitude $A$
    obtained above, we compute $\sigma_{V_{\rm TH}}^2(R_{\rm TH})$ 
    using Eq.~\eqref{eq: sigmaV}, where $R_{\rm TH}$ 
    is the tophat distance scale that effectively matches the $k$-sensitivity of $\mathcal W^{\rm kSZ}_1(k)$.
\end{enumerate}
We discuss these two steps in more detail below. 

\subsubsection{Power spectrum amplitude $A$ 
from the dipole $\mathcal C_1^{vv}$}\label{sec: step 1}
This step closely resembles what is done in \citet[][see their \S II A]{Tegmark_2002} to convert the measured CMB temperature anisotropy angular power spectrum $\hat C_\ell^{TT}$ into constraints on the linear matter power spectrum $P_{mm}(k)$.
For kSZ velocity reconstruction, the analogous angular power spectrum is $\mathcal C_\ell^{vv}$, and we further fix $\ell=1$. 

With the linear parametrization $P_{mm}(k)= Ak$, Eqs.~\eqref{eq: kszv bulk flow variance approx} and \eqref{eq: 9 over 4pi C1vv} can be combined as
\begin{eqnarray}\label{eq: C1vv to Ak}
    \frac{9}{4\pi}\mathcal C_1^{vv} = A \int_{-\infty}^\infty d\ln k\, \big(\mathcal W_1^{\rm kSZ}(k)\big)^2\,k,
\end{eqnarray}
and then rearranged to get the matter power spectrum amplitude
\begin{eqnarray}\label{eq: Aalpha}
A \equiv \frac{\frac{9}{4\pi}\, \mathcal C_1^{vv}}{\mathcal N^{{\rm kSZ}}}
\end{eqnarray}
where $\mathcal{N}^{\rm kSZ}$ is the $\ell=1$ kSZ effective window function normalization [i.e., the integral in Eq.~\eqref{eq: C1vv to Ak}].

\subsubsection{$\sigma_{V_{\rm TH}}^2(R_{\rm TH})$ constraint from the matter power spectrum amplitude $A$}\label{sec: step 2}
We use our estimate of the matter power spectrum amplitude from Eq.~\eqref{eq: Aalpha} in Eq.~\eqref{eq: sigmaV} to get
\begin{eqnarray}
    \label{eq: sigma_V sq upper lim line-1}
    \sigma_{V_{\rm TH}}^2(R_{\rm TH}) &=& A \int_{-\infty}^\infty d\ln k\ \big(\mathcal W^{\rm TH}(k; R_{\rm TH})\big)^2 k\ \ \  \\
    \label{eq: sigma_V sq upper lim line-2}
    &\equiv& \frac{9}{4\pi}\, \mathcal C_1^{vv}\, \frac{\mathcal N^{\rm TH}}{\mathcal N^{\rm kSZ}},
\end{eqnarray}
where $\mathcal N^{\rm TH}$ is the tophat window function normalization, given by the integral in Eq.~\eqref{eq: sigma_V sq upper lim line-1}.
We have chosen the spherical top-hat \WeffTH\ with a smoothing scale $R_{\rm TH}$ as our idealized window function (as opposed to \WeffG). 

Eq.~\eqref{eq: sigma_V sq upper lim line-2} gives the bulk flow upper limits as shown on the vertical axis of Fig.~\ref{fig: vbulk}. To get the horizontal axis coordinate, we need to find $R_{\rm TH}$.
$R_{\rm TH}
$ has to be chosen such that $\big(\WeffTH(k;R_{\rm TH})\big)^2$ is most strongly peaked in the same $k$-range as $\big(\mathcal W^{{\rm kSZ}}_1(k)\big)^2$, as this effectively maps the two weighting functions onto each other. 
There is not a unique way to do so; the convention we follow (adapted from Ref. \cite{Tegmark_2002}) is to find the $R_{\rm TH}
$ that generates a $k$-profile of $\big(\WeffTH(k;R_{\rm TH})\big)^2$ whose 50th percentile $k^{\rm TH}_{50}$ matches the 50th percentile $k_{50}^{\rm kSZ}$ of $\big(\mathcal W^{{\rm kSZ}}_1(k)\big)^2$.
Since we know $\mathcal W_1^{\rm kSZ}$ [Eq.~\eqref{appeq: W1 ksz final}], we can find $k_{50}^{\rm kSZ} = k_{50}^{\rm TH} \equiv k_{50}$. 
Then, obtaining $R_{\rm TH}$ requires scanning a range of tophat smoothing scales while holding fixed the normalized cumulative distribution function (CDF) of the effective tophat window [Eq.~\eqref{eq: tophat Wf}] to be 0.5.
After changing variables to $\kappa\equiv kR$, this amounts to solving the following equation for the median $\kappa_{50} \equiv k_{50}R_{\rm TH}$:
\begin{eqnarray}\label{eq: Reff transcendental eq}
     {\rm CDF}(\kappa_{50}) = \frac{\int_{-\infty}^{\ln \kappa_{50}} d\ln \kappa\ \big(W^{\rm TH}(\kappa)\big)^2\,\kappa^2}{\int_{-\infty}^{\infty} d\ln \kappa\ \big(W^{\rm TH}(\kappa)\big)^2\,\kappa^2} = 0.5,
\end{eqnarray}
where
\begin{eqnarray}\label{eq: WTH kappa}
    W^{\rm TH}(\kappa) \equiv \frac{3(\sin \kappa - \kappa\cos \kappa)}{\kappa^3}
\end{eqnarray}
is the Fourier transform of the real-space spherical tophat window function. 
(Note that $\mathcal W^{\rm TH}(k;R) = H_0 f_0 \sqrt k\big/\big(\sqrt 2\pi\big)\,W^{\rm TH}(\kappa)$---see footnote \ref{ftnote: window_funcs}.) The solution to Eq.~\eqref{eq: Reff transcendental eq} gives 
\begin{eqnarray}\label{eq: Reff soln}
    R_{\rm TH}
    \approx 1.738/k_{50}.
\end{eqnarray}
\begin{figure}
    \centering
    \includegraphics[width=\linewidth]{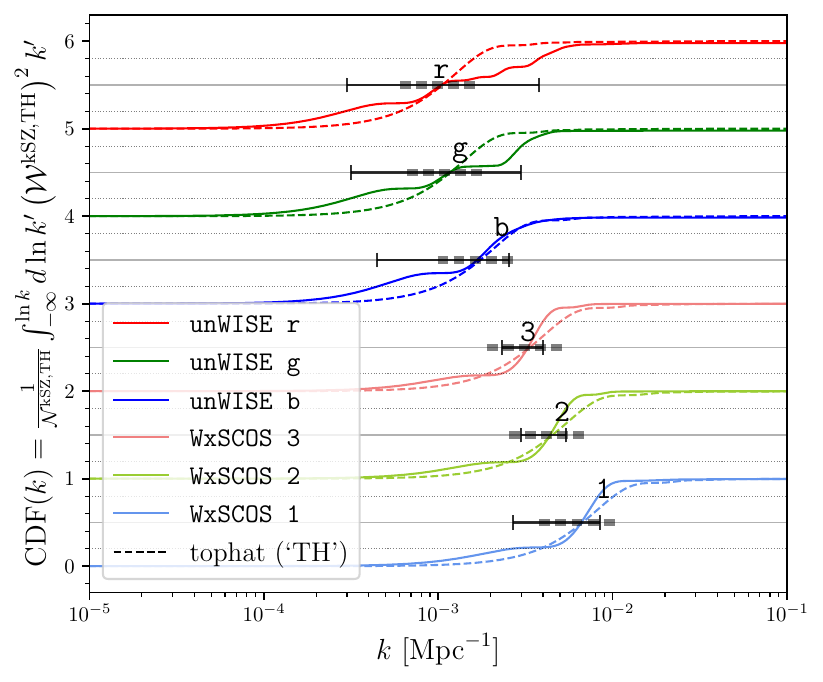}
    \caption{The cumulative distribution functions (CDFs) of normalized $\big(\mathcal W^{{\rm kSZ}}_1(k)\big)^2$ (solid curves) and $\big(\WeffTH(k;R_{\rm TH})\big)^2$ (dashed curves), relevant for mapping the kSZ effective window function onto an equivalent spherical top-hat window (see \S\ref{sec: step 2}). 
    $R_{\rm TH}$ is the smoothing scale of the spherical top-hat window function, chosen such that the 50th percentile of $\big(\WeffTH(k;R_{\rm TH})\big)^2$ matches the 50th percentile of $\big(\mathcal W^{{\rm kSZ}}_1(k)\big)^2$. 
    We do this for the six galaxy subsamples we use (labeled by the tags $\{\texttt{`1'},\texttt{`2'},\texttt{`3'},\texttt{`b'},\texttt{`g'},\texttt{`r'}\}$), the details of which are provided in \S\ref{sec: data}. 
    The CDFs have been vertically offset for clarity.
    To guide the eye, the 20th, 50th, and 80th percentile levels for each galaxy subsample are indicated by light gray horizontal lines.
    The goodness of the mapping from the kSZ to tophat window functions can be gauged by the match between the black and dark gray horizontal line segments. These segments delineate the $k$-band corresponding to the 20th--80th percentile range of the window functions for kSZ (solid, black with caps) and tophat (thick dashed, dark gray).
}
    \label{fig: window func cdfs}
\end{figure}

In Figure \ref{fig: window func cdfs}, we show six normalized CDFs of $\big(\mathcal W^{{\rm kSZ}}_1(k)\big)^2$ (solid curves) for the six galaxy subsamples used in this work. The CDFs have been vertically offset for clarity.
The dashed curves are the corresponding tophat CDFs of $\big(\WeffTH(k;R_{\rm TH})\big)^2$. 
Since the tophat smoothing scale $R_{\rm TH}$ is chosen such that $k^{\rm TH}_{50} = k_{50}^{\rm kSZ}$, the dashed and solid curves for each galaxy subsample coincide when the CDFs cross the 50th percentile mark. 

The goodness of the mapping from the kSZ to tophat window functions can be gauged by the match between the black and dark gray horizontal line segments in Figure \ref{fig: window func cdfs}. 
%

%
\section{Results}\label{sec: results}
The kSZ-reconstructed velocity maps obtained for the masked galaxy subsamples (Figure \ref{fig: velocity maps}) yield the cut-sky angular power spectrum, $\hat C_\ell^{vv;\rm cut}$. 
We find $\hat C_\ell^{vv;\rm cut}$ to simply be consistent with the expected cut-sky reconstruction noise, $N^{\rm cut}_{\rm rec} = f_{\rm sky} N_{\rm rec}$ (Figure \ref{fig: ClvvUL_Nrec} demonstrates this). 
Broad consistency with noise means no signal is detected in the reconstructed $\hat C_\ell^{vv;\rm cut}$, and consequently, we place upper limits using the non-detection. 
%
%

%
\subsection{Theoretical bulk flow variance upper limits (Figure~\ref{fig: vbulk})}\label{sec: bulk flow upper limits}
To constrain the underlying theoretical bulk flow variance, we first need a constraint on $C_1^{vv}$. Thus, we set the placeholder $\mathcal C_1^{vv} = C_1^{vv}$ and $\mathcal V = \mathcal C_1^{vv;\rm cut}(b_v, C_1^{vv;\rm \lcdm}) + N_{\rm rec}^{\rm cut}$ in Eq.~\eqref{eq: C1vv posterior}. 
This corresponds to constraining the underlying theoretical dipole, and thus $\mathcal V$ includes cosmic variance (case (\textit{i}) discussed in \S\ref{sec: C1vv upper lim}). 
With the posterior in Eq.~\eqref{eq: C1vv posterior} adapted as such, we obtain a $1\sigma$ upper limit $C_1^{vv;1\sigma}$ using Eq.~\eqref{eq: C1vv_ul definition}.
Finally, with $\mathcal C_1^{vv}\rightarrow C_1^{vv;1\sigma}$ in Eq.~\eqref{eq: sigma_V sq upper lim line-2}, we obtain 1$\sigma$ upper limits on the tophat bulk flow variance, $\sigma^{1\sigma}_{V_{\rm TH}}$, for six galaxy subsamples. 

Figure~\ref{fig: vbulk} shows these six upper limits as a function of the tophat scale, comprising the main results of this work. 
Each downward arrow in the figure is plotted at its respective smoothing scale $R_{\rm TH}$ of the tophat (`TH') window function that we find to be effectively equivalent to the kSZ window function [see discussion of Eq.~\eqref{eq: Reff soln}].
The width of the horizontal line segment attached to each arrow indicates the 20th--80th percentile width of the respective tophat window. 
Specifically, the left edge (minimum) $R_{\rm min} = R_{\rm TH}\ k_{50}/k^{{\rm TH}}_{80}$, and the right edge (maximum) $R_{\rm max} = R_{\rm TH}\ k_{50}/k_{20}^{{\rm TH}}$. 
The $k$-values $k^{{\rm TH}}_{20}$, $k_{50}$, and $k^{{\rm TH}}_{80}$, are where, in Figure \ref{fig: window func cdfs}, the tophat CDF (dashed) crosses the 20th, 50th, and 80th percentiles, respectively.
The shaded vertical colored bands in Figure~\ref{fig: vbulk} extend up to the 2$\sigma$ upper~limit.

The trend in upper limits seen in Figure~\ref{fig: vbulk} is governed by Eq.~\eqref{eq: sigma_V sq upper lim line-2}. 
The \wscos\ upper limits get tighter (lower) from bin \texttt{`1'}$\rightarrow$bin \texttt{`3'} driven by $C_1^{vv;1\sigma}$ getting progressively smaller, as reconstruction noise decreases (compare the gray horizontal lines across the top panels of Figure~\ref{fig: ClvvUL_Nrec}).
%
The \unwise\ upper limits get weaker (higher) from \texttt{`b'}$\rightarrow$\texttt{`g'}$\rightarrow$\texttt{`r'} dominantly because $C_1^{vv;1\sigma}$ increases significantly due to higher reconstruction noise (compare the gray horizontal lines across the bottom panels of Figure~\ref{fig: ClvvUL_Nrec}). This is because $N_{g\,\rm tot}$ decreases and shot noise ($N_{\rm shot} = 4\pi f_{\rm sky}/N_{g\,\rm tot}$) gets higher.

\begin{figure}
    \centering
    \includegraphics[width=\linewidth]{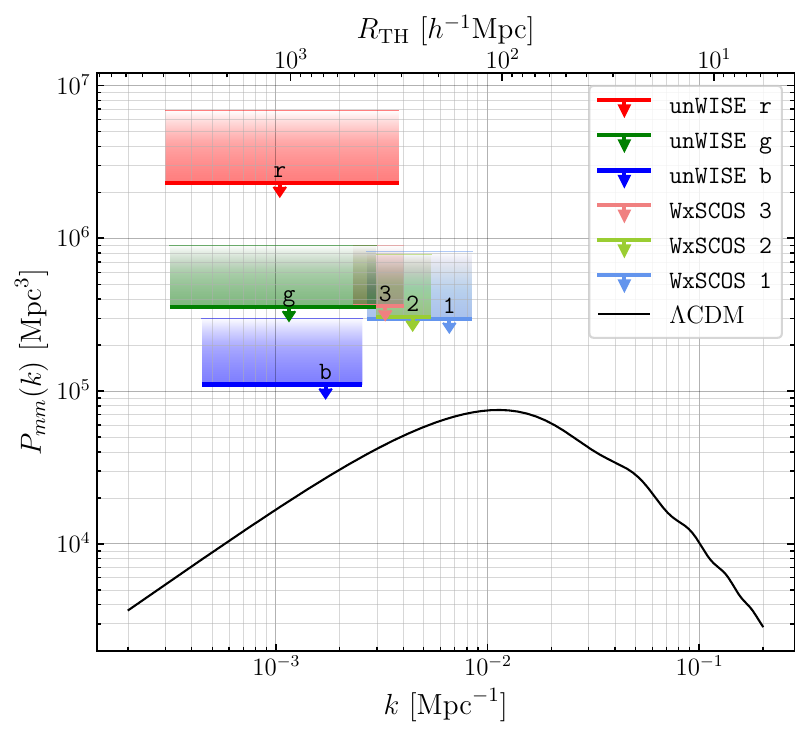}
    \caption{The 1$\sigma$ upper limits (marked by downward arrows) on the matter power spectrum $P_{mm}(k)$ today, derived using Eq.~\eqref{eq: Pmm upper lim} for the six galaxy subsamples used in this work. 
    The vertical shaded colored bands extend up to the 2$\sigma$ level.
    Where each downward arrow falls along the bottom horizontal axis ($k_{50}$) corresponds directly to median of each effective window function, i.e., where the respective CDF in Figure \ref{fig: window func cdfs} crosses the 50th percentile mark.
    The width of each upper limit corresponds to the $k$-range spanning the 20th--80th percentiles of each kSZ effective window function.
    To help map between this figure and Figure \ref{fig: vbulk}, the top horizontal axis is the corresponding smoothing scale of the spherical tophat effective window function, $R_{\rm TH}\ [h^{-1}{\rm Mpc}] \approx 1.738\, h/(k\ [{\rm Mpc}^{-1}])$, obtained by solving Eq.~\eqref{eq: Reff transcendental eq}.
    The black curve is the $\Lambda$CDM matter power spectrum.
    %
    }
    \label{fig: Pmm upper limits}
\end{figure}

For comparison to theory, we also show the $\Lambda$CDM expectation for bulk flow, $\sigma_{V_{\rm TH}}^{\rm\lcdm}$ (black curve). This \lcdm\ theoretical expectation comes from Eq. \eqref{eq: sigmaV}, computed with a tophat window function, and with the matter power spectrum from \texttt{CAMB} \cite{CAMB_site}.
The gray shaded region around the black curve encompasses 68\% of the Maxwellian probability distribution of velocities around its peak value, $V_{p}=\sqrt{2/3}\sigma^{\rm\lcdm}_{V_{\rm TH}}$. To find the interval covered by the gray region we require \cite{Li_2012},
\begin{eqnarray}\label{eq: Delta V interval integral}
    \int_{V_p-\Delta V}^{V_p+\Delta V} p(V) dV = 0.68,
\end{eqnarray}
where the Maxwellian distribution of velocities with a variance $\sigma_V$ is \cite{Bahcall_1994}
\begin{eqnarray}\label{eq: p(V)dV}
    p(V) dV &=& \sqrt{\frac{2}{\pi}}  \left( \frac{3}{\sigma_{V}^2} \right)^{3/2} V^2 \exp\left( -\frac{3 V^2}{2\sigma_V^2} \right) dV.
\end{eqnarray}
This results in the grey shaded range in Figure~\ref{fig: vbulk}
\begin{eqnarray}\label{eq: gray shaded range}
    [V_p-\Delta V,\ V_p+\Delta V] \approx [0.43\sigma^{\rm\lcdm}_{V_{\rm TH}}, 1.20 \sigma^{\rm\lcdm}_{V_{\rm TH}}].
\end{eqnarray}
\subsection{Matter power spectrum upper limits (Figure~\ref{fig: Pmm upper limits})}\label{sec: Pmm upper limits}
A by-product of our procedure (described in \S\ref{sec: procedure}) is that we can also constrain the linear matter power spectrum on large scales. 
To do this, we first need to constrain $C_1^{vv}$. We set the placeholder $\mathcal C_1^{vv} = C_1^{vv}$ and $\mathcal V = \mathcal C_1^{vv;\rm cut}(b_v, C_1^{vv;\rm \lcdm})+N_{\rm rec}^{\rm cut}$ in Eq.~\eqref{eq: C1vv posterior}. 
Once again, as for $\sigma_{V_{\rm TH}}$ constraints above, this corresponds to constraining the underlying theoretical dipole (i.e., case~(\textit{i}) discussed in \S\ref{sec: C1vv upper lim}), and thus $\mathcal V$ includes cosmic variance.
With Eq.~\eqref{eq: C1vv posterior} adjusted to be the appropriate posterior, we get the $1\sigma$ upper limit $C_1^{vv;1\sigma}$ using Eq.~\eqref{eq: C1vv_ul definition}.
For each of the six galaxy catalog bins, we set $\mathcal C_1^{vv} \rightarrow C_1^{vv;1\sigma}$ in Eq.~\eqref{eq: Aalpha} to get an upper limit on the amplitude of the matter power spectrum, $A^{1\sigma}$. Then, we finally compute 
\begin{eqnarray} \label{eq: Pmm upper lim}
    P^{1\sigma}_{mm}(k_{50}) = A^{1\sigma}\, k_{50},
\end{eqnarray}
where $k_{50}$ is the $k$-value corresponding to the 50th percentile of $\big(\mathcal W_1^{\rm kSZ}(k)\big)^2$ (the same as the 50th percentile of $\big(\WeffTH(k;R_{\rm TH})\big)^2$). 

In Figure \ref{fig: Pmm upper limits}, the downward arrows mark the 1$\sigma$ upper limits, $P^{1\sigma}_{mm}(k_{50})$ derived using Eq. \eqref{eq: Pmm upper lim}.
The width of the horizontal line segment on each downward arrow corresponds to the 20th--80th percentile range of the respective kSZ window function. 
Specifically, the line segment extends from the left edge (minimum) $k_{20}^{{\rm kSZ}}$ to the right edge (maximum) $k_{80}^{{\rm kSZ}}$. 
The values of $k_{20}^{{\rm kSZ}},\, k_{50},$ and $k_{80}^{{\rm kSZ}}$ are where, in Figure \ref{fig: window func cdfs}, the kSZ CDFs (solid curves) reach the 20th, 50th, and 80th percentiles, respectively.
The vertical shaded colored bands extend up to the 2$\sigma$ upper limit.
Also shown for comparison with theory is the $\Lambda$CDM matter power spectrum at redshift $z=0$ (black curve), computed with \texttt{CAMB} \cite{CAMB_site}.

The trend seen in the upper limits in Figure~\ref{fig: Pmm upper limits} is determined largely by Eq.~\eqref{eq: Aalpha}. 
The \wscos\ upper limits get weaker (higher) as we go from bin \texttt{`1'}$\rightarrow$bin \texttt{`3'} dominantly because $\mathcal N^{\rm kSZ}$ gets smaller. The decrease in $\mathcal N^{\rm kSZ}$ is stronger than the slight decrease in $C_1^{vv;1\sigma}$ from bin \texttt{`1'}$\rightarrow$bin \texttt{`3'}.
%
The \unwise\ upper limits get weaker (higher) from \texttt{`b'}$\rightarrow$\texttt{`g'}$\rightarrow$\texttt{`r'}, too, but driven most strongly by the increase in $C_1^{vv;1\sigma}$ due to the reconstruction noise getting higher (compare the gray horizontal lines across the lower panels of Figure~\ref{fig: ClvvUL_Nrec}), as $N_{g\,\rm tot}$ decreases and shot noise gets higher. 

We recognize that the window of scales probed by the \wscos~\texttt{1} sample is close to $k_{\rm eq}\sim 10^{-2}\ {\rm Mpc}^{-1}$, where the theoretical parametrization $P_{mm}(k)=A\,k$ begins to break down (turnover of the black curve in Figure \ref{fig: Pmm upper limits}). 
A more accurate treatment in this $k$-regime would be to constrain the amplitude $\tilde A$ of a theoretical transfer function, such that $P_{mm}(k)=\tilde A\,T(k)$. 
However, we do not expect the upper limits for \wscos~\texttt{1} to change by much with this modified parametrization. This is because even though the $\tilde A^{1\sigma}$ inferred from a modified Eq.~\eqref{eq: Aalpha} would increase (as $\mathcal N^{\rm kSZ}$ would decrease), the inferred $P_{mm}^{1\sigma}(k_{50}) = \tilde A^{1\sigma} T(k_{50})$ would in turn be suppressed, thereby keeping the upper limit largely unchanged.
The upper limit for $\sigma_{V_{\rm TH}}(R_{\rm TH})$ is even more robust against the parametrization of the matter power spectrum (as long as $\mathcal W^{\rm TH}$ and $\mathcal W^{\rm kSZ}$ probe similar $k$-ranges) due to the ratio of normalizations, $\mathcal N^{\rm TH}/\mathcal N^{\rm kSZ}$, in Eq.~\eqref{eq: sigma_V sq upper lim line-2}.
%
\section{Discussion}\label{sec: discussion}
\begin{figure}
    \centering
    \includegraphics[width=\linewidth]{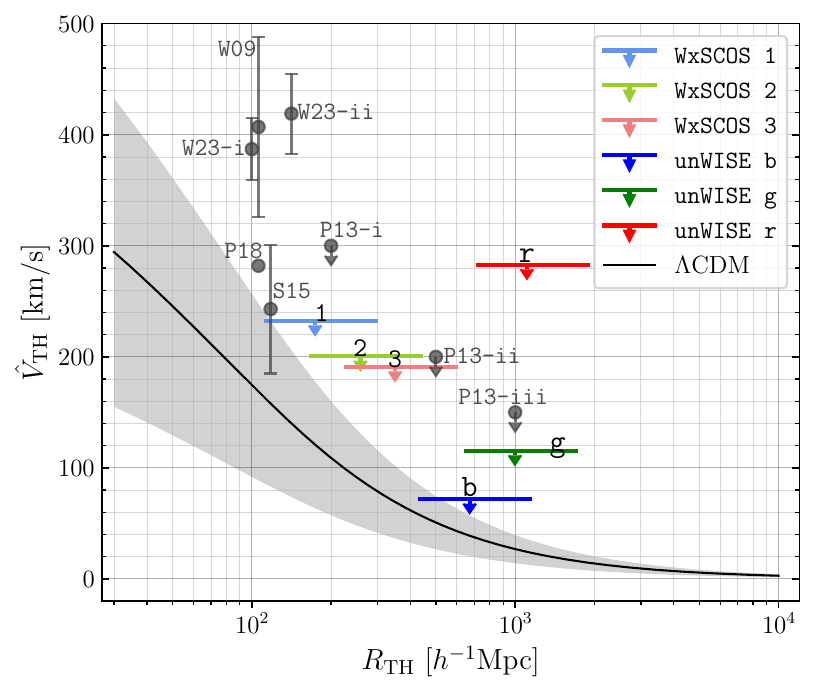}
    \caption{A visual comparison of various 1$\sigma$ measurements and upper limits of bulk flow in the literature (gray circles with error bars) with our kSZ-derived 1$\sigma$ upper limits (marked by colored downward arrows).
    To enable self-consistent comparison with the bulk flow realization ($\hat V$) constraints in the literature, we convert our kSZ-derived tophat $\hat\sigma_{V_{\rm TH}}$ upper limits to $\hat V_{\rm TH}=\sqrt{2/3}\,\hat\sigma_{V_{\rm TH}}$.
    We also translate the various distance scales the other works probe to an equivalent tophat scale, $R_{\rm TH}$. 
    How we do this is described in Appendix~\ref{app: vbulk comparison}.
    The measurements quoted here are from: 
    \citet{Watkins_2023} (`W23-\textit{i}--\textit{ii}'),
    \citet{Watkins_2009} (`W09'),
    %
    %
    \citet{Peery_2018} (`P18'), 
    \citet{Scrimgeour_2015_6dfgs} (`S15'), and
    \citet{Planck_2013} (`P13-\textit{i}--\textit{iii}'). 
    (Error bars are omitted for `P18', as they report a 3D $\chi^2$ rather than the marginalized error on the amplitude $\hat V$.)
    The original values published by these works and their equivalents plotted here are compiled in Table \ref{tab: bulk_flow_lit}.
    There is some tension between our upper limit from \wscos~\texttt{1} and the `W23-\textit{i}--\textit{ii}' \cite{Watkins_2023} and `W09' \cite{Watkins_2009} measurements at $R_{\rm TH}\sim 100\ h^{-1}$Mpc.
    Our upper limits from \unwise\ \blue\ and \green\ are among the tightest in the literature at $\sim$Gpc~scales.
    }
    \label{fig: comparison plot}
\end{figure}

\subsection{Comparison with other bulk flow realization measurements (Figure~\ref{fig: comparison plot})}\label{sec: comparison with other measurements}
Direct comparison with the raw bulk flow measurements of other studies is challenging due to the highly varied survey selection functions and weighting schemes used. 
However, as suggested in Ref. \cite{Li_2012}, we can use the theoretical bulk flow expectation as a standardizing bridge. This expectation (i.e., $\sigma_V$ in Eq.~\eqref{eq: sigmaV} computed with an idealized window function) is the baseline to which studies compare their results.
For instance, if a study reports a bulk velocity measurement compared against a theoretical expectation $\sigma_{V_G}(R_{\rm G})$ based on a Gaussian window function of scale $R_{\rm G}$, we can map this onto an equivalent tophat smoothing scale, $R_{\rm TH}$, by requiring $\sigma_{V_{\rm TH}}(R_{\rm TH})=\sigma_{V_{\rm G}}(R_{\rm G})$. 
The way we map the different scales onto $R_{\rm TH}$ is described in Appendix~\ref{app: vbulk comparison}, and summarized in Table~\ref{tab: bulk_flow_lit}.

In Figure \ref{fig: comparison plot}, we collect several measurements of bulk velocity in the literature, having mapped their respective scales to an equivalent $R_{\rm TH}$. 
All works quoted in this figure constrain the bulk flow realization $\hat V$, which we plot as-is.
To make a comparison with our constraints, we find the tophat equivalent of our kSZ-derived upper limits on the realization-specific bulk flow of our universe, $\hat V_{\rm TH} = \sqrt{2/3}\,\hat \sigma_{V_{\rm TH}}$. 
We use this relationship between the bulk flow $\hat V$ and its variance $\hat \sigma_V$ because for a Maxwellian distribution of velocities Eq.~\eqref{eq: p(V)dV}, the most likely (or peak) velocity $V_p$ is related to the variance as $V_p = \sqrt{2/3}\,\sigma_V$.
To first get $\hat \sigma_{V_{\rm TH}}$, we derive constraints on the realization dipole power, $\hat C_1^{vv}$, by setting the placeholder $\mathcal C_1^{vv}=\hat C_1^{vv}$ and $\mathcal V = N_{\rm rec}^{\rm cut}$ in Eq.~\eqref{eq: C1vv posterior} (case~(\textit{ii}) discussed in \S\ref{sec: C1vv upper lim}).
With the posterior from Eq.~\eqref{eq: C1vv posterior} appropriately adapted to constrain a realization (i.e., no cosmic variance included in $\mathcal V$), we get the $1\sigma$ upper limit $\hat C_1^{vv;1\sigma}$ with Eq.~\eqref{eq: C1vv_ul definition}. 
Finally, using Eq.~\eqref{eq: sigma_V sq upper lim line-2} with $\mathcal C_1^{vv}\rightarrow\hat C_1^{vv;1\sigma}$ gives the requisite $1\sigma$ bulk flow realization upper limit 
\begin{eqnarray}\label{eq: hat V_TH realization}
    \hat V_{\rm TH}^{1\sigma} = \sqrt{2/3}\,\hat \sigma^{1\sigma}_{V_{\rm TH}}.
\end{eqnarray}
This is where we plot our upper limits in Figure~\ref{fig: comparison plot} (colored downward arrows). 

For comparison of all bulk flow realization constraints against underlying theoretical bulk flow, we also plot the \lcdm\ bulk flow expectation (black curve) $V_{\rm TH}^{\rm\lcdm} = \sqrt{2/3}\,\sigma_{V_{\rm TH}}^{\rm \lcdm}$, where $\sigma_{V_{\rm TH}}^{\rm \lcdm}$ is Eq.~\eqref{eq: sigmaV} computed with a tophat window and the \lcdm\ matter power spectrum from \texttt{CAMB} \cite{CAMB_site}. 
The grey shaded region is the $1\sigma$ credible interval of the theoretical bulk flow given by Eqs.~(\ref{eq: Delta V interval integral}--\ref{eq: gray shaded range}).

We note that there is $\sim 2\sigma$ tension between our \wscos~\texttt{1} upper limit and the `W09' \cite{Watkins_2009} and `W23-\textit{i}--\textit{ii}' \cite{Watkins_2023} measurements at $R_{\rm TH}\sim 100 h^{-1}$Mpc. 
Since we constrain the tophat bulk flow realization, our upper limits are weaker than what a more rigorous method of comparing window functions (e.g., the Minimum Variance method mentioned earlier) would yield. Thus, we anticipate that the current upper limits are a conservative indicator of the tension at $\sim 100 h^{-1}$Mpc, and the true tension might be stronger.
Furthermore, with \wscos\ \texttt{`1'}--\texttt{`3'}, we do not find evidence of the large coherent bulk flows of 600-1000 $\kmpers$ out to $\sim 300\, h^{-1}$~Mpc as measured by Refs.~\cite{Kashlinsky_2008, Kashlinsky_2009}.
Lastly, our upper limits from \unwise~\blue\ and \green\ are among the tightest in the literature at $\sim$Gpc scales. 
\subsection{Velocity components $\hat V_x,\, \hat V_y,\, \hat V_z$ (Table~\ref{tab: Vxyz})}\label{sec: Vxyz}
We also constrain the velocity components $\hat V_i$ ($i\in\{x, y, z\}$) in Galactic Cartesian coordinates using the cut-sky dipole power $\hat C_1^{vv;\rm cut}$ measured from our kSZ-reconstructed velocity maps. 
To do so, we create a spatial template for each of the three orthogonal directions, corresponding to the Cartesian components of the radial unit vector $\hvec{n}$, defined as:
\begin{align}
\hat n_x &= \sqrt{\frac{2\pi}{3}} \left( Y_{1,-1} - Y_{1,1} \right) & (l=0^\circ, b=0^\circ) \nonumber\\
\hat n_y &= i \sqrt{\frac{2\pi}{3}} \left( Y_{1,-1} + Y_{1,1} \right) & (l=90^\circ, b=0^\circ) \nonumber\\
\hat n_z &= \sqrt{\frac{4\pi}{3}} Y_{1,0} & (b=90^\circ). \nonumber
\end{align}
These templates are masked with the appropriate galaxy subsample mask, and then we compute their auto-power spectrum dipole $C_1^{ii;\rm cut}$. (On the full sky, these templates would have a dipole power of $C_1^{ii} = 4\pi/9$.) The cut-sky theoretical auto-power dipole of the templates is scaled by the squared model amplitude $b_v^2 V_i^2$.
The posterior for $V_i$ can then be written as,
\begin{eqnarray}\label{eq: Vi posterior}
    && P(V_i | \hat C_1^{vv;\rm cut}) \propto  \nonumber \\
    && \int db_v\, P(b_v) \frac{1}{\mathcal V^{3/2}} \exp\bigg[-\frac{3}{2 \mathcal V}\Big(\hat C_1^{vv;\rm cut} + b_v^2 V_i^2 C_1^{ii;\rm cut} \nonumber \\
    && \ \ \ \ \ \ \ \ \ \ \ \ \ \ \ \ \ \ \ \ \ \ \ \ \ \ \ \ \ \ \ \ \ \ \ \ \ \ \ ~ - 2 b_v V_i \hat C_1^{vi;\rm cut} \Big) \bigg].
\end{eqnarray}
This approach was also taken in Ref.~\cite{Krywonos_2024} to place constraints on the intrinsic CMB dipole by marginalizing over the direction of the template. 
$P(b_v)$ in Eq.~\eqref{eq: Vi posterior} is a prior on the optical depth bias parameter $b_v$ (the same as that used in Eq.~\eqref{eq: C1vv posterior}, shown in Figure~\ref{fig: Pbv_results}), and the variance $\mathcal V = N_{\rm rec}^{\rm cut}$.
$\hat C_1^{vi;\rm cut}$ is the dipole of the cut-sky cross-power spectrum between the kSZ-reconstructed velocity map and the $i$-direction template.
Conceptually, the exponential in Eq. \eqref{eq: Vi posterior} acts as a least-squares template fit. It determines how much of the observed dipole signal ($\hat C_1^{vv;\rm cut}$) can be explained by a model where the flow is oriented solely along the $i$-axis ($b_v^2 V_i^2 C_1^{ii;\rm cut}$). The likelihood is maximized by the preferred amplitude $V_i$ that best accounts for the measured cross-correlation $\hat C_1^{vi;\rm cut}$.

The median values and $1\sigma$ credible intervals are reported in Table~\ref{tab: Vxyz}. 
We note that these constraints are based on the raw kSZ velocity reconstruction, distinct from the tophat-converted velocities presented in Figures~\ref{fig: vbulk}~and~\ref{fig: comparison plot}. 
Consequently, combining these individual 1D component limits in quadrature will systematically yield a lower total velocity magnitude than the formal 3D tophat upper limits shown in those figures. 
\begin{table}
    \centering
    \renewcommand{\arraystretch}{1.5}
    \setlength{\tabcolsep}{4pt}
    
    \begin{tabular}{c|c|c|c|c|} 
        & $R_{\rm TH}$ & $\hat V_x$ & $\hat V_y$ & $\hat V_z$ \\ 
        & $[h^{-1}\ \rm Mpc]$ & ${\rm km}\, s^{-1}$ & ${\rm km}\, s^{-1}$ & ${\rm km}\, s^{-1}$ \\
        \hline
        \hline
        \texttt{WxSCOS 1} & 174 & $-90 \pm 100$ & $-35 \pm 87$ & $-17 \pm 55$ \\ 
        \texttt{WxSCOS 2} & 258 & $-72 \pm 78$ & $33 \pm 67$ & $-40 \pm 42$ \\
        \texttt{WxSCOS 3} & 349 & $120 \pm 75$ & $-25 \pm 64$ & $-13 \pm 40$ \\
        \unwise\ \texttt{`b'} & 703 & $-8 \pm 29$ & $-12 \pm 25$ & $19 \pm 15$ \\
        \unwise\ \texttt{`g'} & 997 & $41 \pm 35$ & $10 \pm 30$ & $-18 \pm 18$ \\
        \unwise\ \texttt{`r'} & 1045 & $77 \pm 99$ & $36 \pm 87$ & $-34 \pm 52$ \\
        \hline
    \end{tabular}
    \caption{Velocity components in Galactic Cartesian coordinates derived from the credible intervals of the posterior in Eq.~\eqref{eq: Vi posterior}. These velocity constraints are based on the raw kSZ velocity reconstruction, distinct from the tophat-converted velocities presented in Figures~\ref{fig: vbulk}~and~\ref{fig: comparison plot}. Consequently, combining these individual 1D component limits in quadrature will systematically yield a lower total velocity magnitude than the formal 3D tophat upper limits shown in those figures.}
    \label{tab: Vxyz}
\end{table}

\subsection{Implications for the CMB dipole vs. number-count dipole anomaly}\label{sec: dipole anomaly}
Several recent studies of quasars and distant radio sources \cite[e.g.,][]{Singal2011, GibelyouHuterer2012, Secrest_2022, Wagenveld_2025} have claimed anomalous dipoles,  $\mathcal D$, in the number counts of these objects. The expectation is that these sources are sufficiently distant that their intrinsic clustering dipole is negligible: they are essentially expected to be isotropic as seen by an observer at rest with respect to the sample. However, as pointed out by \citet{EllisBaldwin1984}, an observer moving with respect to the sample will see a ``kinematic'' dipole due to their own motion.
%
Assuming a purely kinematic origin, this number-count dipole is proportional to the amplitude of the relative velocity between the Solar system (at $R=0$) and the bulk motion of the distant shell of sources, i.e.,
\begin{eqnarray}\label{eq: dipole D definition}
    \mathcal D(R)  \propto |\bm v_\odot - \bm V_{\rm shell}(R)|\,,
\end{eqnarray}
where $\bm v_\odot$ is the velocity of the Sun with respect to the CMB, and $\bm V_{\rm shell}(R)$ is the CMB-frame velocity of a distant shell of sources at a typical distance $R$.
The motion of a distant shell of sources with respect to the cosmic rest frame is precisely the bulk flow we probe with kSZ. Thus, our kSZ-derived upper limits on the tophat bulk flow---$\hat V_{\rm TH}$ in Figure \ref{fig: comparison plot}--- constrain $|\bm V_{\rm shell}(R)|$, and allow us to compare with the number-count dipole measurements.

The dipole in the CMB measured by the \citet{Planck18_2020} gives 
$$|\bm v_\odot| = (369.82 \pm 0.11) ~{\rm km}\ {\rm s}^{-1}.$$
In $\Lambda$CDM, the expectation is that, for large volumes, $\bm V_{\rm shell}(R)$ will be small and uncorrelated with  $\bm v_\odot$.  Because the direction of the bulk flow on very large scales is random relative to $\bm v_\odot$, adding their amplitudes in quadrature gives a \lcdm\ theoretical expectation for the number-count dipole $\mathcal D$ [Eq.~\eqref{eq: dipole D definition}],
\begin{eqnarray}\label{eq: D LCDM}
    \mathcal D^{\Lambda \rm CDM}(R_{\rm TH})\sim \sqrt{|\bm v_\odot|^2 + \big|\bm V_{\rm TH}^{\rm\lcdm}(R_{\rm TH})\big|^2} \approx |\bm v_\odot|,~
\end{eqnarray}
where the second step is follows in $\Lambda$CDM if $R$ is sufficiently large that  $|\bm V_{\rm TH}^{\rm\lcdm}(R_{\rm TH})\big| \ll |\bm v_\odot|$.
The quasar number count dipole anomaly lies in the discrepancy between this \lcdm\ expectation and the measured $\hat{\mathcal D}$, where the latter has been observed to be up to $\sim 2\,\mathcal D^{\Lambda \rm CDM}$ \cite[e.g., with \catwise\ quasars,][]{Secrest_2021}.

Of the six galaxy subsamples used in this work, the one whose redshift distribution overlaps most closely with \catwise\ (`\texttt{CW}') is \unwise\ \green (compare Figure 3 in Ref. \cite{Secrest_2021} with Figure \ref{fig: bdndz} in this work). 
Thus we can set 
$$R_{\rm TH}^{\texttt{CW}}\approx R_{\rm TH}^{\green}\approx 1\, h^{-1} {\rm Gpc}\,.$$

The black curve in Figure \ref{fig: comparison plot} gives the theoretical prediction for the tophat bulk flow, which yields $|\bm V_{\rm TH}^{\Lambda\rm CDM}(R_{\rm TH}^{\texttt{CW}})|\approx 30\ \kmpers$ at  $R_{\rm TH}^{\green}\approx 1\, h^{-1} {\rm Gpc}$, confirming the expectation in Eq.~\eqref{eq: D LCDM}.
However, the \catwise\ observation implies a measured dipole amplitude of 
$$\hat{\mathcal D}(R_{\rm TH}^{\texttt{CW}})\approx 2\,\mathcal D^{\Lambda \rm CDM}(R_{\rm TH}^{\texttt{CW}}) = 740\ {\rm km}\ s^{-1}.$$ 
The direction of this anomalously large dipole is consistent with being aligned with the CMB dipole.
If we interpret this as being due to a bulk motion of the \catwise\ quasar sample with respect to the CMB, the sample's velocity $\hat{\bm V}_{\rm TH}(R_{\rm TH}^{\texttt{CW}})$ must be roughly anti-aligned with the Solar velocity, implying [Eq. \eqref{eq: dipole D definition}]
\begin{eqnarray}
    \hat{\mathcal D}(R_{\rm TH}^{\texttt{CW}}) \sim |\bm v_\odot - \hat{\bm V}_{\rm TH}(R_{\rm TH}^{\texttt{CW}})|
    \approx |\bm v_\odot| + \big|\hat{\bm V}_{\rm TH}(R_{\rm TH}^{\texttt{CW}})\big|,\nonumber
\end{eqnarray}
which gives an implied bulk flow estimate for the \catwise\ quasar shell,
$$ \big|\hat{\bm V}_{\rm TH}(R_{\rm TH}^{\texttt{CW}})\big| \sim \hat{\mathcal D}(R_{\rm TH}^{\texttt{CW}}) - |\bm v_\odot| = 370 \ {\rm km}\ s^{-1}.$$
This bulk flow implied by the \catwise\ number-count dipole measurement would be in strong tension with our \unwise\ \green\ upper limit of $\hat V_{\rm TH}(R^{\green}_{\rm TH})<115\ {\rm km}\ s^{-1}$, and even stronger tension with the \unwise\ \blue\ upper limit of $\hat V_{\rm TH}(R^{\blue}_{\rm TH})<72\ {\rm km}\ s^{-1}$ in Figure~\ref{fig: comparison plot}.
%



Our upper limits on bulk flow do not indicate any tension with $\Lambda$CDM, and establish the sensitivity and reach of kSZ velocity reconstruction on $\sim$Gpc scales. 
\subsection{Low-$k$, low-$z$ $P_{mm}(k)$ with \unwise\ \blue}\label{sec: low-k low-z Pmm}
The upper limit on $P_{mm}(k)$ with \unwise\ \blue\ (labeled `\texttt{b}' in Figure \ref{fig: Pmm upper limits}) is, to our knowledge, one of the few informative constraints obtained from low-redshift ($z\lesssim 1)$ LSS tracers at $k\lesssim 10^{-3}\ {\rm Mpc}^{-1}$ (the other comparable one being in \citet{Macaulay_2011}). Almost all constraints on the matter power spectrum at such low-$k$ come from the CMB, which probes $z\gg 1$. We find that \unwise\ \blue, with a median redshift of $z_{\rm med}\approx 0.6$, can constrain matter clustering at low-$k$. 
The limit is not stringent enough to constrain beyond-$\Lambda$CDM scenarios. However, it demonstrates the promise of kSZ velocity reconstruction as a probe of large scales, which will only improve with upcoming data. 
\subsection{Data-informed optical depth degeneracy error}\label{sec: bv disc}
The optical depth bias is a survey-dependent quantity, which can make it challenging to directly compare $b_v$ constraints or use existing constraints to set a prior for a specific analysis. In Appendix \ref{app: bv analysis}, we present a method to translate ACT$\times$DESI measurements of $b_v$ into constraints on our gas-halo model parameters. This procedure is general and can be applied to any measurement of $b_v$, provided the corresponding fiducial $C_\ell^{\tau g}$, $C_\ell^{gg}$, and $C_\ell^{TT}$ used in that analysis are available.
Once constraints on the gas–halo model parameters are obtained, they can be propagated to compute a survey-specific prior on $b_v$ for any kSZ velocity reconstruction dataset. This method therefore enables constraints from high signal-to-noise measurements to inform analyses with different survey characteristics.

In the context of this work, ACT×DESI provides relatively high signal-to-noise measurements of $b_v$, while the \textit{Planck}×\texttt{unWISE} and \texttt{WxSCOS} combinations probe larger angular scales relevant for bulk-flow constraints. This is a convenient way of leveraging the constraining power of deep surveys like ACT, while also benefiting from the large sky coverage of \textit{Planck} for cosmological measurements. More broadly, this framework provides a consistent way to translate optical depth bias information between different kSZ velocity reconstruction analyses, and is therefore useful for a wide range of cosmological studies using kSZ velocity reconstruction.

\section{Conclusion}\label{sec: conclusion}
    Using the technique of kSZ velocity reconstruction laid out in Ref. \cite{Bloch_2024}, we have derived some of the tightest upper limits on bulk flow at $200 \lesssim R\,[h^{-1}{\rm Mpc}]\lesssim 2000$ (Figures~\ref{fig: vbulk}, \ref{fig: comparison plot}). 
    This not only bridges the theoretical divide between kSZ-reconstructed velocities and bulk flow, but also extends the landscape of bulk flow measurements to distances $\sim 10\times$ larger than those traditionally probed by peculiar velocity~surveys using distance estimators. 
We find that the kSZ-reconstructed velocity field also yields informative upper limits on the large-scale ($k\lesssim 10^{-3}{\rm Mpc}^{-1}$) matter power spectrum, $P_{mm}(k)$ (Figure \ref{fig: Pmm upper limits}), with $z\lesssim1$ tracers.
To obtain both of these types of cosmological constraints, we used two galaxy catalogs that maximize sky coverage and number density (\texttt{WxSCOS} and \unwise) divided into three $z$-bins each, in combination with the \texttt{SMICA} CMB temperature map from \planck.

Given our resulting sensitivity on Gpc scales, the kSZ-reconstructed velocity field is well-suited for testing the Cosmological Principle. 
Specifically, constraints on the bulk flow realization $\hat V$ at $R\sim \rm Gpc$ serve as an independent check of the anomalously large number-count dipole measurement \cite{Secrest_2021, Secrest_2022}. 
Our \unwise\ \blue\ and \green\ upper limits (marked \texttt{`b'} and \texttt{`g'} in Figure~\ref{fig: comparison plot}) do not indicate the presence of any anomalous bulk motion of LSS tracers, and are in strong tension with the \catwise\ number-count dipole measurement. We find results consistent with the large-scale statistical homogeneity and isotropy expected in \lcdm\ (\S\ref{sec: dipole anomaly}).

To ensure the robustness of our constraints, we also implement a novel analysis to account for the optical depth degeneracy (inherent in kSZ velocity reconstruction, parametrized by the bias parameter $b_v$) in a data-informed manner. 
We translate the high signal-to-noise measurement of $b_v$ from ACT$\times$DESI data \cite{McCarthy:2025brx} to a range of $b_v$ values for the data combinations used in this work (i.e., for \planck$\times$\texttt{WxSCOS} and \planck$\times$\unwise; see \S\ref{sec: bv disc}, Appendix \ref{app: bv analysis}).

Although the kSZ velocity reconstruction pipeline yields a full map of velocities (Figure \ref{fig: velocity maps}), the dipole ($\ell=1$) alone suffices for the stringent constraints derived in this work.
We outline the theoretical relationship between the dipole of the kSZ-derived velocity angular power spectrum, $C_1^{vv}$, and the more familiar bulk velocity variance---$\langle|\bm V|^2\rangle$ or $\sigma_V^2$ (\S\ref{sec: ideal Wf}--\ref{sec: BF from kszv theory}). 
We illustrate that the two quantities fit the same template of Eq. \eqref{eq: V auto-correlation template}, and hence differ mainly by their respective effective window functions. 
Leveraging this relationship, we formulate a procedure for mapping between the two types of window functions in \S\ref{sec: procedure}.
We also put our upper limits in the context of other existing bulk flow measurements by mapping various idealized window functions used in the literature to a tophat (see Figure~\ref{fig: comparison plot}, \S\ref{sec: comparison with other measurements}, and Appendix~\ref{app: vbulk comparison}).

\subsection*{Future directions}
We have demonstrated the continued promise of kSZ velocity reconstruction to constrain cosmology on the largest scales with existing datasets. 
As galaxy and CMB surveys with large sky-coverage collect even more pristine data, the reconstruction will become more sensitive to the underlying signal, $C_\ell^{vv}$, especially at low-$\ell$.
The noise can be driven down with the higher galaxy number densities projected for SPHEREx \cite{dore2015_spherex,Bock_2026_SPHEREx}, Euclid \cite{laureijs2011_euclid, Amendola_2018_euclid}, and LSST \cite{Ivezic_2019_lsst}, and higher resolution CMB data from Simons Observatory \cite{Ade_2019_SO} and CMB-HD \cite{sehgal2020_cmb-hd}. Consequently, as current upper limits improve, we can begin to validate or eliminate beyond-$\Lambda$CDM scenarios. 

Future work involving more sophisticated window-function matching, such as the Minimum Variance method, will be essential to investigate the emergent $\sim2\sigma$ tension between our \wscos\,\texttt{`1'} kSZ-derived tophat bulk flow and the `W09' \cite{Watkins_2009} and `W23' \cite{Watkins_2023} measurements at $\sim 100\ h^{-1}$Mpc (Figure~\ref{fig: comparison plot}). Doing so could also help clarify the poorly-understood origin of the $\sim 600\,\kmpers$ bulk flow of the Local Group.

Tighter constraints would be valuable for bulk flow at $\gtrsim$Gpc scales as well, to explore the tensions within existing measurements that test the Cosmological Principle. 
Similarly, for the matter power spectrum, it would be interesting to check for any indications of excess power at low-$k$ in low-$z$ tracers with kSZ.
Furthermore, reaching the necessary sensitivity at $k \sim 10^{-2} \text{ Mpc}^{-1}$ would allow us to probe the turnover scale ($k_{\rm eq}$) of the matter power spectrum. Measuring this primordial feature with low-redshift samples (e.g., with $z_{\rm med} \approx 0.1$ similar to \wscos~\texttt{1}) would provide an independent standard ruler for cosmological constraints.
Ultimately, kSZ velocity reconstruction stands poised to transition from providing competitive upper limits today to acting as a precision probe of fundamental cosmology in the next generation of surveys.
\section*{Acknowledgements}
We thank Roya Mohayaee for conversations that inspired this project, in particular for suggesting a comparison between kSZ remote dipole field reconstruction and other bulk-flow measurements via Figure~\ref{fig: vbulk}. 
We thank Selim Hotinli for helpful conversations, Alex Krolewski for access to \unwise\ galaxy maps, and Dustin Lang for valuable support in optimizing the computational analyses for this work on the Perimeter Institute's Symmetry HPC cluster.
JK acknowledges support from the Natural Sciences and Engineering Research Council of Canada (NSERC) through the Vanier Canada Graduate Scholarship. This research was supported in part by Perimeter Institute for Theoretical Physics. Research at Perimeter Institute is supported by the Government of Canada through the Department of Innovation, Science and Economic Development Canada and by the Province of Ontario through the Ministry of Colleges and Universities. MCJ and MJH are supported through NSERC by Discovery grants.

\subsection*{Software}
The computational analyses for this research utilized the \texttt{Python} programming language, specifically the \texttt{NumPy} \citep{numpy}, \texttt{SciPy} \citep{scipy}, and \texttt{Matplotlib} \citep{matplotlib} libraries.
\subsection*{AI Usage Declaration}
During the preparation of this manuscript, we used ChatGPT (OpenAI) and Gemini (Google) to facilitate literature searches, assist with conceptual ideation, coding, debugging code, formatting, and copyediting. The authors have reviewed and edited the output from these tools and take full responsibility for the accuracy and integrity of the content of this work.
\appendix
\counterwithin{figure}{section}
\section{kSZ-derived bulk flow variance}\label{app: derivation}
%
%
In this appendix, we work out how the kSZ-derived bulk flow variance [Eq.~\eqref{eq: kszv bulk flow variance}] can be expressed in terms of the quantities analogous to the more standard bulk flow variance [e.g., Eq.~\eqref{eq: sigmaV}]. 

We start by using Eq.~\eqref{eq: Vbulk with vpec(k)} to write Eq.~\eqref{eq: Vbulk projected} as
\begin{eqnarray}\label{appeq: Vbulk projected-1}
    \bm V(\hvec n)\cdot \hvec n &=& \int d\chi\, W(\chi \hvec n)\int \frac{d^3\bm k}{(2\pi)^3}v_{\rm pec}(\bm k,\chi)\,\nonumber \\
    && \ \ \ \ \ \ \ \ \ \ \ \ \ \ \ \ \ \ \ \ \times\ (\hvec k \cdot \hvec n) e^{ik\chi(\hvec k \cdot \hvec n)},
\end{eqnarray}
where we have expressed $\bm v_{\rm pec}(\bm k, \eta(\chi)) = v_{\rm pec}(\bm k, \chi)\hvec k$, and $\bm v_{\rm pec}(\bm k, \eta(\chi))$ is explicitly defined in Eq.~\eqref{eq: vpec(k) vector}. 
With the plane-wave expansion, we can write the combination in the last line of Eq.~\eqref{appeq: Vbulk projected-1} in terms of spherical harmonics
\begin{eqnarray}\label{appeq: plane-wave expansion}
    (\hvec k\cdot \hvec n)e^{i k \chi (\hvec k\cdot \hvec n)} &=& 4\pi \sum_{\ell m} i^{\ell-1} Y_{\ell m}^*(\hvec k) Y_{\ell m}(\hvec n) j'_\ell(k\chi), \nonumber\\
\end{eqnarray}
where 
\begin{eqnarray}\label{appeq: jl'}
    j'_\ell(k\chi) \equiv \frac{\partial j_{\ell}(k\chi)}{\partial(k\chi)} = \frac{\ell j_{\ell-1}(k\chi) - (\ell+1) j_{\ell+1}(k\chi)}{(2\ell+1)}.
\end{eqnarray}
This gives
\begin{eqnarray}\label{appeq: Vbulk projected full sum}
    \bm V(\hvec n)\cdot\hvec n &=& 4\pi \sum_{\ell m} i^{\ell-1} Y_{\ell m}(\hvec n) \int \frac{d^3\bm k}{(2\pi)^3} Y_{\ell m}^*(\hvec k) \nonumber \\
    && \times \int d\chi\, W(\chi \hvec n)\, v_{\rm pec}(\bm k,\chi)\, j_{\ell}'(k\chi).
\end{eqnarray}
The variance of Eq.~\eqref{appeq: Vbulk projected full sum} is then,
\begin{eqnarray}\label{appeq: Vbulk projected variance full sum}
    &&\big\langle |\bm V(\hvec n)\cdot\hvec n|^2\big\rangle = (4\pi)^2 \sum_{\ell m} \sum_{\ell'm'} i^{\ell-\ell'} Y_{\ell m}(\hvec n) Y_{\ell'm'}^*(\hvec n) \nonumber \\
    &&~~~~~~~\times \int \frac{dk}{(2\pi)^3} k^2 \int d^2\hvec k\, Y_{\ell m}^*(\hvec k) \int \frac{d^3\bm k'}{(2\pi)^3} Y_{\ell'm'}(\hvec k') \nonumber \\
    &&~~~~~~~~~~\times \int d\chi\, W(\chi\hvec n) j'_\ell(k\chi) \int d\chi'\, W(\chi'\hvec n) j'_{\ell'}(k'\chi') \nonumber \\
    &&~~~~~~~~~~~~~~~\times \big\langle v_{\rm pec}(\bm k,\chi) v_{\rm pec}^*(\bm k', \chi') \big\rangle.
\end{eqnarray}
The auto-correlation of peculiar velocities that appears in the last line of Eq.~\eqref{appeq: Vbulk projected variance full sum} can be written as
\begin{eqnarray}\label{appeq: vpec power spectrum}
    \left\langle v_{\rm pec}(\bm k,\chi) v^*_{\rm pec}(\bm k',\chi') \right\rangle &=& (2\pi)^3 D_v(k,\chi) D_v(k',\chi')\nonumber \\
    && \times\ \delta_D^{(3)}(\bm k - \bm k')\, P_{vv}(k),
\end{eqnarray}
where we have defined 
$$D_v(k,\chi) \equiv v_{\rm pec}(\bm k,\chi)/v_{\rm pec}(\bm k, \chi=0).$$
The velocity power spectrum $P_{vv}(k)$ is evaluated at $\chi = \chi(z=0) = 0$, and can be expressed in terms of the matter power spectrum today [using Eq.~\eqref{eq: vpec(k) vector}] as 
$$P_{vv}(k) = H_0^2 f_0^2 P_{mm}(k)/k^2,$$
with $H_0 = H(\chi=0)$, and $f_0 = f(\chi=0)$. 
The Dirac delta function allows us to integrate over $\bm k'$ when computing $\langle|\bm V(\hvec n)\cdot\hvec n|^2\rangle$, set $\int d^2\hvec k\, Y_{\ell m}^*(\hvec k) Y_{\ell'm'}(\hvec k) = \delta_{\ell\ell'}\delta_{mm'}$, and then $\sum_m|Y_{\ell m}(\hvec n)|^2= (2\ell+1)/(4\pi)$. 
Plugging all of this into Eq.~\eqref{appeq: Vbulk projected variance full sum}, we can write
\begin{eqnarray}
    \label{appeq: V los variance full sum-1}
    &&\big\langle |\bm V(\hvec n)\cdot\hvec n|^2\big\rangle = \sum_\ell \frac{2\ell+1}{4\pi} \int dk\, P_{mm}(k) \nonumber\\
    &&~~~~~~\times\,\frac{2}{\pi}H_0^2 f_0^2 \bigg[\int d\chi\, W(\chi\hvec n)\, D_v(k,\chi)\, j'_\ell(k\chi)\bigg]^2 \\
    \label{appeq: 1/3 ksz Wf full sum}%
    &&~~~~~~= \frac{1}{3}\int_{-\infty}^\infty d\ln k\, \sum_\ell \big(\mathcal W_\ell^{\rm kSZ}(k)\big)^2 P_{mm}(k) \\
    &&~~~~~~=\sum_\ell \frac{2\ell+1}{4\pi} C_\ell^{vv}.
\end{eqnarray}
Eq.~\eqref{appeq: 1/3 ksz Wf full sum} ties back to Eq.~\eqref{eq: kszv bulk flow variance}. By comparing Eqs. \eqref{appeq: V los variance full sum-1} and \eqref{appeq: 1/3 ksz Wf full sum}, we can read off the $\ell$-component of the full kSZ effective window function to~be,
\begin{eqnarray}\label{appeq: ksz Wf any ell}
    \big(\mathcal W_\ell^{\rm kSZ}(k)\big)^2 &=& 3(2\ell+1)\frac{ H_0^2 f_0^2}{2\pi^2} k \nonumber \\
    && \times \bigg[\int d\chi\, W(\chi\hvec n)\, D_v(k,\chi)\, j'_\ell(k\chi)\bigg]^2. \nonumber \\
\end{eqnarray}

The integral on the right-hand side of Eq.~\eqref{appeq: ksz Wf any ell} contains the same generic weighting or window function $W(\chi\hvec n)$ that appears in Eq.~\eqref{eq: Vbulk projected}. For kSZ velocity reconstruction specifically, $W(\chi\hvec n)$ reduces to a purely radial velocity window function, $W_v(\chi)$ [defined below in Eq.~\eqref{appeq: Wv}].

When the window function $W(\chi\hvec n)$ is wide enough compared to the coherence length of cosmic velocities (i.e., $\gtrsim 100 h^{-1}$Mpc), as is the case for our galaxy subsamples, $\bm V(\hvec n) \cdot \hvec n$ is dominated by the largest scale anisotropy---the dipole ($\ell=1$). 
This is because integrating over the wide window function averages out the small-scale peculiar velocity field. 
We make this approximation in computing $\bm V(\hvec n) \cdot \hvec n$ throughout this work.

Setting $\ell =1$ simplifies Eqs.~(\ref{appeq: 1/3 ksz Wf full sum}--\ref{appeq: ksz Wf any ell}) to give 
\begin{eqnarray}
    \big\langle |\bm V(\hvec n)\cdot\hvec n|^2\big\rangle &\approx& \frac{1}{3}\int_{-\infty}^\infty d\ln k\, \big(\mathcal W_1^{\rm kSZ}(k)\big)^2 P_{mm}(k)~~~~~~\\
    &\approx& \frac{3}{4\pi} C_1^{vv},
\end{eqnarray}
where
\begin{eqnarray}\label{appeq: W1 ksz final}
    (\mathcal W_1^{\rm kSZ}(k))^2 &=& 9\frac{H_0^2 f_0^2}{2\pi^2} k \nonumber \\
    && \times \bigg[ \int d\chi W(\chi\hvec n) D_v(k,\chi) j_1'(k\chi) \bigg]^2.\nonumber \\ 
\end{eqnarray} 
These approximate forms enter in Eqs.~\eqref{eq: kszv bulk flow variance approx} and \eqref{eq: 9 over 4pi C1vv}.
Figure~\ref{fig: ClvvUL_Nrec} also demonstrates the validity of this approximation: the theoretical signal $C_\ell^{vv;\rm\lcdm}$ (black curves) is dominated by $\ell=1$; for all galaxy subsamples we use, the theory quadrupole is at least $2\times$ smaller than the dipole.
%

%
\section{kSZ-reconstructed velocity maps}\label{app: vel recon}
\begin{figure*}
    \begin{minipage}{0.94\linewidth}
        \centering
        \includegraphics[width=\linewidth]{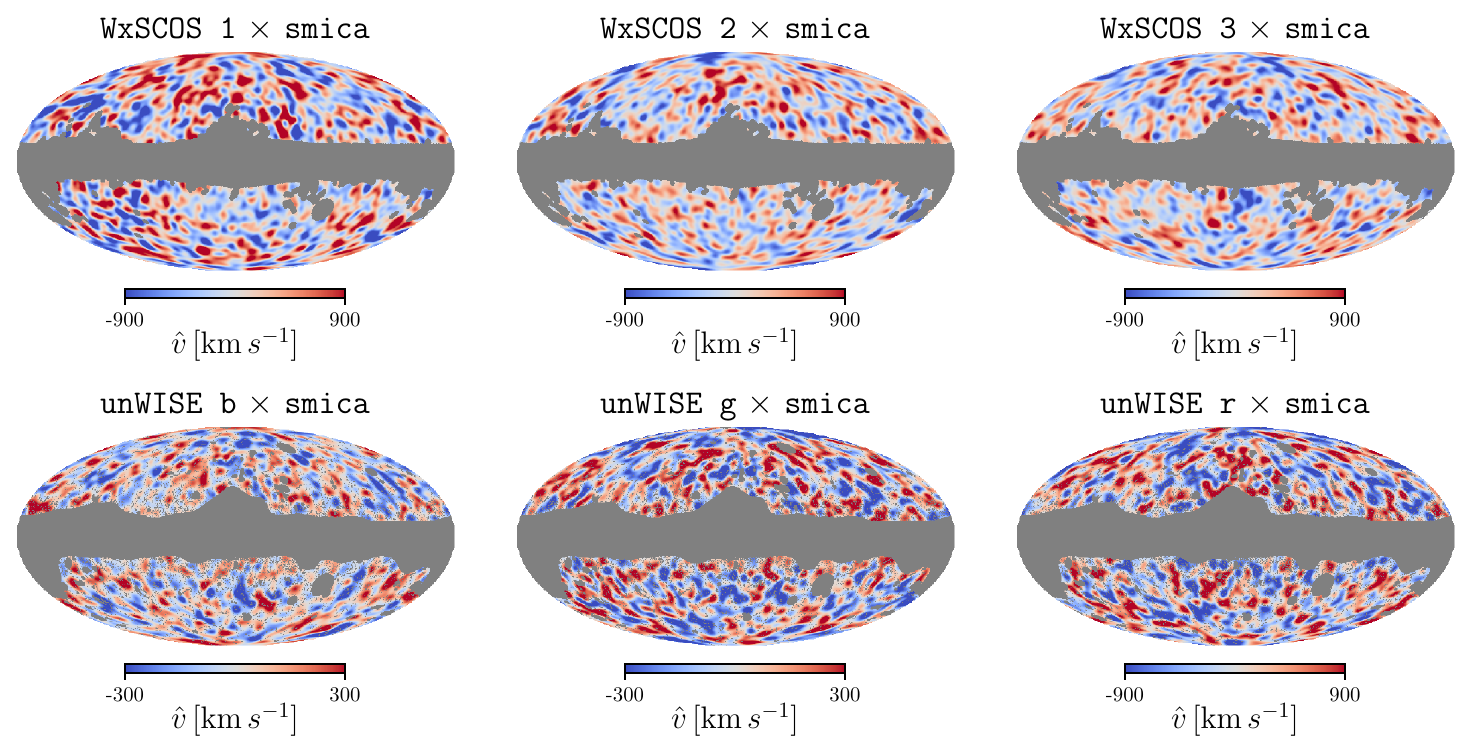}
         \vspace{-2.5em}
        \caption{The six kSZ-reconstructed velocity maps with weights that account for systematics applied. We have removed the monopole, and smoothed with a $5^\circ$ Gaussian kernel.}
        \label{fig: velocity maps}
    \end{minipage}
    
    
    \begin{minipage}{\linewidth}
        \centering
        \includegraphics[width=0.9\linewidth]{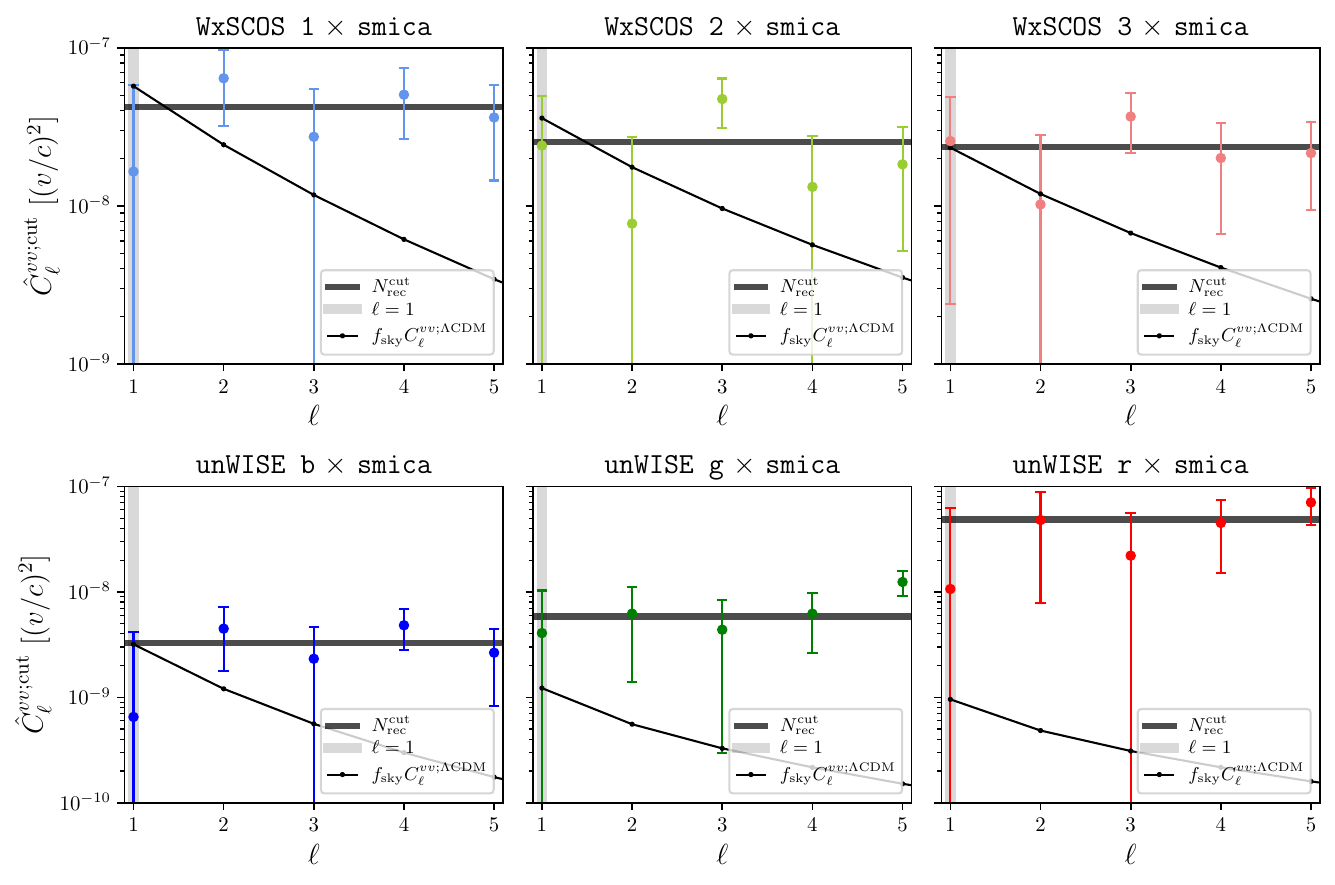}
         \vspace{-1.5em}
        \caption{The main point of this figure is to show that the reconstructed cut-sky $\hat C_\ell^{vv;\rm cut}$ is consistent with the expected cut-sky reconstruction noise, $N^{\rm cut}_{\rm rec} = f_{\rm sky}N_{\rm rec}$ [Eq. \eqref{appeq: Nrec}], at the $\sim1\sigma$ level. 
        This is evident in all the colored 1$\sigma$ error-bars being consistent with the dark gray horizontal $N^{\rm cut}_{\rm rec}$ line. 
        The colored circles mark $\hat C_\ell^{vv;\rm cut}$, and the error-bars extend out to $\hat C_\ell^{vv;\rm cut} \pm \sigma_{\hat C_\ell^{vv;\rm cut}}$, where $\sigma_{\hat C_\ell^{vv;\rm cut}} \equiv \sqrt{2/\big((2\ell+1)f_{\rm sky}\big)}N_{\rm rec}^{\rm cut}$ is the 1$\sigma$ cosmic variance error on the cut-sky reconstruction noise.
        We highlight the $\ell=1$ dipole with a vertical light gray band as that is the only multipole used in this work for deriving $\sigma_V(R)$ and $P_{mm}(k)$ upper limits.
        For reference, we show the cut-sky \lcdm\ angular power spectrum signal $f_{\rm sky}C_\ell^{vv;\rm\lcdm}$ (thin black curves), where $C_\ell^{vv;\rm\lcdm}$ is given by Eq.~\eqref{eq: lcdm C1vv}.}
        \label{fig: ClvvUL_Nrec}
    \end{minipage}
\end{figure*}

Our velocity reconstruction procedure closely follows what Ref. \cite{Bloch_2024} developed for \unwise\ \blue\ $\times$ \planck~\texttt{SMICA}, which we review below. 
In addition to \unwise\ \blue, we apply the same procedure to \unwise\ \green\ and \red, and to the three \texttt{WxSCOS} galaxy subsamples as well, while using \planck's \texttt{SMICA} map in all six cases.

The line-of-sight velocity field is reconstructed from the CMB temperature field $\Theta(\hvec n)$ and the galaxy overdensity field $\delta^g(\hvec n)$ using a quadratic estimator
\begin{eqnarray}\label{appeq: quadratic estimator}
    \hat{v} (\hvec n)&=& - N_{\rm rec} \left[ \sum_{\ell=\ell_{\rm min}}^{\ell_{\rm max}}\sum_{m=-\ell}^{\ell} \frac{\Theta_{\ell m}}{C_\ell^{TT,\rm obs}} Y_{\ell m}(\hvec n)\right] \nonumber\\
    && \ \ \ \ \ \ \ \ \ \ \times \left[\sum_{\ell=\ell_{\rm min}}^{\ell_{\rm max}}\sum_{m=-\ell}^\ell \frac{C_\ell^{\tau g}\delta^g_{\ell m}}{C_\ell^{g g,\rm obs}} Y_{\ell m}(\hvec n)\right], \nonumber\\    
\end{eqnarray}
where $\ell\in(\ell_{\rm min},\,\ell_{\rm max})=(500,\,4000)$ imposes a high-pass filter, $C_\ell^{TT,{\rm obs}}$ and $C_\ell^{gg,{\rm obs}}$ are the observed CMB temperature and observed galaxy overdensity power spectra, respectively. 
$C_\ell^{\tau g}$ is directly related to the optical depth-galaxy cross-spectrum ($C_\ell^{\dot\tau g}$) as follows 
\begin{eqnarray}\label{appeq: Cltaug}
    C_\ell^{\tau g} = C_\ell^{\dot \tau g}(\bar\chi) \frac{\int d\chi\ C_{\bar\ell}^{\dot \tau g}(\chi)}{C_{\bar\ell}^{\dot\tau g}(\bar\chi)},
\end{eqnarray}
where $\bar\chi=\chi(z_{\rm med})$ is the comoving distance at the median redshift of the galaxy subsample, and $\bar \ell=2000$.
$C_\ell^{\dot\tau g}$ [Eq. \eqref{eq: Cltaudotg}] uses the optical depth window function $W_\tau(\chi)$ [Eq. \eqref{eq: Wtau}], and the galaxy window function $W_g(\chi)$ [Eq. \eqref{appeq: Wg}]. 
The normalization coefficient in Eq. \eqref{appeq: quadratic estimator} is given by the full-sky reconstruction noise,
\begin{eqnarray}\label{appeq: Nrec}
    N_{\rm rec} \equiv \left[ \sum_{\ell'=\ell_{\rm min}}^{\ell_{\rm max}} \frac{2 \ell' + 1}{4 \pi} \frac{({C}_{\ell'}^{\tau g})^2  }{C_{\ell'}^{TT,{\rm obs}} C_{\ell'}^{gg,{\rm obs}}}\right]^{-1} \ ,
\end{eqnarray}
where, once again, the range $\ell' \in (\ell_{\rm min},\ell_{\rm max})=(500, 4000)$ imposes a high-pass filter. 

The output from the reconstruction comprises six 2-D skymaps of velocities, corresponding to the six galaxy subsamples used in this work. We show these reconstructed velocity maps in Figure \ref{fig: velocity maps}.

From these 2-D velocity maps, we obtain the cut-sky angular power spectrum, $\hat C_\ell^{vv;\rm cut}$.
We find that the reconstructed $\hat C_\ell^{vv;\rm cut}$ is consistent within $\sim 1\sigma$ with the expected cut-sky reconstruction noise, $N^{\rm cut}_{\rm rec}$, as illustrated by Figure~\ref{fig: ClvvUL_Nrec} for $1\leq\ell\leq 5$.
This implies that we do not detect the signal $C_1^{vv}$, which is why our constraints on bulk flow and the matter power spectrum are in the form of upper limits, and not measurements.
 Only the $\ell=1$ multipole is used for the $\sigma_V(R)$ and $P_{mm}(k)$ upper limits derived in this~work.
\subsection{Corrective weights}\label{sec: corrective wts}
As pointed out in Ref. \cite[][]{Bloch_2024}, since the quadratic estimator is built to detect statistically anisotropic correlations between the galaxy and CMB fields, it is highly sensitive to spatially varying survey systematics (e.g., extinction, redshift calibration errors, survey depth variations, bright stars). 
These anisotropic physical effects [$P(\hvec n)$] introduce a spurious fluctuation into the observed galaxy overdensity, $\delta^P_g(\hvec n) = P(\hvec n)\Theta^P(\hvec n)$, where $\Theta^P(\hvec n)$ is the underlying component correlated with the CMB. 
This directly alters the true galaxy-CMB correlation and contaminates the velocity signals that the estimator measures.

The correlation between the galaxy maps and the CMB data is the strongest in the high frequency maps from \planck\ (e.g., 545 GHz, 857 GHz), which are overwhelmingly dominated by the Cosmic Infrared Background (CIB) rather than the primary CMB.
While the standard reconstruction procedure uses cleaned CMB data and the galaxy map to estimate the velocity field [Eq. \eqref{appeq: quadratic estimator}], we assess potential contamination by applying the same estimator to the \planck\ 857 GHz (CIB dominated) map  instead to recover a map of systematics. 
This 857 GHz$\times$galaxy reconstruction of velocities [denoted by $\hat v^{857}(\hvec n)$] captures the strong correlation between the CIB and the large-scale structure, and the output traces the spatial variations of the systematics [$P(\hvec n)$] that we want to target.

To isolate $P(\hvec n)$ from the reconstructed $\hat v^{857}(\hvec n)$ map, we divide by the monopole (see Eq. (59) in \cite[][]{Bloch_2024}), 
\begin{eqnarray}
    P(\hvec n) = \frac{\hat v^{857}(\hvec n)}{\hat v_{00}^{857}} - 1.
\end{eqnarray}
As $P(\hvec n)$ is noise-dominated on small scales, we smooth it with a $28^\circ$ FWHM Gaussian kernel to isolate the large-scale features.

The smoothed $P(\hvec n)$ can then be converted to a map of corrective weights,
\begin{eqnarray}
    W^{\rm corr}(\hvec n) = \frac{1}{1+P(\hvec n)}.
\end{eqnarray}

\begin{figure*}
    \centering
    \includegraphics[width=0.9\linewidth]{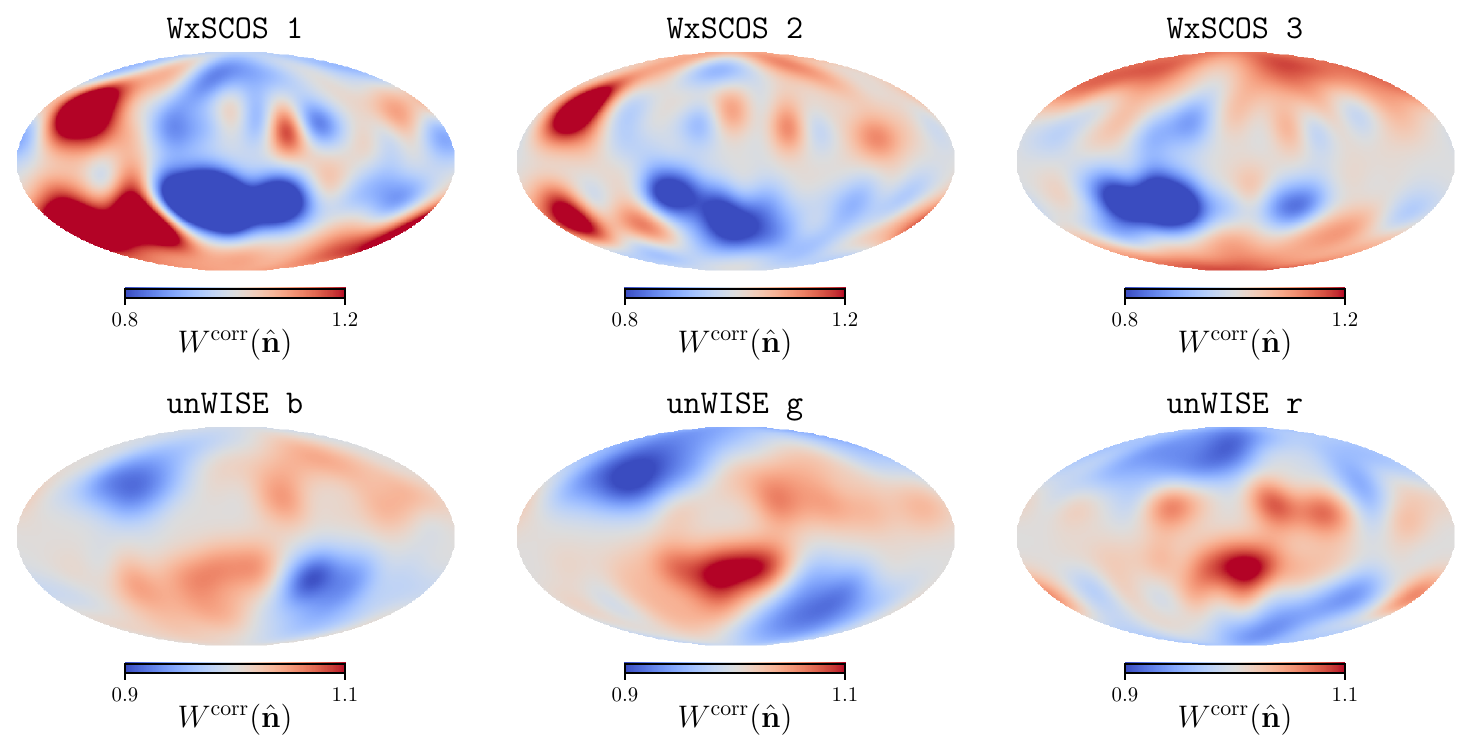}
    \caption{The empirical corrective weights applied to the galaxy subsample maps to correct for systematics. See \S\ref{app: vel recon}\ref{sec: corrective wts} for details.}
    \label{fig: corr wts maps}
\end{figure*}

Figure \ref{fig: corr wts maps} shows the $W^{\rm corr}(\hvec n)$ maps for all six galaxy subsamples.

These maps are then applied to the galaxy number-count maps as follows,
\begin{eqnarray}
    N_g^{\rm corr}(\hvec n) = W^{\rm corr}(\hvec n)\, N_g(\hvec n).
\end{eqnarray}

As mentioned in \S\ref{sec: WxSCOS}--\ref{sec: unWISE}, it from is these corrected maps that we obtain the galaxy overdensity maps, 
$$\delta_g(\hvec n) = N_g^{\rm corr}(\hvec n)/\bar N^{\rm corr}_g - 1,$$
where $\bar N^{\rm corr}_g = N^{\rm corr}_{g\,{\rm tot}}/(f_{\rm sky}N_{\rm pix})$ is the average number of galaxies per unmasked pixel in the corrected number-count map. 
The $\delta_g(\hvec n)$ maps enter the velocity estimator [Eq.~\eqref{appeq: quadratic estimator}] to create the velocity maps in Figure~\ref{fig: velocity maps}.
\section{Galaxy catalog redshift distributions and window functions}\label{app: galaxy distributions}
\subsection{$b_g\ dN/dz$}
\begin{figure}
    \centering
    \includegraphics[width=0.9\linewidth]{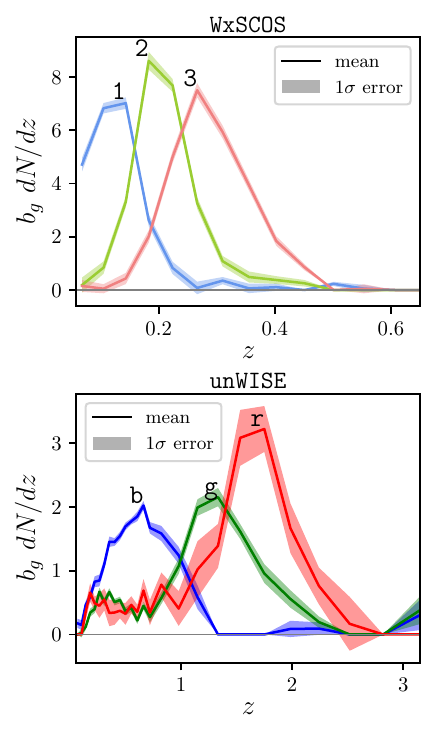}
    \vspace{-0.7cm}
    \caption{The redshift distributions (rescaled by galaxy bias $b_g$) from \texttt{tomographer.org} \cite{tomographer_website} of the \texttt{WxSCOS} (top) and \unwise\ (bottom) subsamples used in this work. The solid curves are the mean distributions, and the colored bands represent the 1$\sigma$ errors. These errors propagate to the posteriors on the full-sky dipole $\mathcal C_1^{vv}$ [Eq.~\eqref{eq: C1vv posterior}] and velocity components $V_i$ [Eq.~\eqref{eq: Vi posterior}] via the prior $P(b_v)$, as explained in Eqs.~\eqref{eq: Pbv with bgdndz integral} and \eqref{eq: Pbv with bgdndz approx}.}
    \label{fig: bdndz}
\end{figure}

Here we discuss the galaxy bias-limited redshift distributions---shown in Figure \ref{fig: bdndz}---for the galaxy subsamples used in this work.
We use the clustering-based redshift inference tool, \texttt{tomographer.org} \cite{tomographer_website}, which is built on the concepts and techniques developed in Refs. \cite{menard_2013_tomog, chiang_2019_tomog}.

Given a 2-D \texttt{healpy} map of a galaxy subsample, \texttt{tomographer.org} uses a reference sample of $\sim 2\times 10^6$ spectroscopic objects from the Sloan Digital Sky Survey to compute the galaxy bias-limited, intensity-weighted redshift distribution of the input sample, $b_I\, dI/dz$. 
The ``intensity'', $I$, is in map units of number count per pixel.
The bias-limited number density distribution we are interested in is obtained by rescaling as follows:
\begin{eqnarray}\label{appeq: bgdNdz}
    b_g \frac{dN}{dz} = \frac{1}{\bar n_g}\left(b_I \frac{dI}{dz}\right) = \frac{f_{\rm sky}N_{\rm pix}}{N_{g\,\rm tot}}\left(b_I \frac{dI}{dz}\right),
\end{eqnarray}
where $f_{\rm sky}$ is the unmasked fraction of the sky in the input galaxy map, $N_{\rm pix}$ is the number of pixels in the input map, and $N_{g\,\rm tot}$ is the total number of galaxies in the subsample. The combination $\bar n_g = N_{g\,\rm tot}/(f_{\rm sky}N_{\rm pix})$ is the average number of galaxies per unmasked pixel. 
\subsection{$W_g(\chi)$ and $W_v(\chi)$}
The quantity $b_g\, dN/dz$ [Eq. \eqref{appeq: bgdNdz}] is used for computing two crucial window functions that we describe next: the galaxy window function $W_g(\chi)$, and the velocity window function $W_v(\chi)$. The galaxy window function,
\begin{eqnarray}\label{appeq: Wg}
    W_g(\chi) = \left(b_g\frac{dN}{dz}\right)H(\chi),
\end{eqnarray}
enters into the reconstruction through $C_\ell^{\tau g}$ [Eq. \eqref{appeq: Cltaug}]. 
$W_g(\chi)$ (and hence $b_g\,dN/dz$) also enters the velocity window function (through $C_\ell^{\dot \tau g}$) \cite{Bloch_2024},
\begin{eqnarray}\label{appeq: Wv}
    W_v(\chi) \approx \frac{C_{\bar \ell}^{\dot \tau g}(\chi)}{\int d\chi\ C_{\bar \ell}^{\dot \tau g}(\chi)},
\end{eqnarray}
 where $C_\ell^{\dot \tau g}$ is defined in Eq. \eqref{eq: Cltaudotg}, and $\bar\ell=2000$. 
 $W_v(\chi)$ is what effectively determines the volume over which peculiar velocities are averaged over to obtain the bulk velocity of a galaxy subsample, e.g., in Eq. \eqref{eq: Vbulk projected}. It also determines the effective kSZ window function defined in Eq. \eqref{appeq: ksz Wf any ell}.

\section{Data-informed prior on optical depth bias parameter $b_v$}\label{app: bv analysis}
This appendix is relevant to \S\ref{sec: C1vv upper lim}, \S\ref{sec: Vxyz}, and \S\ref{sec: bv disc}.
\subsection{Optical depth bias modeling}
Any modeling uncertainties in kSZ velocity reconstruction results in a bias of the reconstructed velocities, known as the optical depth bias $b_v$:
\begin{eqnarray}
    \hat{C}_\ell^{vv} = b_v^2 \big[C_\ell^{vv}\big]^t, 
\end{eqnarray}
where the superscript ``$t$'' denotes the unknown ``true'' velocity angular power spectrum. 
This discrepancy between the measurement and truth can arise due to incorrect modeling of the gas-halo connection, galaxy-halo connection,  uncertainties in the redshift distribution of the galaxy survey, and environmental/selection effects, but has been shown to be scale-independent on large-scales \cite{Battaglia_2016,Smith_2018,Cayuso_2023, Kvasiuk:2023nje, Giri:2020pkk}.
The optical depth bias is given by \cite{Bloch_2024}
\begin{equation}\label{eq:bv}
    b_v(\chi)\simeq \frac{\sum_{\ell_1}\frac{2\ell_1+1}{4\pi}\frac{C^{\tau g}_{\ell_1}[C^{\tau g}_{\ell_1}]^t}{C^{TT}_{\ell_1}C^{gg}_{\ell_1}}}{\sum_{\ell_2}\frac{2\ell_2+1}{4\pi}\frac{(C^{\tau g}_{\ell_2})^2}{C^{TT}_{\ell_2}C^{gg}_{\ell_2}}} 
     \frac{[C^{\dot{\tau} g}_{\ell=\bar{\ell}}(\chi)]^t}{C^{\dot{\tau}g}_{\ell=\bar{\ell}}(\chi)} \frac{C^{\tau g}_{\ell=\bar{\ell}}}{[C^{\tau g}_{\ell=\bar{\ell}}]^t}
\end{equation}
in the squeezed limit where the reconstructed velocities ($\ell) $ are on much larger scales than the primary CMB ($\ell_1$) and galaxy spectra ($\ell_2$)---i.e., $\ell \ll \ell_1, \ell_2$---making $b_v$ scale-independent \cite{Bloch_2024}. 
The $C_{\ell}$'s marked with a superscript ``$t$'' are our range of models for what the unknown ``true'' quantities are, and all the other angular power spectra in Eq.~\eqref{eq:bv} represent fiducial models.
The galaxy angular power spectra are modeled as
\begin{eqnarray}\label{eq: clgg}
     C_\ell^{gg}=\int \frac{d\chi}{\chi^2}P_{mm}\left(\chi,k=\frac{\ell+\frac{1}{2}}{\chi}\right)W_g^2(\chi)+N_{\rm shot}
 \end{eqnarray}
 where the shot noise is $N_{\rm shot} = \big(N_{g\,\rm tot}/(4\pi f_{\rm sky})\big)^{-1}$, the galaxy window function $W_g(\chi)$ is given by Eq.~(\ref{appeq: Wg}), and we calculate the non-linear matter power spectrum $P_{mm}$ using \texttt{CAMB}. We calculate $b_g(z)$ and $dN/dz$ using \texttt{tomographer.org} as outlined in Appendix~\ref{app: galaxy distributions}. Since the velocity reconstruction only uses the small angular scales of the galaxy survey, 
 we neglect additional contributions from redshift space distortions and magnification. 

 The fiducial $C_\ell^{TT}$ is modeled as $C_\ell^{TT,\Lambda \rm CDM}+N_\ell$ where \texttt{CAMB} calculates the theory spectra and $N_\ell$ is the survey specific noise curve. As we do not have theoretical noise curves for the temperature maps we consider, we end up using the temperature power spectra computed directly from the data. 
%

 
 
The $C_\ell^{\tau g}$ in Eq.~\eqref{eq:bv} is defined in Eq.~\eqref{appeq: Cltaug}, and we explain how $C_\ell^{\dot \tau g}(\chi)$ is constructed next.
Our model for how the electrons trace the underlying dark matter field is given by a scale-dependent linear bias $\delta_e(\bm{k},\chi)=b_e(k,\chi)\delta_m(\bm{k},\chi)$, following Ref.~\cite{Takahashi2020}:
\begin{equation}\label{eq:be}
    b_e(z,k) = b_*(z)\left[1+\left(\frac{k}{k_*(z)}\right)^{\gamma(z)} \right]^{-1/2}.
\end{equation}
Here, the parameters are
\begin{eqnarray}\label{eq: be model parameters}
    b_*(z) &=& \sqrt{-0.013z+0.971} \nonumber \\
    \gamma(z) &=& 0.10z^2-0.59z+1.91 \\
    k_*(z) &=& (-0.42z^3 + 3.10z^2 -3.24z + 4.36) \,{\rm Mpc^{-1}}. \nonumber
\end{eqnarray}
This then allows us to compute the optical depth-galaxy cross-power spectrum as
\begin{eqnarray}\label{eq: Cltaudotg} 
    &&C^{\dot{\tau}g}_\ell(\chi) = \frac{1}{\chi^2} W_g(\chi)\, W_\tau(\chi)\, b_e\left(\chi,k=\frac{\ell + 1/2}{\chi}\right)~~~ \\
    &&~~~~~~~~~~~~~~\times~ P_{mm}\left( \chi, k=\frac{\ell + 1/2}{\chi} \right) \nonumber
\end{eqnarray}
where the optical depth window function is
\begin{eqnarray}\label{eq: Wtau} 
    W_\tau(\chi) \equiv \sigma_T \bar{n}_{e,0}(1+z(\chi))^2
\end{eqnarray}
with $\sigma_T$ as the Thomson scattering cross section and $\bar{n}_{e,0}$ as the present-day average number density of electrons. We model the latter by
\begin{eqnarray}\label{eq: n_e bar}
    \bar{n}_{e,0} = \frac{f_{\rm gas}\, X\, \Omega_{b,0}\, \rho_{\rm crit,0}}{\mu_e\, m_p},
\end{eqnarray}
where $f_{\rm gas}$ is the ionized gas mass fraction of baryons, $X$ is the fraction of the total electrons that are ionized, $\Omega_{b,0}$ is the baryon density parameter today,  $\rho_{\rm crit,0}$ is the critical density today, and $\mu_em_p$ is the mean baryon mass per electron. For our fiducial model, we follow the modeling assumptions of Ref.~\cite{Bloch_2024}, setting $X=1$, $\mu_e=1.14$, and $f_{\rm gas} = 0.9$. We also use a reference scale $\ell=\bar{\ell}=2000$ and reference redshift $\chi=\bar{\chi}$ which is the median redshift of the $dN/dz$. This allows us to calculate $C_\ell^{\tau g}$ given by Eq.~(\ref{appeq: Cltaug}). 

%
\subsection{Obtain data-informed $b_v$ prior, $P(b_v)$}

To obtain cosmological constraints from kSZ velocity reconstruction, we need to marginalize over the optical depth bias parameter, $b_v$.
We use the high signal-to-noise measurements of $b_v$ from the recent ACT×DESI analysis \cite{McCarthy:2025brx} to inform the prior, $P(b_v)$, in Eqs.~\eqref{eq: C1vv posterior} and \eqref{eq: Vi posterior}. Since $b_v$ depends on the specific galaxy sample and CMB experiment, these measurements cannot be directly applied to other surveys. Instead, we use them to constrain our gas-halo connection model parameters [in Eqs.~(\ref{eq:be}--\ref{eq: n_e bar})], which can then be used to determine a range of allowed $b_v$ for the surveys we use. 

The ACT×DESI analysis of Ref. \cite{McCarthy:2025brx} provides measurements of $b_v$ for 16 redshift bins, evaluated with their model of $C_\ell^{\tau g}$. 
In Eq.~(\ref{eq:bv}), we use their $C_\ell^{\tau g}$ as the fiducial model. In the limit of narrow redshift bins, $[C^{\dot{\tau} g}_{\ell=\bar{\ell}}(\chi)]^t/C^{\dot{\tau}g}_{\ell=\bar{\ell}}(\chi)$ and
$C^{\tau g}_{\ell=\bar{\ell}}/[C^{\tau g}_{\ell=\bar{\ell}}]^t \rightarrow 1$, so we can ignore these terms in Eq.~(\ref{eq:bv}) for DESI. The fiducial $C_\ell^{gg}$ is modeled as in Eq.~\eqref{eq: clgg}.
$C_\ell^{TT}$ is calculated using the public ACT DR6+\textit{Planck} needlet internal linear combination (NILC) blackbody temperature map%
\footnote{\url{https://lambda.gsfc.nasa.gov/product/act/act_dr6.02/act_dr6.02_nilc_info.html}} %
made using the method of Ref.~\cite{ACT:2023wcq}. We use the data power spectrum here since we do not have theoretical noise curves for the temperature. The quantities in Eq.~(\ref{eq:bv}) with superscript ``$t$'' are the unknown ``true'' quantities we want to constrain. We use our gas-halo model described above in Eqs. (\ref{eq:be}--\ref{eq: n_e bar}) for the true quantities, but leave the parameters $f_{\rm gas}$ and $k_*$ free. 
Our aim is to determine which $f_{\rm gas}$ and $k_*$ values yield true spectra consistent with the observed $b_v$ values. We vary $f_{\rm gas}$ because it represents the fraction of gas in the universe which is ionized, so it is physical, and $k_*$ is varied because $b_e$ is very sensitive to changes in $k_*$ as it shifts the scale at which feedback suppresses structure. We vary $k_*$ [Eq. \eqref{eq: be model parameters}] through an additive parameter, $k_*'$, as 
$$k_* = -0.42z^3 + 3.10z^2 -3.24z + 4.36 + k_*',$$
in order to keep the same redshift dependent functional form for all redshift bins.

\begin{figure}
    \centering
\includegraphics[width=\linewidth]{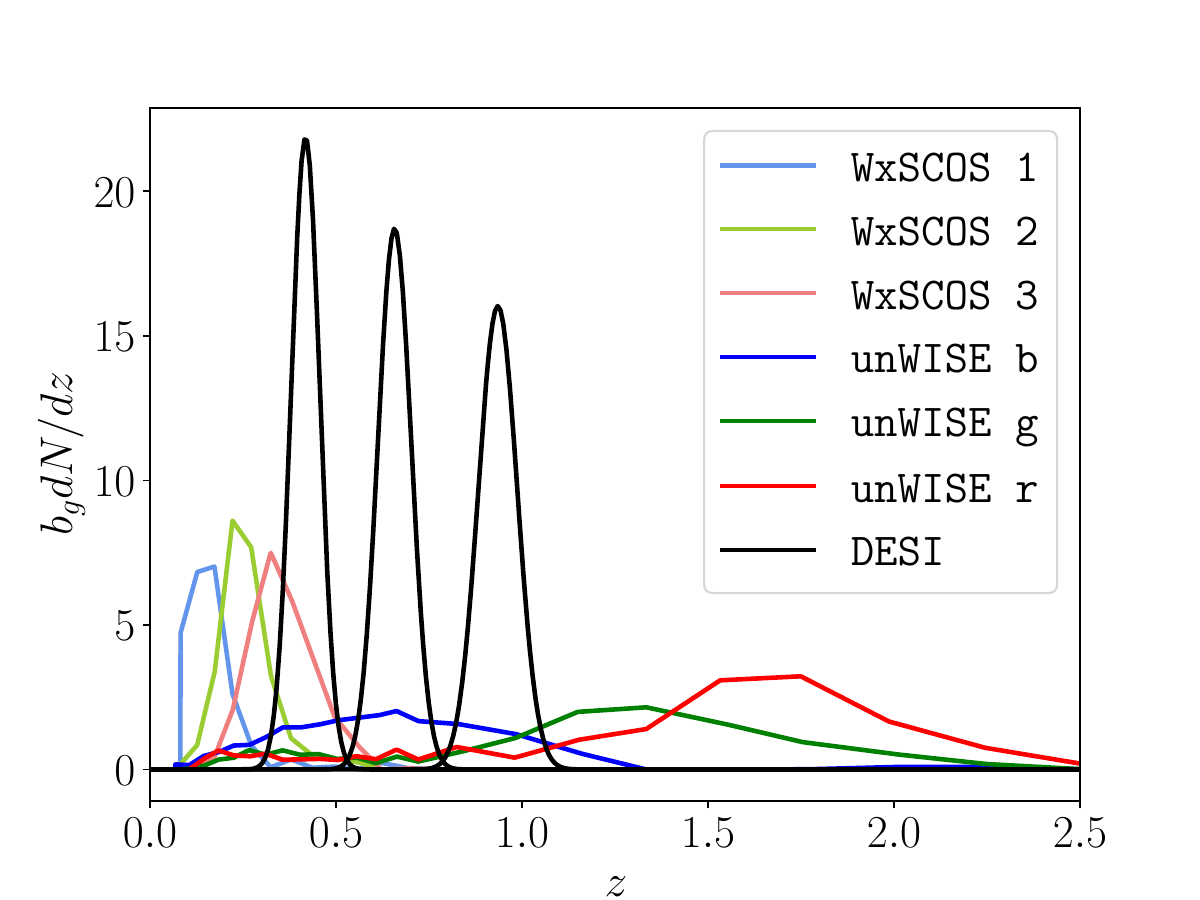}
    \caption{The galaxy bias scaled redshift distributions for \texttt{unWISE} and \texttt{WxSCOS} compared to three of the DESI bins.
    } 
    \label{fig:bdndz_with_DESI}
\end{figure}

\begin{figure}
    \centering
\includegraphics[width=\linewidth]{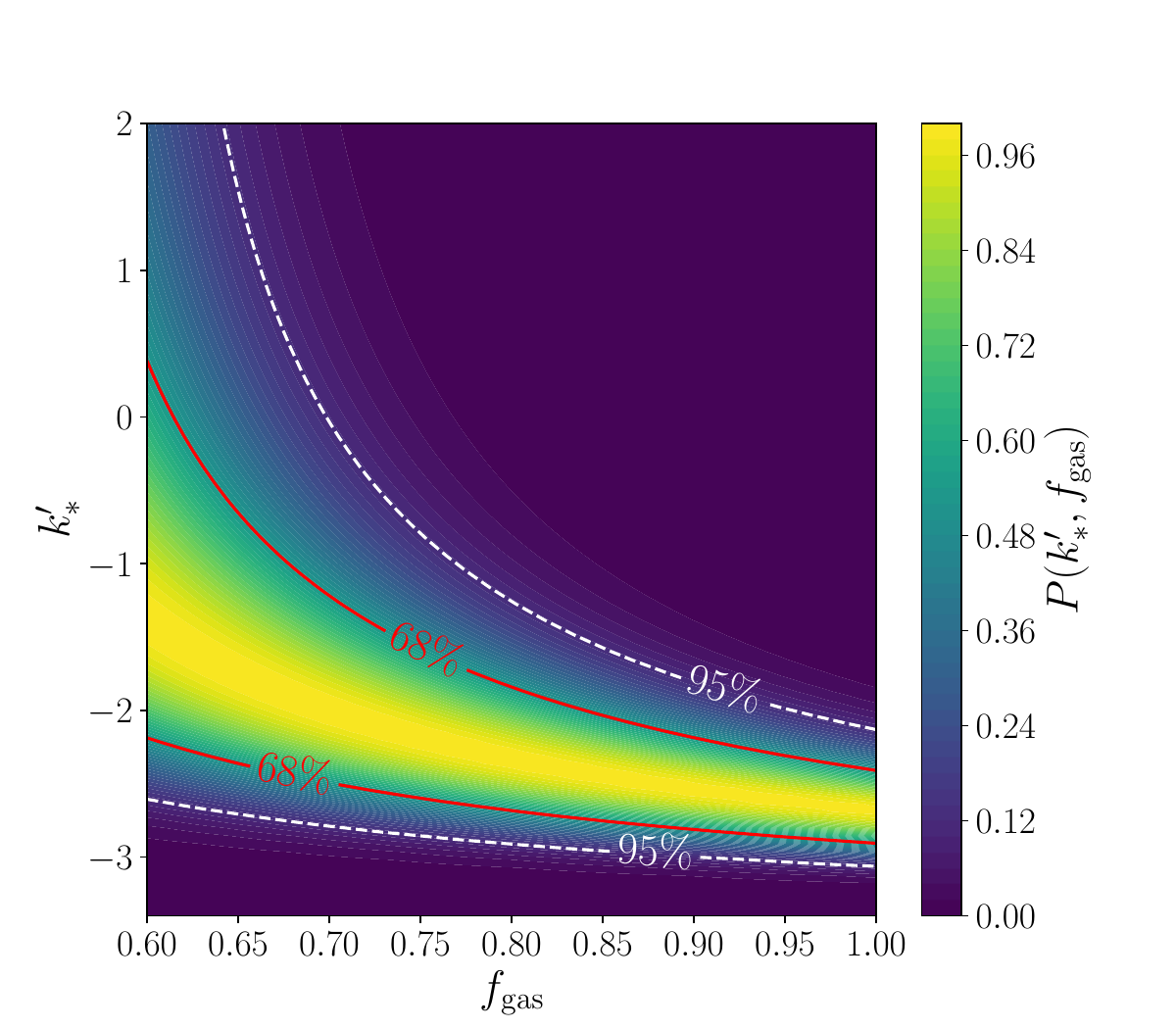}
    \caption{The ACTxDESI posterior over $k_*'$ and $f_{\rm gas}$, including the 68\% (red) and 95\% (white) credible regions. } 
    \label{fig:ACTxDESI_posterior}
\end{figure}

For each DESI redshift bin $\alpha$,
we compute a model prediction $b_v^{\rm pred}(k_*', f_{\rm gas})$ using Eq.~(\ref{eq:bv}). We assume Gaussian errors and define the joint likelihood
\begin{eqnarray}
    {\rm log}\, \mathcal{L}(k_*',f_{\rm gas}) = -\frac{1}{2} \sum_\alpha \left[ \frac{b_v^{\rm pred}(k_*',f_{\rm gas}) - b_v^{\rm obs,\alpha}}{\sigma_{b_v^{\rm obs, \alpha}}} \right]^2.
\end{eqnarray}
We use 3 out of the 16 redshift bins (specifically indices 0, 6, 14 in Ref. \cite{McCarthy:2025brx}) to minimize overlap in their redshift distributions and hence reduce bin-to-bin covariance.  These bin choices, and their overlap with the \texttt{unWISE} and \texttt{WxSCOS} bins we use, is illustrated in Figure~(\ref{fig:bdndz_with_DESI}). We evaluate the likelihood on a 2D grid of ($k_*',f_{\rm gas}$), using flat priors over the ranges $f_{\rm gas}\in [0.6,1]$ and $k_*'\in[-3,<2]$. 
The resultant 2D posterior is shown in Figure~(\ref{fig:ACTxDESI_posterior}) with the 68\% and 95\% credible regions.

To determine the prior on $b_v$ appropriate for \unwise×\textit{Planck}  and \texttt{WxSCOS}×\textit{Planck}, we propagate the 2D posterior on the gas parameters through to the survey-specific predictions for $b_v$. 
We draw $10^5$ pairs of $\{k_*',f_{\rm gas}\}$ values from the posterior $P(k_*',f_{\rm gas})$ and for each pair $i$, compute $b_v^{(i)}=b_v(k_*'^{(i)},f_{\rm gas}^{(i)})$. 
%
%
Here, $b_v$ is again calculated using Eq.~(\ref{eq:bv}). $C_\ell^{TT}$ is calculated using the public \textit{Planck} \texttt{SMICA} temperature map, not a theory model, since we do not have theoretical noise curves for \texttt{SMICA}. 
We use our fiducial $C^{\dot{\tau}g}_\ell(\chi)$ model defined by Eqs.~(\ref{eq:be}--\ref{eq: Cltaudotg}), and for the unknown ``truth'' we use the sampled $\{k_*'^{(i)},f_{\rm gas}^{(i)}\}$ values in the gas-halo model. This procedure yields a Monte Carlo representation of $P(b_v)$ which effectively performs: 
\begin{eqnarray}\label{eq: Pbv integral}
    P(b_v) &=& \int dk_*'\, df_{\rm gas}\, P(k_*',f_{\rm gas})\, P(b_v|k_*',f_{\rm gas}).~~
\end{eqnarray}

\begin{figure*}
    \centering
    \includegraphics[width=\linewidth]{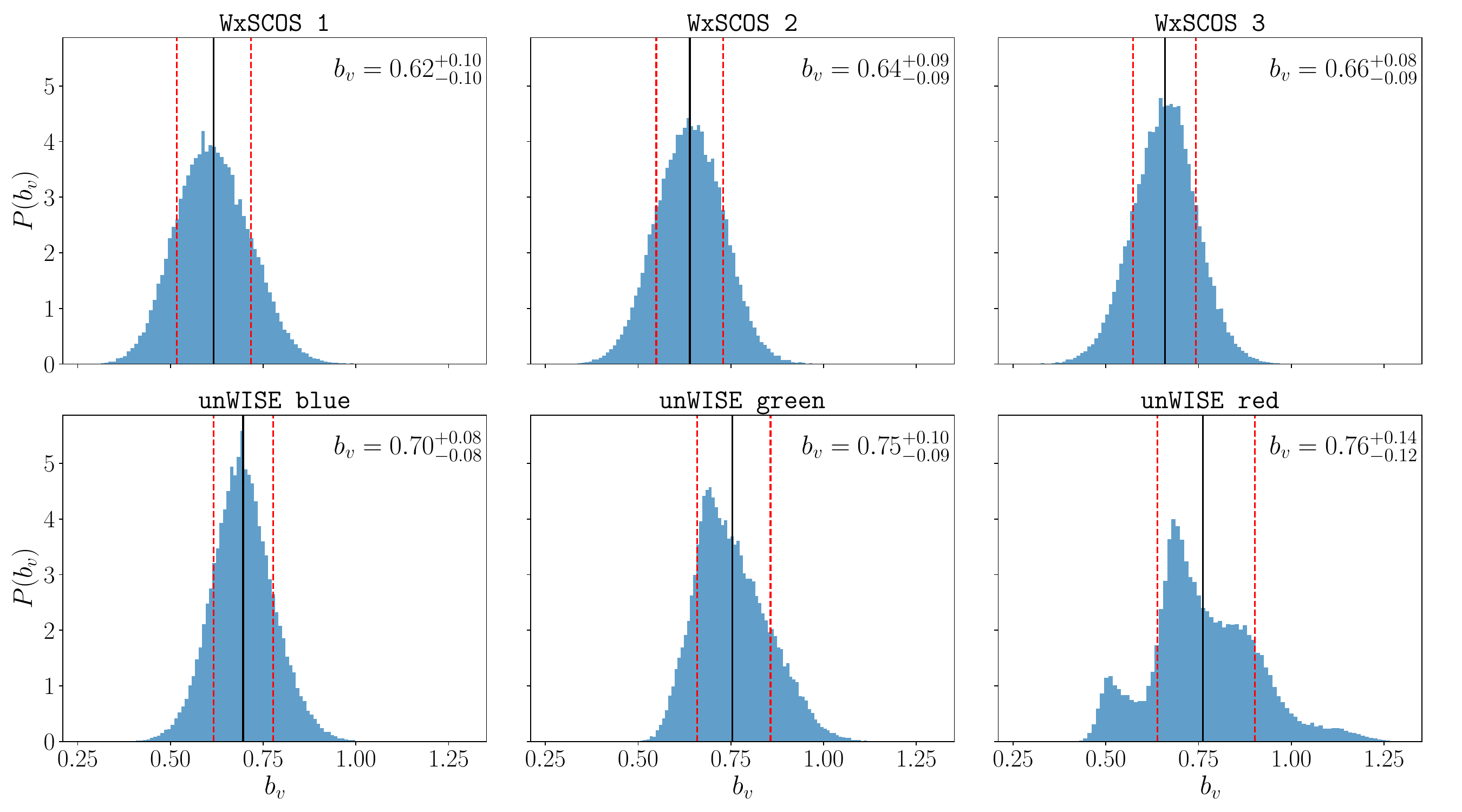}
    \caption{The optical depth bias posteriors for \texttt{WxSCOS} (upper row) and \texttt{unWISE} (lower row) redshift bins. The mean $b_v$ is shown in black and the 68\% credible intervals in red. }
    \label{fig: Pbv_results}
\end{figure*}

The optical depth bias encompasses not just astrophysical modeling uncertainty, but also redshift distribution uncertainties. To incorporate these errors into our $b_v$ prior, we use the 1$\sigma$  errors on $b_g dN/dz$ from \texttt{tomographer.org} (shown in Figure~\ref{fig: bdndz}). We follow the above procedure for calculating $P(b_v)$ given samples $\{k_*'^{(i)},f_{\rm gas}^{(i)}\}$, but repeat it three times: once using the mean $b_g dN/dz$, and then twice using the mean $(b_g dN/dz)\pm 1\sigma$ error. 
The marginalized posterior distribution,
\begin{eqnarray}\label{eq: Pbv with bgdndz integral}
    P(b_v) &=& \int dk_*'\, df_{\rm gas}\, d(b_g dN/dz)\, P(k_*',f_{\rm gas})\, P(b_g dN/dz)\nonumber \\
    &&\ \ \ \times\ P(b_v|k_*',f_{\rm gas},b_g dN/dz) \,  
    \nonumber\\
    &=& \int P(b_v|b_g dN/dz)\,
    P(b_g dN/dz)\, d(b_g dN/dz), \nonumber \\
\end{eqnarray}
is approximated by resampling the three $P(b_v)$'s with Gaussian weights $w_i \in [0.16, 0.68, 0.16]$ as
\begin{eqnarray}\label{eq: Pbv with bgdndz approx}
    P(b_v) &\approx& \sum_i w_i P(b_v|b_g dN/dz_i).
\end{eqnarray}
From this distribution, we compute the mean and credible intervals on $b_v$ that are used as priors in our cosmological analysis. The resulting posteriors for each \texttt{unWISE} and \texttt{WxSCOS} redshift bin are shown in Figure~\ref{fig: Pbv_results}. 
\section{Comparison with other bulk flow measurements}\label{app: vbulk comparison}
\begin{table*}
    \centering
    \renewcommand{\arraystretch}{1.4} 
    \caption{Summary of literature bulk flow realization measurements mapped to equivalent tophat scales. 
    The original measurements denote the quoted scale and realized bulk flow ($\hat V$) under each work's assumed window function. We plot these in Figure \ref{fig: comparison plot} at their equivalent tophat scale ($R_{\rm TH}$).}
    \vspace{0.3cm}
    \begin{tabular}{ll ccc c}
        \toprule
        & & \multicolumn{2}{c}{\textbf{Published Values}} & \multicolumn{2}{c}{\textbf{Tophat Equivalent}} \\
        \cmidrule(lr){3-4} \cmidrule(lr){5-6}
        \textbf{Reference} & \textbf{Fig. \ref{fig: comparison plot} Tag}\ \ \ \ \  & \textbf{Scale} [$h^{-1}$ Mpc] & $\hat V$ [km/s] & \textbf{$R_{\rm TH}$} [$h^{-1}$ Mpc] & $\hat V$ [km/s] \\
        
        \midrule
        \multicolumn{4}{l}{\small{\textbf{Gaussian Window}}} & \multicolumn{1}{c}{\small [$R_{\rm TH} \approx 2.12 R_G $, Eq.~\eqref{appeq: Reff from RG}]} &  (unchanged) \\
        Watkins (2009) \cite{Watkins_2009}    & \texttt{`W09'} & $R_{\rm G} = 50$ & $407 \pm 81$ & $106$ & $407 \pm 81$ \\
        %
        %
        Scrimgeour (2015) \cite{Scrimgeour_2015_6dfgs} & \texttt{`S15'} & $R_{\rm G} = 55.6$  & $243 \pm 58$ & $118$ & $243 \pm 58$ \\

        \midrule
        \multicolumn{4}{l}{\small{\textbf{$1/r^2$ Weighting}}} & \multicolumn{1}{c}{\small [$R_{\rm TH} \approx 0.707R_{1/r^2}$, Eq.~\eqref{appeq: Reff from Rinvr2}]} &  (unchanged) \\
        Peery (2018) \cite{Peery_2018}      & \texttt{`P18'} & $R_{1/r^2} = 150$ & $282$ & $106$ & $282$ \\
        Watkins (2023) \cite{Watkins_2023}    & \texttt{`W23-i'} & $R_{1/r^2} = 150$ & $387 \pm 28$ & $106$ & $387 \pm 28$ \\
                          & \texttt{`W23-ii'} & $R_{1/r^2} = 200$ & $419 \pm 36$ & $141$ & $419 \pm 36$ \\
        
        \midrule
        \multicolumn{4}{l}{\small{\textbf{Tophat Window}}} & \multicolumn{1}{c}{\small (unchanged)} & \multicolumn{1}{c}{\small (95\% $\rightarrow$ 68\% UL)} \\
        Planck (2013) \cite{Planck_2013}     & \texttt{`P13-i'} & $R_{\rm TH} = 200$  & $< 600$ (95\%) & $200$  & $< 300$ \\
             & \texttt{`P13-ii'} & $R_{\rm TH} = 500$  & $< 400$ (95\%) & $500$  & $< 200$ \\
             & \texttt{`P13-iii'} & $R_{\rm TH} = 1000$ & $< 300$ (95\%) & $1000$ & $< 150$ \\
        \bottomrule
    \end{tabular}
    \label{tab: bulk_flow_lit}
\end{table*}
This appendix is relevant to \S\ref{sec: comparison with other measurements} and Figure~\ref{fig: comparison plot}.

Since different survey window functions filter the cosmic density field differently, the physical scales probed by each study can vary~significantly.
One way around this hurdle was proposed by Ref. \cite{Li_2012}--- to use the theoretical bulk flow expectation ($\sigma_V$) as a standardizing bridge. 
If a study reports a bulk velocity measurement to be compared against a theoretical expectation $\sigma_{V_{\rm I}}(R_{\rm I})$ based on some idealized window function of scale $R_{\rm I}$, we can map this onto an equivalent tophat smoothing scale, $R_{\rm TH}$, by requiring $\sigma_{V_{\rm TH}}(R_{\rm TH})=\sigma_{V_{\rm I}}(R_{\rm I})$. 
By matching the theoretical variances $\sigma_V$, we map disparate literature measurements onto a common tophat equivalent, enabling a consistent comparison against our kSZ-derived tophat upper limits. 

In the subsections below, we describe how we map the scales of the idealized window functions used by various studies onto a tophat scale,~$R_{\rm TH}$, in order to make Figure~\ref{fig: comparison plot}. 
All works quoted in this figure constrain bulk flow amplitude realization $\hat V$, which we plot as-is. 
For a self-consistent comparison, we find tophat equivalents of our kSZ-derived upper limits [$\hat V^{1\sigma}_{\rm TH}$, Eq.~\eqref{eq: hat V_TH realization}] as described in \S\ref{sec: comparison with other measurements}.
Table \ref{tab: bulk_flow_lit} summarizes the original data published by these works, and the values that we plot in Figure \ref{fig: comparison plot}. 

\subsection{From Gaussian scale $R_{\rm G}$ to tophat scale $R_{\rm TH}$}
To do this, we need to solve for $R_{\rm TH}$ that satisfies
\begin{eqnarray}\label{appeq: sigma_vTH = sigma_vG}
    \sigma^2_{V_{\rm TH}}(R_{\rm TH}) = \sigma^2_{V_{\rm G}}(R_{\rm G}),
\end{eqnarray}
provided the study we want to compare to quotes a Gaussian window function scale, $R_{\rm G}$.
Using Eq. \eqref{eq: sigmaV}, writing $P_{mm}(k) = A\,k$,%
\footnote{The form $P_{mm}(k)=A k$ is only valid for $R_{\rm TH}>100\,h^{-1}$Mpc. Thus we only select measurements from works that probe an equivalent scale that meets this criterion.}
and defining $\kappa=k R$, the equivalence in Eq.~\eqref{appeq: sigma_vTH = sigma_vG} implies
\begin{eqnarray}\label{appeq: Reff from RG-integrals}
    &&\frac{1}{R_{\rm TH}^2}\int_{-\infty}^\infty d\ln\kappa\ \kappa^2 \big(W^{\rm TH}(\kappa)\big)^2 \nonumber \\
    &&\ \ \ \ \ \ \ \ \ \ \ \ \ \ \ \ \ \ = \frac{1}{R_{\rm G}^2}\int_{-\infty}^\infty d\ln\kappa\ \kappa^2 \big(W^{\rm G}(\kappa)\big)^2,
\end{eqnarray}
where $W^{\rm TH}(\kappa)$ is defined in Eq. \eqref{eq: WTH kappa}, and 
\begin{eqnarray} \label{appeq: WG window function}
    W^{\rm G}(\kappa) = \exp(-\kappa^2/2).
\end{eqnarray}
The integrals in Eq.~\eqref{appeq: Reff from RG-integrals} are analytically computable, and lead to the solution
\begin{eqnarray}\label{appeq: Reff from RG}
    R_{\rm TH} = \frac{3}{\sqrt2} R_{\rm G} \approx 2.12\,R_{\rm G}.
\end{eqnarray}
This is the conversion we use for Refs. \cite{Watkins_2009, Scrimgeour_2015_6dfgs} (`W09', `S15' in Figure~\ref{fig: comparison plot} and Table~\ref{tab: bulk_flow_lit}).

\subsection{From $R_{1/r^2}$ scale to tophat $R_{\rm TH}$}
Some of the works we compare to (Refs. \cite{Peery_2018, Watkins_2023}) compute their theoretical expectation of $\sigma_V$ using the idealized window function
\begin{eqnarray}
    W(r;R_{1/r^2}) = R^2_{1/r^2}/r^2.
\end{eqnarray}
As Ref. \cite{Peery_2018} shows [see their Figure 3 and Eq. (34)], this window function matches fairly well with a Gaussian window with the Gaussian averaging scale $R_{\rm G}\rightarrow R_{1/r^2}/3$. 
Thus, with this substitution in Eq. \eqref{appeq: Reff from RG}, we can map between their reported $R_{1/r^2}$ and our target $R_{\rm TH}$ as
\begin{eqnarray}\label{appeq: Reff from Rinvr2}
    R_{\rm TH} = \frac{3}{\sqrt2}R_{\rm G} \approx 0.707\, R_{1/r^2}.
\end{eqnarray}
This is the conversion we use when reporting values (`P18', `W23-\textit{i}', `W23-\textit{ii}') in Figure \ref{fig: comparison plot} and Table \ref{tab: bulk_flow_lit}.

\subsection{From the \citet{Planck_2013} tophat-based 95\% UL to 68\%}
The `P13-\textit{i}--\textit{iii}' points in Figure \ref{fig: comparison plot} and Table \ref{tab: bulk_flow_lit} from Ref.~\cite{Planck_2013} are already reported at a tophat scale equivalent to $R_{\rm TH}$. However, Ref. \cite{Planck_2013} reports 95\% upper limits (UL) on bulk velocity realization $\hat V$. For consistency with the our 68\% ULs and the other $1\sigma$ measurements that we compare with, we halve their reported 95\% UL bulk velocity values to get a 68\% UL.
%


%

\clearpage 
\bibliography{main.bib}

\begin{thebibliography}{80}%
\makeatletter
\providecommand \@ifxundefined [1]{%
 \@ifx{#1\undefined}
}%
\providecommand \@ifnum [1]{%
 \ifnum #1\expandafter \@firstoftwo
 \else \expandafter \@secondoftwo
 \fi
}%
\providecommand \@ifx [1]{%
 \ifx #1\expandafter \@firstoftwo
 \else \expandafter \@secondoftwo
 \fi
}%
\providecommand \natexlab [1]{#1}%
\providecommand \enquote  [1]{``#1''}%
\providecommand \bibnamefont  [1]{#1}%
\providecommand \bibfnamefont [1]{#1}%
\providecommand \citenamefont [1]{#1}%
\providecommand \href@noop [0]{\@secondoftwo}%
\providecommand \href [0]{\begingroup \@sanitize@url \@href}%
\providecommand \@href[1]{\@@startlink{#1}\@@href}%
\providecommand \@@href[1]{\endgroup#1\@@endlink}%
\providecommand \@sanitize@url [0]{\catcode `\\12\catcode `\$12\catcode `\&12\catcode `\#12\catcode `\^12\catcode `\_12\catcode `\%12\relax}%
\providecommand \@@startlink[1]{}%
\providecommand \@@endlink[0]{}%
\providecommand \url  [0]{\begingroup\@sanitize@url \@url }%
\providecommand \@url [1]{\endgroup\@href {#1}{\urlprefix }}%
\providecommand \urlprefix  [0]{URL }%
\providecommand \Eprint [0]{\href }%
\providecommand \doibase [0]{https://doi.org/}%
\providecommand \selectlanguage [0]{\@gobble}%
\providecommand \bibinfo  [0]{\@secondoftwo}%
\providecommand \bibfield  [0]{\@secondoftwo}%
\providecommand \translation [1]{[#1]}%
\providecommand \BibitemOpen [0]{}%
\providecommand \bibitemStop [0]{}%
\providecommand \bibitemNoStop [0]{.\EOS\space}%
\providecommand \EOS [0]{\spacefactor3000\relax}%
\providecommand \BibitemShut  [1]{\csname bibitem#1\endcsname}%
\let\auto@bib@innerbib\@empty
\bibitem [{\citenamefont {{Zeldovich}}\ and\ \citenamefont {{Sunyaev}}(1980)}]{Sunyaev_Zeldovich_1980}%
  \BibitemOpen
  \bibfield  {author} {\bibinfo {author} {\bibfnamefont {Y.~B.}\ \bibnamefont {{Zeldovich}}}\ and\ \bibinfo {author} {\bibfnamefont {S.~R.~A.}\ \bibnamefont {{Sunyaev}}},\ }\bibfield  {title} {\bibinfo {title} {{Peculiar velocities of galaxy clusters, and the mean density of matter in the universe}},\ }\href@noop {} {\bibfield  {journal} {\bibinfo  {journal} {Pisma v Astronomicheskii Zhurnal}\ }\textbf {\bibinfo {volume} {6}},\ \bibinfo {pages} {737} (\bibinfo {year} {1980})}\BibitemShut {NoStop}%
\bibitem [{\citenamefont {{Peebles}}(1980)}]{Peebles_1980}%
  \BibitemOpen
  \bibfield  {author} {\bibinfo {author} {\bibfnamefont {P.~J.~E.}\ \bibnamefont {{Peebles}}},\ }\href@noop {} {\emph {\bibinfo {title} {{The large-scale structure of the universe}}}}\ (\bibinfo {year} {1980})\BibitemShut {NoStop}%
\bibitem [{\citenamefont {Scrimgeour}\ \emph {et~al.}(2015)\citenamefont {Scrimgeour}, \citenamefont {Davis}, \citenamefont {Blake}, \citenamefont {Staveley-Smith}, \citenamefont {Magoulas}, \citenamefont {Springob}, \citenamefont {Beutler}, \citenamefont {Colless}, \citenamefont {Johnson}, \citenamefont {Jones}, \citenamefont {Koda}, \citenamefont {Lucey}, \citenamefont {Ma}, \citenamefont {Mould},\ and\ \citenamefont {Poole}}]{Scrimgeour_2015_6dfgs}%
  \BibitemOpen
  \bibfield  {author} {\bibinfo {author} {\bibfnamefont {M.~I.}\ \bibnamefont {Scrimgeour}}, \bibinfo {author} {\bibfnamefont {T.~M.}\ \bibnamefont {Davis}}, \bibinfo {author} {\bibfnamefont {C.}~\bibnamefont {Blake}}, \bibinfo {author} {\bibfnamefont {L.}~\bibnamefont {Staveley-Smith}}, \bibinfo {author} {\bibfnamefont {C.}~\bibnamefont {Magoulas}}, \bibinfo {author} {\bibfnamefont {C.~M.}\ \bibnamefont {Springob}}, \bibinfo {author} {\bibfnamefont {F.}~\bibnamefont {Beutler}}, \bibinfo {author} {\bibfnamefont {M.}~\bibnamefont {Colless}}, \bibinfo {author} {\bibfnamefont {A.}~\bibnamefont {Johnson}}, \bibinfo {author} {\bibfnamefont {D.~H.}\ \bibnamefont {Jones}}, \bibinfo {author} {\bibfnamefont {J.}~\bibnamefont {Koda}}, \bibinfo {author} {\bibfnamefont {J.~R.}\ \bibnamefont {Lucey}}, \bibinfo {author} {\bibfnamefont {Y.-Z.}\ \bibnamefont {Ma}}, \bibinfo {author} {\bibfnamefont {J.}~\bibnamefont {Mould}},\ and\ \bibinfo {author} {\bibfnamefont {G.~B.}\ \bibnamefont {Poole}},\ }\bibfield  {title} {\bibinfo
  {title} {The 6df galaxy survey: bulk flows on 50-70h-1mpc scales},\ }\href {https://doi.org/10.1093/mnras/stv2146} {\bibfield  {journal} {\bibinfo  {journal} {Monthly Notices of the Royal Astronomical Society}\ }\textbf {\bibinfo {volume} {455}},\ \bibinfo {pages} {386–401} (\bibinfo {year} {2015})}\BibitemShut {NoStop}%
\bibitem [{\citenamefont {Planck~Collaboration}(2014)}]{Planck_2013}%
  \BibitemOpen
  \bibfield  {author} {\bibinfo {author} {\bibfnamefont {a.}~\bibnamefont {Planck~Collaboration}},\ }\bibfield  {title} {\bibinfo {title} {Planck2013 results. xvi. cosmological parameters},\ }\href {https://doi.org/10.1051/0004-6361/201321591} {\bibfield  {journal} {\bibinfo  {journal} {Astronomy \& Astrophysics}\ }\textbf {\bibinfo {volume} {571}},\ \bibinfo {pages} {A16} (\bibinfo {year} {2014})}\BibitemShut {NoStop}%
\bibitem [{\citenamefont {Lavaux}\ \emph {et~al.}(2013)\citenamefont {Lavaux}, \citenamefont {Afshordi},\ and\ \citenamefont {Hudson}}]{Lavaux_2013}%
  \BibitemOpen
  \bibfield  {author} {\bibinfo {author} {\bibfnamefont {G.}~\bibnamefont {Lavaux}}, \bibinfo {author} {\bibfnamefont {N.}~\bibnamefont {Afshordi}},\ and\ \bibinfo {author} {\bibfnamefont {M.~J.}\ \bibnamefont {Hudson}},\ }\bibfield  {title} {\bibinfo {title} {First measurement of the bulk flow of nearby galaxies using the cosmic microwave background},\ }\href {https://doi.org/10.1093/mnras/sts698} {\bibfield  {journal} {\bibinfo  {journal} {Monthly Notices of the Royal Astronomical Society}\ }\textbf {\bibinfo {volume} {430}},\ \bibinfo {pages} {1617–1635} (\bibinfo {year} {2013})}\BibitemShut {NoStop}%
\bibitem [{\citenamefont {Watkins}\ and\ \citenamefont {Feldman}(2025)}]{Watkins_2025}%
  \BibitemOpen
  \bibfield  {author} {\bibinfo {author} {\bibfnamefont {R.}~\bibnamefont {Watkins}}\ and\ \bibinfo {author} {\bibfnamefont {H.~A.}\ \bibnamefont {Feldman}},\ }\href {https://arxiv.org/abs/2512.03168} {\bibinfo {title} {The origins of the bulk flow}} (\bibinfo {year} {2025}),\ \Eprint {https://arxiv.org/abs/2512.03168} {arXiv:2512.03168 [astro-ph.CO]} \BibitemShut {NoStop}%
\bibitem [{\citenamefont {Kashlinsky}\ \emph {et~al.}(2008)\citenamefont {Kashlinsky}, \citenamefont {Atrio-Barandela}, \citenamefont {Kocevski},\ and\ \citenamefont {Ebeling}}]{Kashlinsky_2008}%
  \BibitemOpen
  \bibfield  {author} {\bibinfo {author} {\bibfnamefont {A.}~\bibnamefont {Kashlinsky}}, \bibinfo {author} {\bibfnamefont {F.}~\bibnamefont {Atrio-Barandela}}, \bibinfo {author} {\bibfnamefont {D.}~\bibnamefont {Kocevski}},\ and\ \bibinfo {author} {\bibfnamefont {H.}~\bibnamefont {Ebeling}},\ }\bibfield  {title} {\bibinfo {title} {A measurement of large-scale peculiar velocities of clusters of galaxies: Results and cosmological implications},\ }\href {https://doi.org/10.1086/592947} {\bibfield  {journal} {\bibinfo  {journal} {The Astrophysical Journal}\ }\textbf {\bibinfo {volume} {686}},\ \bibinfo {pages} {L49–L52} (\bibinfo {year} {2008})}\BibitemShut {NoStop}%
\bibitem [{\citenamefont {{Watkins}}\ and\ \citenamefont {{Feldman}}(2015)}]{Watkins_2015a}%
  \BibitemOpen
  \bibfield  {author} {\bibinfo {author} {\bibfnamefont {R.}~\bibnamefont {{Watkins}}}\ and\ \bibinfo {author} {\bibfnamefont {H.~A.}\ \bibnamefont {{Feldman}}},\ }\bibfield  {title} {\bibinfo {title} {{Large-scale bulk flows from the Cosmicflows-2 catalogue}},\ }\href {https://doi.org/10.1093/mnras/stu2414} {\bibfield  {journal} {\bibinfo  {journal} {Monthly Notices of the Royal Astronomical Society}\ }\textbf {\bibinfo {volume} {447}},\ \bibinfo {pages} {132} (\bibinfo {year} {2015})},\ \Eprint {https://arxiv.org/abs/1407.6940} {arXiv:1407.6940 [astro-ph.CO]} \BibitemShut {NoStop}%
\bibitem [{\citenamefont {{Peery}}\ \emph {et~al.}(2018)\citenamefont {{Peery}}, \citenamefont {{Watkins}},\ and\ \citenamefont {{Feldman}}}]{Peery_2018}%
  \BibitemOpen
  \bibfield  {author} {\bibinfo {author} {\bibfnamefont {S.}~\bibnamefont {{Peery}}}, \bibinfo {author} {\bibfnamefont {R.}~\bibnamefont {{Watkins}}},\ and\ \bibinfo {author} {\bibfnamefont {H.~A.}\ \bibnamefont {{Feldman}}},\ }\bibfield  {title} {\bibinfo {title} {{Easily interpretable bulk flows: continuing tension with the standard cosmological model}},\ }\href {https://doi.org/10.1093/mnras/sty2332} {\bibfield  {journal} {\bibinfo  {journal} {Monthly Notices of the Royal Astronomical Society}\ }\textbf {\bibinfo {volume} {481}},\ \bibinfo {pages} {1368} (\bibinfo {year} {2018})},\ \Eprint {https://arxiv.org/abs/1808.07772} {arXiv:1808.07772 [astro-ph.CO]} \BibitemShut {NoStop}%
\bibitem [{\citenamefont {{Whitford}}\ \emph {et~al.}(2023)\citenamefont {{Whitford}}, \citenamefont {{Howlett}},\ and\ \citenamefont {{Davis}}}]{Whitford_2023}%
  \BibitemOpen
  \bibfield  {author} {\bibinfo {author} {\bibfnamefont {A.~M.}\ \bibnamefont {{Whitford}}}, \bibinfo {author} {\bibfnamefont {C.}~\bibnamefont {{Howlett}}},\ and\ \bibinfo {author} {\bibfnamefont {T.~M.}\ \bibnamefont {{Davis}}},\ }\bibfield  {title} {\bibinfo {title} {{Evaluating bulk flow estimators for CosmicFlows-4 measurements}},\ }\href {https://doi.org/10.1093/mnras/stad2764} {\bibfield  {journal} {\bibinfo  {journal} {Monthly Notices of the Royal Astronomical Society}\ }\textbf {\bibinfo {volume} {526}},\ \bibinfo {pages} {3051} (\bibinfo {year} {2023})},\ \Eprint {https://arxiv.org/abs/2306.11269} {arXiv:2306.11269 [astro-ph.CO]} \BibitemShut {NoStop}%
\bibitem [{\citenamefont {Watkins}\ \emph {et~al.}(2023)\citenamefont {Watkins}, \citenamefont {Allen}, \citenamefont {Bradford}, \citenamefont {Ramon}, \citenamefont {Walker}, \citenamefont {Feldman}, \citenamefont {Cionitti}, \citenamefont {Al-Shorman}, \citenamefont {Kourkchi},\ and\ \citenamefont {Tully}}]{Watkins_2023}%
  \BibitemOpen
  \bibfield  {author} {\bibinfo {author} {\bibfnamefont {R.}~\bibnamefont {Watkins}}, \bibinfo {author} {\bibfnamefont {T.}~\bibnamefont {Allen}}, \bibinfo {author} {\bibfnamefont {C.~J.}\ \bibnamefont {Bradford}}, \bibinfo {author} {\bibfnamefont {A.}~\bibnamefont {Ramon}}, \bibinfo {author} {\bibfnamefont {A.}~\bibnamefont {Walker}}, \bibinfo {author} {\bibfnamefont {H.~A.}\ \bibnamefont {Feldman}}, \bibinfo {author} {\bibfnamefont {R.}~\bibnamefont {Cionitti}}, \bibinfo {author} {\bibfnamefont {Y.}~\bibnamefont {Al-Shorman}}, \bibinfo {author} {\bibfnamefont {E.}~\bibnamefont {Kourkchi}},\ and\ \bibinfo {author} {\bibfnamefont {R.~B.}\ \bibnamefont {Tully}},\ }\bibfield  {title} {\bibinfo {title} {Analysing the large-scale bulk flow using cosmicflows4: increasing tension with the standard cosmological model},\ }\href {https://doi.org/10.1093/mnras/stad1984} {\bibfield  {journal} {\bibinfo  {journal} {Monthly Notices of the Royal Astronomical Society}\ }\textbf {\bibinfo {volume} {524}},\ \bibinfo {pages}
  {1885–1892} (\bibinfo {year} {2023})}\BibitemShut {NoStop}%
\bibitem [{\citenamefont {{Turner}}(2024)}]{Turner_2024}%
  \BibitemOpen
  \bibfield  {author} {\bibinfo {author} {\bibfnamefont {R.~J.}\ \bibnamefont {{Turner}}},\ }\bibfield  {title} {\bibinfo {title} {{Cosmology with Peculiar Velocity Surveys}},\ }\href {https://doi.org/10.48550/arXiv.2411.19484} {\bibfield  {journal} {\bibinfo  {journal} {arXiv e-prints}\ ,\ \bibinfo {eid} {arXiv:2411.19484}} (\bibinfo {year} {2024})},\ \Eprint {https://arxiv.org/abs/2411.19484} {arXiv:2411.19484 [astro-ph.CO]} \BibitemShut {NoStop}%
\bibitem [{\citenamefont {{Tully}}\ and\ \citenamefont {{Fisher}}(1977)}]{Tully_1977}%
  \BibitemOpen
  \bibfield  {author} {\bibinfo {author} {\bibfnamefont {R.~B.}\ \bibnamefont {{Tully}}}\ and\ \bibinfo {author} {\bibfnamefont {J.~R.}\ \bibnamefont {{Fisher}}},\ }\bibfield  {title} {\bibinfo {title} {{A new method of determining distances to galaxies.}},\ }\href@noop {} {\bibfield  {journal} {\bibinfo  {journal} {\aap}\ }\textbf {\bibinfo {volume} {54}},\ \bibinfo {pages} {661} (\bibinfo {year} {1977})}\BibitemShut {NoStop}%
\bibitem [{\citenamefont {{Dressler}}\ \emph {et~al.}(1987)\citenamefont {{Dressler}}, \citenamefont {{Lynden-Bell}}, \citenamefont {{Burstein}}, \citenamefont {{Davies}}, \citenamefont {{Faber}}, \citenamefont {{Terlevich}},\ and\ \citenamefont {{Wegner}}}]{Dressler_1987_FP}%
  \BibitemOpen
  \bibfield  {author} {\bibinfo {author} {\bibfnamefont {A.}~\bibnamefont {{Dressler}}}, \bibinfo {author} {\bibfnamefont {D.}~\bibnamefont {{Lynden-Bell}}}, \bibinfo {author} {\bibfnamefont {D.}~\bibnamefont {{Burstein}}}, \bibinfo {author} {\bibfnamefont {R.~L.}\ \bibnamefont {{Davies}}}, \bibinfo {author} {\bibfnamefont {S.~M.}\ \bibnamefont {{Faber}}}, \bibinfo {author} {\bibfnamefont {R.}~\bibnamefont {{Terlevich}}},\ and\ \bibinfo {author} {\bibfnamefont {G.}~\bibnamefont {{Wegner}}},\ }\bibfield  {title} {\bibinfo {title} {{Spectroscopy and Photometry of Elliptical Galaxies. I. New Distance Estimator}},\ }\href {https://doi.org/10.1086/164947} {\bibfield  {journal} {\bibinfo  {journal} {\apj}\ }\textbf {\bibinfo {volume} {313}},\ \bibinfo {pages} {42} (\bibinfo {year} {1987})}\BibitemShut {NoStop}%
\bibitem [{\citenamefont {{Phillips}}(1993)}]{Phillips_1993_SNeIa}%
  \BibitemOpen
  \bibfield  {author} {\bibinfo {author} {\bibfnamefont {M.~M.}\ \bibnamefont {{Phillips}}},\ }\bibfield  {title} {\bibinfo {title} {{The Absolute Magnitudes of Type IA Supernovae}},\ }\href {https://doi.org/10.1086/186970} {\bibfield  {journal} {\bibinfo  {journal} {\apjl}\ }\textbf {\bibinfo {volume} {413}},\ \bibinfo {pages} {L105} (\bibinfo {year} {1993})}\BibitemShut {NoStop}%
\bibitem [{\citenamefont {{Leavitt}}\ and\ \citenamefont {{Pickering}}(1912)}]{Leavitt_1912_cepheids}%
  \BibitemOpen
  \bibfield  {author} {\bibinfo {author} {\bibfnamefont {H.~S.}\ \bibnamefont {{Leavitt}}}\ and\ \bibinfo {author} {\bibfnamefont {E.~C.}\ \bibnamefont {{Pickering}}},\ }\bibfield  {title} {\bibinfo {title} {{Periods of 25 Variable Stars in the Small Magellanic Cloud.}},\ }\href@noop {} {\bibfield  {journal} {\bibinfo  {journal} {Harvard College Observatory Circular}\ }\textbf {\bibinfo {volume} {173}},\ \bibinfo {pages} {1} (\bibinfo {year} {1912})}\BibitemShut {NoStop}%
\bibitem [{\citenamefont {{Lee}}\ \emph {et~al.}(1993)\citenamefont {{Lee}}, \citenamefont {{Freedman}},\ and\ \citenamefont {{Madore}}}]{Lee_1993_TRGB}%
  \BibitemOpen
  \bibfield  {author} {\bibinfo {author} {\bibfnamefont {M.~G.}\ \bibnamefont {{Lee}}}, \bibinfo {author} {\bibfnamefont {W.~L.}\ \bibnamefont {{Freedman}}},\ and\ \bibinfo {author} {\bibfnamefont {B.~F.}\ \bibnamefont {{Madore}}},\ }\bibfield  {title} {\bibinfo {title} {{The Tip of the Red Giant Branch as a Distance Indicator for Resolved Galaxies}},\ }\href {https://doi.org/10.1086/173334} {\bibfield  {journal} {\bibinfo  {journal} {\apj}\ }\textbf {\bibinfo {volume} {417}},\ \bibinfo {pages} {553} (\bibinfo {year} {1993})}\BibitemShut {NoStop}%
\bibitem [{\citenamefont {{Sunyaev}}\ and\ \citenamefont {{Zeldovich}}(1980)}]{Sunyaev_1980_ksz}%
  \BibitemOpen
  \bibfield  {author} {\bibinfo {author} {\bibfnamefont {R.~A.}\ \bibnamefont {{Sunyaev}}}\ and\ \bibinfo {author} {\bibfnamefont {Y.~B.}\ \bibnamefont {{Zeldovich}}},\ }\bibfield  {title} {\bibinfo {title} {{The velocity of clusters of galaxies relative to the microwave background - The possibility of its measurement.}},\ }\href {https://doi.org/10.1093/mnras/190.3.413} {\bibfield  {journal} {\bibinfo  {journal} {Monthly Notices of the Royal Astronomical Society}\ }\textbf {\bibinfo {volume} {190}},\ \bibinfo {pages} {413} (\bibinfo {year} {1980})}\BibitemShut {NoStop}%
\bibitem [{\citenamefont {Shao}\ \emph {et~al.}(2011)\citenamefont {Shao}, \citenamefont {Zhang}, \citenamefont {Lin}, \citenamefont {Jing},\ and\ \citenamefont {Pan}}]{Shao_2011}%
  \BibitemOpen
  \bibfield  {author} {\bibinfo {author} {\bibfnamefont {J.}~\bibnamefont {Shao}}, \bibinfo {author} {\bibfnamefont {P.}~\bibnamefont {Zhang}}, \bibinfo {author} {\bibfnamefont {W.}~\bibnamefont {Lin}}, \bibinfo {author} {\bibfnamefont {Y.}~\bibnamefont {Jing}},\ and\ \bibinfo {author} {\bibfnamefont {J.}~\bibnamefont {Pan}},\ }\bibfield  {title} {\bibinfo {title} {Kinetic sunyaev-zel’dovich tomography with spectroscopic redshift surveys: Kinetic sz tomography},\ }\href {https://doi.org/10.1111/j.1365-2966.2011.18166.x} {\bibfield  {journal} {\bibinfo  {journal} {Monthly Notices of the Royal Astronomical Society}\ }\textbf {\bibinfo {volume} {413}},\ \bibinfo {pages} {628–642} (\bibinfo {year} {2011})}\BibitemShut {NoStop}%
\bibitem [{\citenamefont {Munshi}\ \emph {et~al.}(2016)\citenamefont {Munshi}, \citenamefont {Iliev}, \citenamefont {Dixon},\ and\ \citenamefont {Coles}}]{Munshi_2016}%
  \BibitemOpen
  \bibfield  {author} {\bibinfo {author} {\bibfnamefont {D.}~\bibnamefont {Munshi}}, \bibinfo {author} {\bibfnamefont {I.~T.}\ \bibnamefont {Iliev}}, \bibinfo {author} {\bibfnamefont {K.~L.}\ \bibnamefont {Dixon}},\ and\ \bibinfo {author} {\bibfnamefont {P.}~\bibnamefont {Coles}},\ }\bibfield  {title} {\bibinfo {title} {Extracting the late-time kinetic sunyaev–zel’dovich effect},\ }\href {https://doi.org/10.1093/mnras/stw2067} {\bibfield  {journal} {\bibinfo  {journal} {Monthly Notices of the Royal Astronomical Society}\ }\textbf {\bibinfo {volume} {463}},\ \bibinfo {pages} {2425–2442} (\bibinfo {year} {2016})}\BibitemShut {NoStop}%
\bibitem [{\citenamefont {Terrana}\ \emph {et~al.}(2017)\citenamefont {Terrana}, \citenamefont {Harris},\ and\ \citenamefont {Johnson}}]{Terrana_2017}%
  \BibitemOpen
  \bibfield  {author} {\bibinfo {author} {\bibfnamefont {A.}~\bibnamefont {Terrana}}, \bibinfo {author} {\bibfnamefont {M.-J.}\ \bibnamefont {Harris}},\ and\ \bibinfo {author} {\bibfnamefont {M.~C.}\ \bibnamefont {Johnson}},\ }\bibfield  {title} {\bibinfo {title} {Analyzing the cosmic variance limit of remote dipole measurements of the cosmic microwave background using the large-scale kinetic sunyaev zel’dovich effect},\ }\href {https://doi.org/10.1088/1475-7516/2017/02/040} {\bibfield  {journal} {\bibinfo  {journal} {Journal of Cosmology and Astroparticle Physics}\ }\textbf {\bibinfo {volume} {2017}}\bibinfo  {number} { (02)},\ \bibinfo {pages} {040–040}}\BibitemShut {NoStop}%
\bibitem [{\citenamefont {Deutsch}\ \emph {et~al.}(2018)\citenamefont {Deutsch}, \citenamefont {Dimastrogiovanni}, \citenamefont {Johnson}, \citenamefont {Münchmeyer},\ and\ \citenamefont {Terrana}}]{Deutsch_2018}%
  \BibitemOpen
\bibfield  {number} {  }\bibfield  {author} {\bibinfo {author} {\bibfnamefont {A.-S.}\ \bibnamefont {Deutsch}}, \bibinfo {author} {\bibfnamefont {E.}~\bibnamefont {Dimastrogiovanni}}, \bibinfo {author} {\bibfnamefont {M.~C.}\ \bibnamefont {Johnson}}, \bibinfo {author} {\bibfnamefont {M.}~\bibnamefont {Münchmeyer}},\ and\ \bibinfo {author} {\bibfnamefont {A.}~\bibnamefont {Terrana}},\ }\bibfield  {title} {\bibinfo {title} {Reconstruction of the remote dipole and quadrupole fields from the kinetic sunyaev zel’dovich and polarized sunyaev zel’dovich effects},\ }\bibfield  {journal} {\bibinfo  {journal} {Physical Review D}\ }\textbf {\bibinfo {volume} {98}},\ \href {https://doi.org/10.1103/physrevd.98.123501} {10.1103/physrevd.98.123501} (\bibinfo {year} {2018})\BibitemShut {NoStop}%
\bibitem [{\citenamefont {Smith}\ \emph {et~al.}(2018)\citenamefont {Smith}, \citenamefont {Madhavacheril}, \citenamefont {Münchmeyer}, \citenamefont {Ferraro}, \citenamefont {Giri},\ and\ \citenamefont {Johnson}}]{Smith_2018}%
  \BibitemOpen
  \bibfield  {author} {\bibinfo {author} {\bibfnamefont {K.~M.}\ \bibnamefont {Smith}}, \bibinfo {author} {\bibfnamefont {M.~S.}\ \bibnamefont {Madhavacheril}}, \bibinfo {author} {\bibfnamefont {M.}~\bibnamefont {Münchmeyer}}, \bibinfo {author} {\bibfnamefont {S.}~\bibnamefont {Ferraro}}, \bibinfo {author} {\bibfnamefont {U.}~\bibnamefont {Giri}},\ and\ \bibinfo {author} {\bibfnamefont {M.~C.}\ \bibnamefont {Johnson}},\ }\href {https://arxiv.org/abs/1810.13423} {\bibinfo {title} {Ksz tomography and the bispectrum}} (\bibinfo {year} {2018}),\ \Eprint {https://arxiv.org/abs/1810.13423} {arXiv:1810.13423 [astro-ph.CO]} \BibitemShut {NoStop}%
\bibitem [{\citenamefont {Cayuso}\ \emph {et~al.}(2023)\citenamefont {Cayuso}, \citenamefont {Bloch}, \citenamefont {Hotinli}, \citenamefont {Johnson},\ and\ \citenamefont {McCarthy}}]{Cayuso_2023}%
  \BibitemOpen
  \bibfield  {author} {\bibinfo {author} {\bibfnamefont {J.}~\bibnamefont {Cayuso}}, \bibinfo {author} {\bibfnamefont {R.}~\bibnamefont {Bloch}}, \bibinfo {author} {\bibfnamefont {S.~C.}\ \bibnamefont {Hotinli}}, \bibinfo {author} {\bibfnamefont {M.~C.}\ \bibnamefont {Johnson}},\ and\ \bibinfo {author} {\bibfnamefont {F.}~\bibnamefont {McCarthy}},\ }\bibfield  {title} {\bibinfo {title} {Velocity reconstruction with the cosmic microwave background and galaxy surveys},\ }\href {https://doi.org/10.1088/1475-7516/2023/02/051} {\bibfield  {journal} {\bibinfo  {journal} {Journal of Cosmology and Astroparticle Physics}\ }\textbf {\bibinfo {volume} {2023}}\bibinfo  {number} { (02)},\ \bibinfo {pages} {051}}\BibitemShut {NoStop}%
\bibitem [{\citenamefont {Bloch}\ and\ \citenamefont {Johnson}(2024)}]{Bloch_2024}%
  \BibitemOpen
\bibfield  {number} {  }\bibfield  {author} {\bibinfo {author} {\bibfnamefont {R.}~\bibnamefont {Bloch}}\ and\ \bibinfo {author} {\bibfnamefont {M.~C.}\ \bibnamefont {Johnson}},\ }\href {https://arxiv.org/abs/2405.00809} {\bibinfo {title} {Kinetic sunyaev zel'dovich velocity reconstruction from planck and unwise}} (\bibinfo {year} {2024}),\ \Eprint {https://arxiv.org/abs/2405.00809} {arXiv:2405.00809 [astro-ph.CO]} \BibitemShut {NoStop}%
\bibitem [{\citenamefont {Zhang}(2010)}]{Zhang_2010}%
  \BibitemOpen
  \bibfield  {author} {\bibinfo {author} {\bibfnamefont {P.}~\bibnamefont {Zhang}},\ }\bibfield  {title} {\bibinfo {title} {The dark flow induced small-scale kinetic sunyaev–zel’dovich effect},\ }\href {https://doi.org/10.1111/j.1745-3933.2010.00899.x} {\bibfield  {journal} {\bibinfo  {journal} {Monthly Notices of the Royal Astronomical Society: Letters}\ }\textbf {\bibinfo {volume} {407}},\ \bibinfo {pages} {L36–L40} (\bibinfo {year} {2010})}\BibitemShut {NoStop}%
\bibitem [{\citenamefont {Zhang}\ and\ \citenamefont {Stebbins}(2011)}]{Zhang_2011}%
  \BibitemOpen
  \bibfield  {author} {\bibinfo {author} {\bibfnamefont {P.}~\bibnamefont {Zhang}}\ and\ \bibinfo {author} {\bibfnamefont {A.}~\bibnamefont {Stebbins}},\ }\bibfield  {title} {\bibinfo {title} {Confirmation of the copernican principle at gpc radial scale and above from the kinetic sunyaev-zel’dovich effect power spectrum},\ }\bibfield  {journal} {\bibinfo  {journal} {Physical Review Letters}\ }\textbf {\bibinfo {volume} {107}},\ \href {https://doi.org/10.1103/physrevlett.107.041301} {10.1103/physrevlett.107.041301} (\bibinfo {year} {2011})\BibitemShut {NoStop}%
\bibitem [{\citenamefont {Cayuso}\ and\ \citenamefont {Johnson}(2020)}]{Cayuso_2020}%
  \BibitemOpen
  \bibfield  {author} {\bibinfo {author} {\bibfnamefont {J.~I.}\ \bibnamefont {Cayuso}}\ and\ \bibinfo {author} {\bibfnamefont {M.~C.}\ \bibnamefont {Johnson}},\ }\bibfield  {title} {\bibinfo {title} {Towards testing cmb anomalies using the kinetic and polarized sunyaev-zel’dovich effects},\ }\bibfield  {journal} {\bibinfo  {journal} {Physical Review D}\ }\textbf {\bibinfo {volume} {101}},\ \href {https://doi.org/10.1103/physrevd.101.123508} {10.1103/physrevd.101.123508} (\bibinfo {year} {2020})\BibitemShut {NoStop}%
\bibitem [{\citenamefont {Krywonos}\ \emph {et~al.}(2024)\citenamefont {Krywonos}, \citenamefont {Hotinli},\ and\ \citenamefont {Johnson}}]{Krywonos_2024}%
  \BibitemOpen
  \bibfield  {author} {\bibinfo {author} {\bibfnamefont {J.}~\bibnamefont {Krywonos}}, \bibinfo {author} {\bibfnamefont {S.~C.}\ \bibnamefont {Hotinli}},\ and\ \bibinfo {author} {\bibfnamefont {M.~C.}\ \bibnamefont {Johnson}},\ }\href {https://arxiv.org/abs/2408.05264} {\bibinfo {title} {Constraints on cosmology beyond $\lambda$cdm with kinetic sunyaev zel'dovich velocity reconstruction}} (\bibinfo {year} {2024}),\ \Eprint {https://arxiv.org/abs/2408.05264} {arXiv:2408.05264 [astro-ph.CO]} \BibitemShut {NoStop}%
\bibitem [{\citenamefont {Schlafly}\ \emph {et~al.}(2019)\citenamefont {Schlafly}, \citenamefont {Meisner},\ and\ \citenamefont {Green}}]{Schlafly_2019unwise}%
  \BibitemOpen
  \bibfield  {author} {\bibinfo {author} {\bibfnamefont {E.~F.}\ \bibnamefont {Schlafly}}, \bibinfo {author} {\bibfnamefont {A.~M.}\ \bibnamefont {Meisner}},\ and\ \bibinfo {author} {\bibfnamefont {G.~M.}\ \bibnamefont {Green}},\ }\bibfield  {title} {\bibinfo {title} {The unwise catalog: Two billion infrared sources from five years of wise imaging},\ }\href {https://doi.org/10.3847/1538-4365/aafbea} {\bibfield  {journal} {\bibinfo  {journal} {The Astrophysical Journal Supplement Series}\ }\textbf {\bibinfo {volume} {240}},\ \bibinfo {pages} {30} (\bibinfo {year} {2019})}\BibitemShut {NoStop}%
\bibitem [{\citenamefont {Krolewski}\ \emph {et~al.}(2020)\citenamefont {Krolewski}, \citenamefont {Ferraro}, \citenamefont {Schlafly},\ and\ \citenamefont {White}}]{Krolewski_2020unwise}%
  \BibitemOpen
  \bibfield  {author} {\bibinfo {author} {\bibfnamefont {A.}~\bibnamefont {Krolewski}}, \bibinfo {author} {\bibfnamefont {S.}~\bibnamefont {Ferraro}}, \bibinfo {author} {\bibfnamefont {E.~F.}\ \bibnamefont {Schlafly}},\ and\ \bibinfo {author} {\bibfnamefont {M.}~\bibnamefont {White}},\ }\bibfield  {title} {\bibinfo {title} {unwise tomography of planck cmb lensing},\ }\href {https://doi.org/10.1088/1475-7516/2020/05/047} {\bibfield  {journal} {\bibinfo  {journal} {Journal of Cosmology and Astroparticle Physics}\ }\textbf {\bibinfo {volume} {2020}}\bibinfo  {number} { (05)},\ \bibinfo {pages} {047–047}}\BibitemShut {NoStop}%
\bibitem [{\citenamefont {{Planck Collaboration}}\ \emph {et~al.}(2020)\citenamefont {{Planck Collaboration}}, \citenamefont {{Aghanim}}, \citenamefont {{Akrami}}, \citenamefont {{Arroja}}, \citenamefont {{Ashdown}},\ and\ \citenamefont {{Aumont}}}]{Planck18_2020}%
  \BibitemOpen
\bibfield  {number} {  }\bibfield  {author} {\bibinfo {author} {\bibnamefont {{Planck Collaboration}}}, \bibinfo {author} {\bibfnamefont {N.}~\bibnamefont {{Aghanim}}}, \bibinfo {author} {\bibfnamefont {Y.}~\bibnamefont {{Akrami}}}, \bibinfo {author} {\bibfnamefont {F.}~\bibnamefont {{Arroja}}}, \bibinfo {author} {\bibfnamefont {M.}~\bibnamefont {{Ashdown}}},\ and\ \bibinfo {author} {\bibnamefont {{Aumont}}},\ }\bibfield  {title} {\bibinfo {title} {{Planck 2018 results. I. Overview and the cosmological legacy of Planck}},\ }\href {https://doi.org/10.1051/0004-6361/201833880} {\bibfield  {journal} {\bibinfo  {journal} {Astron. Astrophys.}\ }\textbf {\bibinfo {volume} {641}},\ \bibinfo {eid} {A1} (\bibinfo {year} {2020})},\ \Eprint {https://arxiv.org/abs/1807.06205} {arXiv:1807.06205 [astro-ph.CO]} \BibitemShut {NoStop}%
\bibitem [{\citenamefont {{Bilicki}}\ \emph {et~al.}(2016)\citenamefont {{Bilicki}}, \citenamefont {{Peacock}}, \citenamefont {{Jarrett}}, \citenamefont {{Cluver}}, \citenamefont {{Maddox}}, \citenamefont {{Brown}}, \citenamefont {{Taylor}}, \citenamefont {{Hambly}}, \citenamefont {{Solarz}}, \citenamefont {{Holwerda}}, \citenamefont {{Baldry}}, \citenamefont {{Loveday}}, \citenamefont {{Moffett}}, \citenamefont {{Hopkins}}, \citenamefont {{Driver}}, \citenamefont {{Alpaslan}},\ and\ \citenamefont {{Bland-Hawthorn}}}]{Bilicki_2016wscos}%
  \BibitemOpen
  \bibfield  {author} {\bibinfo {author} {\bibfnamefont {M.}~\bibnamefont {{Bilicki}}}, \bibinfo {author} {\bibfnamefont {J.~A.}\ \bibnamefont {{Peacock}}}, \bibinfo {author} {\bibfnamefont {T.~H.}\ \bibnamefont {{Jarrett}}}, \bibinfo {author} {\bibfnamefont {M.~E.}\ \bibnamefont {{Cluver}}}, \bibinfo {author} {\bibfnamefont {N.}~\bibnamefont {{Maddox}}}, \bibinfo {author} {\bibfnamefont {M.~J.~I.}\ \bibnamefont {{Brown}}}, \bibinfo {author} {\bibfnamefont {E.~N.}\ \bibnamefont {{Taylor}}}, \bibinfo {author} {\bibfnamefont {N.~C.}\ \bibnamefont {{Hambly}}}, \bibinfo {author} {\bibfnamefont {A.}~\bibnamefont {{Solarz}}}, \bibinfo {author} {\bibfnamefont {B.~W.}\ \bibnamefont {{Holwerda}}}, \bibinfo {author} {\bibfnamefont {I.}~\bibnamefont {{Baldry}}}, \bibinfo {author} {\bibfnamefont {J.}~\bibnamefont {{Loveday}}}, \bibinfo {author} {\bibfnamefont {A.}~\bibnamefont {{Moffett}}}, \bibinfo {author} {\bibfnamefont {A.~M.}\ \bibnamefont {{Hopkins}}}, \bibinfo {author} {\bibfnamefont {S.~P.}\ \bibnamefont
  {{Driver}}}, \bibinfo {author} {\bibfnamefont {M.}~\bibnamefont {{Alpaslan}}},\ and\ \bibinfo {author} {\bibfnamefont {J.}~\bibnamefont {{Bland-Hawthorn}}},\ }\bibfield  {title} {\bibinfo {title} {{WISE {\texttimes} SuperCOSMOS Photometric Redshift Catalog: 20 Million Galaxies over 3/pi Steradians}},\ }\href {https://doi.org/10.3847/0067-0049/225/1/5} {\bibfield  {journal} {\bibinfo  {journal} {Astrophys. J. Suppl.}\ }\textbf {\bibinfo {volume} {225}},\ \bibinfo {eid} {5} (\bibinfo {year} {2016})},\ \Eprint {https://arxiv.org/abs/1607.01182} {arXiv:1607.01182 [astro-ph.CO]} \BibitemShut {NoStop}%
\bibitem [{\citenamefont {Haehnelt}\ and\ \citenamefont {Tegmark}(1996)}]{Haehnelt_1996}%
  \BibitemOpen
  \bibfield  {author} {\bibinfo {author} {\bibfnamefont {M.~G.}\ \bibnamefont {Haehnelt}}\ and\ \bibinfo {author} {\bibfnamefont {M.}~\bibnamefont {Tegmark}},\ }\bibfield  {title} {\bibinfo {title} {Using the kinematic sunyaev-zeldovich effect to determine the peculiar velocities of clusters of galaxies},\ }\href {https://doi.org/10.1093/mnras/279.2.545} {\bibfield  {journal} {\bibinfo  {journal} {Monthly Notices of the Royal Astronomical Society}\ }\textbf {\bibinfo {volume} {279}},\ \bibinfo {pages} {545–556} (\bibinfo {year} {1996})}\BibitemShut {NoStop}%
\bibitem [{\citenamefont {{Kashlinsky}}\ and\ \citenamefont {{Atrio-Barandela}}(2000)}]{Kashlinsky_2000}%
  \BibitemOpen
  \bibfield  {author} {\bibinfo {author} {\bibfnamefont {A.}~\bibnamefont {{Kashlinsky}}}\ and\ \bibinfo {author} {\bibfnamefont {F.}~\bibnamefont {{Atrio-Barandela}}},\ }\bibfield  {title} {\bibinfo {title} {{Measuring Cosmological Bulk Flows via the Kinematic Sunyaev-Zeldovich Effect in the Upcoming Cosmic Microwave Background Maps}},\ }\href {https://doi.org/10.1086/312735} {\bibfield  {journal} {\bibinfo  {journal} {Astrophys. J. Lett.}\ }\textbf {\bibinfo {volume} {536}},\ \bibinfo {pages} {L67} (\bibinfo {year} {2000})},\ \Eprint {https://arxiv.org/abs/astro-ph/0005197} {arXiv:astro-ph/0005197 [astro-ph]} \BibitemShut {NoStop}%
\bibitem [{\citenamefont {Kashlinsky}\ \emph {et~al.}(2009)\citenamefont {Kashlinsky}, \citenamefont {Atrio-Barandela}, \citenamefont {Kocevski},\ and\ \citenamefont {Ebeling}}]{Kashlinsky_2009}%
  \BibitemOpen
  \bibfield  {author} {\bibinfo {author} {\bibfnamefont {A.}~\bibnamefont {Kashlinsky}}, \bibinfo {author} {\bibfnamefont {F.}~\bibnamefont {Atrio-Barandela}}, \bibinfo {author} {\bibfnamefont {D.}~\bibnamefont {Kocevski}},\ and\ \bibinfo {author} {\bibfnamefont {H.}~\bibnamefont {Ebeling}},\ }\bibfield  {title} {\bibinfo {title} {A measurement of large-scale peculiar velocities of clusters of galaxies: Technical details},\ }\href {https://doi.org/10.1088/0004-637x/691/2/1479} {\bibfield  {journal} {\bibinfo  {journal} {The Astrophysical Journal}\ }\textbf {\bibinfo {volume} {691}},\ \bibinfo {pages} {1479–1493} (\bibinfo {year} {2009})}\BibitemShut {NoStop}%
\bibitem [{\citenamefont {Kashlinsky}\ \emph {et~al.}(2010)\citenamefont {Kashlinsky}, \citenamefont {Atrio-Barandela}, \citenamefont {Ebeling}, \citenamefont {Edge},\ and\ \citenamefont {Kocevski}}]{Kashlinsky_2010}%
  \BibitemOpen
  \bibfield  {author} {\bibinfo {author} {\bibfnamefont {A.}~\bibnamefont {Kashlinsky}}, \bibinfo {author} {\bibfnamefont {F.}~\bibnamefont {Atrio-Barandela}}, \bibinfo {author} {\bibfnamefont {H.}~\bibnamefont {Ebeling}}, \bibinfo {author} {\bibfnamefont {A.}~\bibnamefont {Edge}},\ and\ \bibinfo {author} {\bibfnamefont {D.}~\bibnamefont {Kocevski}},\ }\bibfield  {title} {\bibinfo {title} {A new measurement of the bulk flow of x-ray luminous clusters of galaxies},\ }\href {https://doi.org/10.1088/2041-8205/712/1/l81} {\bibfield  {journal} {\bibinfo  {journal} {The Astrophysical Journal}\ }\textbf {\bibinfo {volume} {712}},\ \bibinfo {pages} {L81–L85} (\bibinfo {year} {2010})}\BibitemShut {NoStop}%
\bibitem [{\citenamefont {Osborne}\ \emph {et~al.}(2011)\citenamefont {Osborne}, \citenamefont {Mak}, \citenamefont {Church},\ and\ \citenamefont {Pierpaoli}}]{Osborne_2011}%
  \BibitemOpen
  \bibfield  {author} {\bibinfo {author} {\bibfnamefont {S.~J.}\ \bibnamefont {Osborne}}, \bibinfo {author} {\bibfnamefont {D.~S.~Y.}\ \bibnamefont {Mak}}, \bibinfo {author} {\bibfnamefont {S.~E.}\ \bibnamefont {Church}},\ and\ \bibinfo {author} {\bibfnamefont {E.}~\bibnamefont {Pierpaoli}},\ }\bibfield  {title} {\bibinfo {title} {Measuring the galaxy cluster bulk flow fromwmapdata},\ }\href {https://doi.org/10.1088/0004-637x/737/2/98} {\bibfield  {journal} {\bibinfo  {journal} {The Astrophysical Journal}\ }\textbf {\bibinfo {volume} {737}},\ \bibinfo {pages} {98} (\bibinfo {year} {2011})}\BibitemShut {NoStop}%
\bibitem [{\citenamefont {Lavaux}\ and\ \citenamefont {Hudson}(2011)}]{Lavaux_2011_2M++}%
  \BibitemOpen
  \bibfield  {author} {\bibinfo {author} {\bibfnamefont {G.}~\bibnamefont {Lavaux}}\ and\ \bibinfo {author} {\bibfnamefont {M.~J.}\ \bibnamefont {Hudson}},\ }\bibfield  {title} {\bibinfo {title} {The 2m++ galaxy redshift catalogue: The 2m++ galaxy redshift catalogue},\ }\href {https://doi.org/10.1111/j.1365-2966.2011.19233.x} {\bibfield  {journal} {\bibinfo  {journal} {Monthly Notices of the Royal Astronomical Society}\ }\textbf {\bibinfo {volume} {416}},\ \bibinfo {pages} {2840–2856} (\bibinfo {year} {2011})}\BibitemShut {NoStop}%
\bibitem [{\citenamefont {Jarosik}\ \emph {et~al.}(2011)\citenamefont {Jarosik}, \citenamefont {Bennett}, \citenamefont {Dunkley}, \citenamefont {Gold}, \citenamefont {Greason}, \citenamefont {Halpern}, \citenamefont {Hill}, \citenamefont {Hinshaw}, \citenamefont {Kogut}, \citenamefont {Komatsu}, \citenamefont {Larson}, \citenamefont {Limon}, \citenamefont {Meyer}, \citenamefont {Nolta}, \citenamefont {Odegard}, \citenamefont {Page}, \citenamefont {Smith}, \citenamefont {Spergel}, \citenamefont {Tucker}, \citenamefont {Weiland}, \citenamefont {Wollack},\ and\ \citenamefont {Wright}}]{Jarosik_2011_WMAP7}%
  \BibitemOpen
  \bibfield  {author} {\bibinfo {author} {\bibfnamefont {N.}~\bibnamefont {Jarosik}}, \bibinfo {author} {\bibfnamefont {C.~L.}\ \bibnamefont {Bennett}}, \bibinfo {author} {\bibfnamefont {J.}~\bibnamefont {Dunkley}}, \bibinfo {author} {\bibfnamefont {B.}~\bibnamefont {Gold}}, \bibinfo {author} {\bibfnamefont {M.~R.}\ \bibnamefont {Greason}}, \bibinfo {author} {\bibfnamefont {M.}~\bibnamefont {Halpern}}, \bibinfo {author} {\bibfnamefont {R.~S.}\ \bibnamefont {Hill}}, \bibinfo {author} {\bibfnamefont {G.}~\bibnamefont {Hinshaw}}, \bibinfo {author} {\bibfnamefont {A.}~\bibnamefont {Kogut}}, \bibinfo {author} {\bibfnamefont {E.}~\bibnamefont {Komatsu}}, \bibinfo {author} {\bibfnamefont {D.}~\bibnamefont {Larson}}, \bibinfo {author} {\bibfnamefont {M.}~\bibnamefont {Limon}}, \bibinfo {author} {\bibfnamefont {S.~S.}\ \bibnamefont {Meyer}}, \bibinfo {author} {\bibfnamefont {M.~R.}\ \bibnamefont {Nolta}}, \bibinfo {author} {\bibfnamefont {N.}~\bibnamefont {Odegard}}, \bibinfo {author} {\bibfnamefont {L.}~\bibnamefont
  {Page}}, \bibinfo {author} {\bibfnamefont {K.~M.}\ \bibnamefont {Smith}}, \bibinfo {author} {\bibfnamefont {D.~N.}\ \bibnamefont {Spergel}}, \bibinfo {author} {\bibfnamefont {G.~S.}\ \bibnamefont {Tucker}}, \bibinfo {author} {\bibfnamefont {J.~L.}\ \bibnamefont {Weiland}}, \bibinfo {author} {\bibfnamefont {E.}~\bibnamefont {Wollack}},\ and\ \bibinfo {author} {\bibfnamefont {E.~L.}\ \bibnamefont {Wright}},\ }\bibfield  {title} {\bibinfo {title} {Seven-year wilkinson microwave anisotropy probe ( wmap ) observations: Sky maps, systematic errors, and basic results},\ }\href {https://doi.org/10.1088/0067-0049/192/2/14} {\bibfield  {journal} {\bibinfo  {journal} {The Astrophysical Journal Supplement Series}\ }\textbf {\bibinfo {volume} {192}},\ \bibinfo {pages} {14} (\bibinfo {year} {2011})}\BibitemShut {NoStop}%
\bibitem [{\citenamefont {Battaglia}(2016)}]{Battaglia_2016}%
  \BibitemOpen
  \bibfield  {author} {\bibinfo {author} {\bibfnamefont {N.}~\bibnamefont {Battaglia}},\ }\bibfield  {title} {\bibinfo {title} {The tau of galaxy clusters},\ }\href {https://doi.org/10.1088/1475-7516/2016/08/058} {\bibfield  {journal} {\bibinfo  {journal} {Journal of Cosmology and Astroparticle Physics}\ }\textbf {\bibinfo {volume} {2016}}\bibinfo  {number} { (08)},\ \bibinfo {pages} {058–058}}\BibitemShut {NoStop}%
\bibitem [{\citenamefont {{Flender}}\ \emph {et~al.}(2017)\citenamefont {{Flender}}, \citenamefont {{Nagai}},\ and\ \citenamefont {{McDonald}}}]{Flender_2017}%
  \BibitemOpen
\bibfield  {number} {  }\bibfield  {author} {\bibinfo {author} {\bibfnamefont {S.}~\bibnamefont {{Flender}}}, \bibinfo {author} {\bibfnamefont {D.}~\bibnamefont {{Nagai}}},\ and\ \bibinfo {author} {\bibfnamefont {M.}~\bibnamefont {{McDonald}}},\ }\bibfield  {title} {\bibinfo {title} {{Constraints on the Optical Depth of Galaxy Groups and Clusters}},\ }\href {https://doi.org/10.3847/1538-4357/aa60bf} {\bibfield  {journal} {\bibinfo  {journal} {Astrophys. J.}\ }\textbf {\bibinfo {volume} {837}},\ \bibinfo {eid} {124} (\bibinfo {year} {2017})},\ \Eprint {https://arxiv.org/abs/1610.08029} {arXiv:1610.08029 [astro-ph.CO]} \BibitemShut {NoStop}%
\bibitem [{\citenamefont {Madhavacheril}\ \emph {et~al.}(2019)\citenamefont {Madhavacheril}, \citenamefont {Battaglia}, \citenamefont {Smith},\ and\ \citenamefont {Sievers}}]{Madhavacheril_2019}%
  \BibitemOpen
  \bibfield  {author} {\bibinfo {author} {\bibfnamefont {M.~S.}\ \bibnamefont {Madhavacheril}}, \bibinfo {author} {\bibfnamefont {N.}~\bibnamefont {Battaglia}}, \bibinfo {author} {\bibfnamefont {K.~M.}\ \bibnamefont {Smith}},\ and\ \bibinfo {author} {\bibfnamefont {J.~L.}\ \bibnamefont {Sievers}},\ }\bibfield  {title} {\bibinfo {title} {Cosmology with the kinematic sunyaev-zeldovich effect: Breaking the optical depth degeneracy with fast radio bursts},\ }\bibfield  {journal} {\bibinfo  {journal} {Physical Review D}\ }\textbf {\bibinfo {volume} {100}},\ \href {https://doi.org/10.1103/physrevd.100.103532} {10.1103/physrevd.100.103532} (\bibinfo {year} {2019})\BibitemShut {NoStop}%
\bibitem [{\citenamefont {McCarthy}\ \emph {et~al.}(2025)\citenamefont {McCarthy} \emph {et~al.}}]{McCarthy:2025brx}%
  \BibitemOpen
  \bibfield  {author} {\bibinfo {author} {\bibfnamefont {F.}~\bibnamefont {McCarthy}} \emph {et~al.},\ }\bibfield  {title} {\bibinfo {title} {{The Atacama Cosmology Telescope: Cross-correlation of kSZ and continuity equation velocity reconstruction with photometric DESI LRGs}},\ }\href@noop {} {\  (\bibinfo {year} {2025})},\ \Eprint {https://arxiv.org/abs/2511.15701} {arXiv:2511.15701 [astro-ph.CO]} \BibitemShut {NoStop}%
\bibitem [{\citenamefont {Li}\ \emph {et~al.}(2012)\citenamefont {Li}, \citenamefont {Pan}, \citenamefont {Gao}, \citenamefont {Jing}, \citenamefont {Yang}, \citenamefont {Chi}, \citenamefont {Feng}, \citenamefont {Kang}, \citenamefont {Lin}, \citenamefont {Shan}, \citenamefont {Wang}, \citenamefont {Zhao},\ and\ \citenamefont {Zhang}}]{Li_2012}%
  \BibitemOpen
  \bibfield  {author} {\bibinfo {author} {\bibfnamefont {M.}~\bibnamefont {Li}}, \bibinfo {author} {\bibfnamefont {J.}~\bibnamefont {Pan}}, \bibinfo {author} {\bibfnamefont {L.}~\bibnamefont {Gao}}, \bibinfo {author} {\bibfnamefont {Y.}~\bibnamefont {Jing}}, \bibinfo {author} {\bibfnamefont {X.}~\bibnamefont {Yang}}, \bibinfo {author} {\bibfnamefont {X.}~\bibnamefont {Chi}}, \bibinfo {author} {\bibfnamefont {L.}~\bibnamefont {Feng}}, \bibinfo {author} {\bibfnamefont {X.}~\bibnamefont {Kang}}, \bibinfo {author} {\bibfnamefont {W.}~\bibnamefont {Lin}}, \bibinfo {author} {\bibfnamefont {G.}~\bibnamefont {Shan}}, \bibinfo {author} {\bibfnamefont {L.}~\bibnamefont {Wang}}, \bibinfo {author} {\bibfnamefont {D.}~\bibnamefont {Zhao}},\ and\ \bibinfo {author} {\bibfnamefont {P.}~\bibnamefont {Zhang}},\ }\bibfield  {title} {\bibinfo {title} {Bulk flow of halos in lcdm simulation},\ }\href {https://doi.org/10.1088/0004-637x/761/2/151} {\bibfield  {journal} {\bibinfo  {journal} {The Astrophysical Journal}\ }\textbf
  {\bibinfo {volume} {761}},\ \bibinfo {pages} {151} (\bibinfo {year} {2012})}\BibitemShut {NoStop}%
\bibitem [{\citenamefont {Watkins}\ \emph {et~al.}(2009)\citenamefont {Watkins}, \citenamefont {Feldman},\ and\ \citenamefont {Hudson}}]{Watkins_2009}%
  \BibitemOpen
  \bibfield  {author} {\bibinfo {author} {\bibfnamefont {R.}~\bibnamefont {Watkins}}, \bibinfo {author} {\bibfnamefont {H.~A.}\ \bibnamefont {Feldman}},\ and\ \bibinfo {author} {\bibfnamefont {M.~J.}\ \bibnamefont {Hudson}},\ }\bibfield  {title} {\bibinfo {title} {Consistently large cosmic flows on scales of 100 mpc/h: a challenge for the standard lcdm cosmology},\ }\href {https://doi.org/10.1111/j.1365-2966.2008.14089.x} {\bibfield  {journal} {\bibinfo  {journal} {Monthly Notices of the Royal Astronomical Society}\ }\textbf {\bibinfo {volume} {392}},\ \bibinfo {pages} {743–756} (\bibinfo {year} {2009})}\BibitemShut {NoStop}%
\bibitem [{\citenamefont {{Feldman}}\ \emph {et~al.}(2010)\citenamefont {{Feldman}}, \citenamefont {{Watkins}},\ and\ \citenamefont {{Hudson}}}]{Feldman_2010_MV_ppr2}%
  \BibitemOpen
  \bibfield  {author} {\bibinfo {author} {\bibfnamefont {H.~A.}\ \bibnamefont {{Feldman}}}, \bibinfo {author} {\bibfnamefont {R.}~\bibnamefont {{Watkins}}},\ and\ \bibinfo {author} {\bibfnamefont {M.~J.}\ \bibnamefont {{Hudson}}},\ }\bibfield  {title} {\bibinfo {title} {{Cosmic flows on 100 h$^{-1}$ Mpc scales: standardized minimum variance bulk flow, shear and octupole moments}},\ }\href {https://doi.org/10.1111/j.1365-2966.2010.17052.x} {\bibfield  {journal} {\bibinfo  {journal} {Monthly Notices of the Royal Astronomical Society}\ }\textbf {\bibinfo {volume} {407}},\ \bibinfo {pages} {2328} (\bibinfo {year} {2010})},\ \Eprint {https://arxiv.org/abs/0911.5516} {arXiv:0911.5516 [astro-ph.CO]} \BibitemShut {NoStop}%
\bibitem [{\citenamefont {{Kaiser}}(1988)}]{Kaiser_1988_MLE}%
  \BibitemOpen
  \bibfield  {author} {\bibinfo {author} {\bibfnamefont {N.}~\bibnamefont {{Kaiser}}},\ }\bibfield  {title} {\bibinfo {title} {{Theoretical implications of deviations from Hubble flow}},\ }\href {https://doi.org/10.1093/mnras/231.2.149} {\bibfield  {journal} {\bibinfo  {journal} {\mnras}\ }\textbf {\bibinfo {volume} {231}},\ \bibinfo {pages} {149} (\bibinfo {year} {1988})}\BibitemShut {NoStop}%
\bibitem [{\citenamefont {{G{\'o}rski}}\ \emph {et~al.}(2005)\citenamefont {{G{\'o}rski}}, \citenamefont {{Hivon}}, \citenamefont {{Banday}}, \citenamefont {{Wandelt}}, \citenamefont {{Hansen}}, \citenamefont {{Reinecke}},\ and\ \citenamefont {{Bartelmann}}}]{healpy2005}%
  \BibitemOpen
  \bibfield  {author} {\bibinfo {author} {\bibfnamefont {K.~M.}\ \bibnamefont {{G{\'o}rski}}}, \bibinfo {author} {\bibfnamefont {E.}~\bibnamefont {{Hivon}}}, \bibinfo {author} {\bibfnamefont {A.~J.}\ \bibnamefont {{Banday}}}, \bibinfo {author} {\bibfnamefont {B.~D.}\ \bibnamefont {{Wandelt}}}, \bibinfo {author} {\bibfnamefont {F.~K.}\ \bibnamefont {{Hansen}}}, \bibinfo {author} {\bibfnamefont {M.}~\bibnamefont {{Reinecke}}},\ and\ \bibinfo {author} {\bibfnamefont {M.}~\bibnamefont {{Bartelmann}}},\ }\bibfield  {title} {\bibinfo {title} {{HEALPix: A Framework for High-Resolution Discretization and Fast Analysis of Data Distributed on the Sphere}},\ }\href {https://doi.org/10.1086/427976} {\bibfield  {journal} {\bibinfo  {journal} {\apj}\ }\textbf {\bibinfo {volume} {622}},\ \bibinfo {pages} {759} (\bibinfo {year} {2005})},\ \Eprint {https://arxiv.org/abs/arXiv:astro-ph/0409513} {arXiv:astro-ph/0409513} \BibitemShut {NoStop}%
\bibitem [{\citenamefont {Zonca}\ \emph {et~al.}(2019)\citenamefont {Zonca}, \citenamefont {Singer}, \citenamefont {Lenz}, \citenamefont {Reinecke}, \citenamefont {Rosset}, \citenamefont {Hivon},\ and\ \citenamefont {Gorski}}]{Zonca2019_healpy}%
  \BibitemOpen
  \bibfield  {author} {\bibinfo {author} {\bibfnamefont {A.}~\bibnamefont {Zonca}}, \bibinfo {author} {\bibfnamefont {L.}~\bibnamefont {Singer}}, \bibinfo {author} {\bibfnamefont {D.}~\bibnamefont {Lenz}}, \bibinfo {author} {\bibfnamefont {M.}~\bibnamefont {Reinecke}}, \bibinfo {author} {\bibfnamefont {C.}~\bibnamefont {Rosset}}, \bibinfo {author} {\bibfnamefont {E.}~\bibnamefont {Hivon}},\ and\ \bibinfo {author} {\bibfnamefont {K.}~\bibnamefont {Gorski}},\ }\bibfield  {title} {\bibinfo {title} {healpy: equal area pixelization and spherical harmonics transforms for data on the sphere in python},\ }\href {https://doi.org/10.21105/joss.01298} {\bibfield  {journal} {\bibinfo  {journal} {Journal of Open Source Software}\ }\textbf {\bibinfo {volume} {4}},\ \bibinfo {pages} {1298} (\bibinfo {year} {2019})}\BibitemShut {NoStop}%
\bibitem [{\citenamefont {Akrami}\ \emph {et~al.}(2020)\citenamefont {Akrami}, \citenamefont {Ashdown},\ and\ \citenamefont {Aumont}}]{Planck_2020_maps}%
  \BibitemOpen
  \bibfield  {author} {\bibinfo {author} {\bibfnamefont {Y.}~\bibnamefont {Akrami}}, \bibinfo {author} {\bibfnamefont {M.}~\bibnamefont {Ashdown}},\ and\ \bibinfo {author} {\bibfnamefont {J.}~\bibnamefont {Aumont}},\ }\bibfield  {title} {\bibinfo {title} {Planck2018 results: Iv. diffuse component separation},\ }\href {https://doi.org/10.1051/0004-6361/201833881} {\bibfield  {journal} {\bibinfo  {journal} {Astronomy \&; Astrophysics}\ }\textbf {\bibinfo {volume} {641}},\ \bibinfo {pages} {A4} (\bibinfo {year} {2020})}\BibitemShut {NoStop}%
\bibitem [{\citenamefont {Tegmark}\ \emph {et~al.}(2003)\citenamefont {Tegmark}, \citenamefont {de~Oliveira-Costa},\ and\ \citenamefont {Hamilton}}]{Tegmark_2003_compsepmap}%
  \BibitemOpen
  \bibfield  {author} {\bibinfo {author} {\bibfnamefont {M.}~\bibnamefont {Tegmark}}, \bibinfo {author} {\bibfnamefont {A.}~\bibnamefont {de~Oliveira-Costa}},\ and\ \bibinfo {author} {\bibfnamefont {A.~J.~S.}\ \bibnamefont {Hamilton}},\ }\bibfield  {title} {\bibinfo {title} {High resolution foreground cleaned cmb map from wmap},\ }\href {https://doi.org/10.1103/PhysRevD.68.123523} {\bibfield  {journal} {\bibinfo  {journal} {Phys. Rev. D}\ }\textbf {\bibinfo {volume} {68}},\ \bibinfo {pages} {123523} (\bibinfo {year} {2003})}\BibitemShut {NoStop}%
\bibitem [{\citenamefont {Hivon}\ \emph {et~al.}(2002)\citenamefont {Hivon}, \citenamefont {Gorski}, \citenamefont {Netterfield}, \citenamefont {Crill}, \citenamefont {Prunet},\ and\ \citenamefont {Hansen}}]{Hivon_2002_master}%
  \BibitemOpen
  \bibfield  {author} {\bibinfo {author} {\bibfnamefont {E.}~\bibnamefont {Hivon}}, \bibinfo {author} {\bibfnamefont {K.~M.}\ \bibnamefont {Gorski}}, \bibinfo {author} {\bibfnamefont {C.~B.}\ \bibnamefont {Netterfield}}, \bibinfo {author} {\bibfnamefont {B.~P.}\ \bibnamefont {Crill}}, \bibinfo {author} {\bibfnamefont {S.}~\bibnamefont {Prunet}},\ and\ \bibinfo {author} {\bibfnamefont {F.}~\bibnamefont {Hansen}},\ }\bibfield  {title} {\bibinfo {title} {Master of the cosmic microwave background anisotropy power spectrum: A fast method for statistical analysis of large and complex cosmic microwave background data sets},\ }\href {https://doi.org/10.1086/338126} {\bibfield  {journal} {\bibinfo  {journal} {The Astrophysical Journal}\ }\textbf {\bibinfo {volume} {567}},\ \bibinfo {pages} {2–17} (\bibinfo {year} {2002})}\BibitemShut {NoStop}%
\bibitem [{\citenamefont {Lewis}\ and\ \citenamefont {Challinor}(2002)}]{CAMB_site}%
  \BibitemOpen
  \bibfield  {author} {\bibinfo {author} {\bibfnamefont {A.}~\bibnamefont {Lewis}}\ and\ \bibinfo {author} {\bibfnamefont {A.}~\bibnamefont {Challinor}},\ }\href {https://camb.info} {\bibinfo {title} {{CAMB}: Code for anisotropies in the microwave background}} (\bibinfo {year} {2002}),\ \bibinfo {note} {maintained by Antony Lewis}\BibitemShut {NoStop}%
\bibitem [{\citenamefont {Tegmark}\ and\ \citenamefont {Zaldarriaga}(2002)}]{Tegmark_2002}%
  \BibitemOpen
  \bibfield  {author} {\bibinfo {author} {\bibfnamefont {M.}~\bibnamefont {Tegmark}}\ and\ \bibinfo {author} {\bibfnamefont {M.}~\bibnamefont {Zaldarriaga}},\ }\bibfield  {title} {\bibinfo {title} {Separating the early universe from the late universe: Cosmological parameter estimation beyond the black box},\ }\bibfield  {journal} {\bibinfo  {journal} {Physical Review D}\ }\textbf {\bibinfo {volume} {66}},\ \href {https://doi.org/10.1103/physrevd.66.103508} {10.1103/physrevd.66.103508} (\bibinfo {year} {2002})\BibitemShut {NoStop}%
\bibitem [{\citenamefont {{Bahcall}}\ \emph {et~al.}(1994)\citenamefont {{Bahcall}}, \citenamefont {{Cen}},\ and\ \citenamefont {{Gramann}}}]{Bahcall_1994}%
  \BibitemOpen
  \bibfield  {author} {\bibinfo {author} {\bibfnamefont {N.~A.}\ \bibnamefont {{Bahcall}}}, \bibinfo {author} {\bibfnamefont {R.}~\bibnamefont {{Cen}}},\ and\ \bibinfo {author} {\bibfnamefont {M.}~\bibnamefont {{Gramann}}},\ }\bibfield  {title} {\bibinfo {title} {{Probing the Large-Scale Velocity Field with Clusters of Galaxies}},\ }\href {https://doi.org/10.1086/187426} {\bibfield  {journal} {\bibinfo  {journal} {\apjl}\ }\textbf {\bibinfo {volume} {430}},\ \bibinfo {pages} {L13} (\bibinfo {year} {1994})},\ \Eprint {https://arxiv.org/abs/astro-ph/9405075} {arXiv:astro-ph/9405075 [astro-ph]} \BibitemShut {NoStop}%
\bibitem [{\citenamefont {{Singal}}(2011)}]{Singal2011}%
  \BibitemOpen
  \bibfield  {author} {\bibinfo {author} {\bibfnamefont {A.~K.}\ \bibnamefont {{Singal}}},\ }\bibfield  {title} {\bibinfo {title} {{Large Peculiar Motion of the Solar System from the Dipole Anisotropy in Sky Brightness due to Distant Radio Sources}},\ }\href {https://doi.org/10.1088/2041-8205/742/2/L23} {\bibfield  {journal} {\bibinfo  {journal} {The Astrophysical Journal}\ }\textbf {\bibinfo {volume} {742}},\ \bibinfo {eid} {L23} (\bibinfo {year} {2011})},\ \Eprint {https://arxiv.org/abs/1110.6260} {arXiv:1110.6260 [astro-ph.CO]} \BibitemShut {NoStop}%
\bibitem [{\citenamefont {{Gibelyou}}\ and\ \citenamefont {{Huterer}}(2012)}]{GibelyouHuterer2012}%
  \BibitemOpen
  \bibfield  {author} {\bibinfo {author} {\bibfnamefont {C.}~\bibnamefont {{Gibelyou}}}\ and\ \bibinfo {author} {\bibfnamefont {D.}~\bibnamefont {{Huterer}}},\ }\bibfield  {title} {\bibinfo {title} {{Dipoles in the sky}},\ }\href {https://doi.org/10.1111/j.1365-2966.2012.22032.x} {\bibfield  {journal} {\bibinfo  {journal} {Monthly Notices of the Royal Astronomical Society}\ }\textbf {\bibinfo {volume} {427}},\ \bibinfo {pages} {1994} (\bibinfo {year} {2012})},\ \Eprint {https://arxiv.org/abs/1205.6476} {arXiv:1205.6476 [astro-ph.CO]} \BibitemShut {NoStop}%
\bibitem [{\citenamefont {Secrest}\ \emph {et~al.}(2022)\citenamefont {Secrest}, \citenamefont {von Hausegger}, \citenamefont {Rameez}, \citenamefont {Mohayaee},\ and\ \citenamefont {Sarkar}}]{Secrest_2022}%
  \BibitemOpen
  \bibfield  {author} {\bibinfo {author} {\bibfnamefont {N.~J.}\ \bibnamefont {Secrest}}, \bibinfo {author} {\bibfnamefont {S.}~\bibnamefont {von Hausegger}}, \bibinfo {author} {\bibfnamefont {M.}~\bibnamefont {Rameez}}, \bibinfo {author} {\bibfnamefont {R.}~\bibnamefont {Mohayaee}},\ and\ \bibinfo {author} {\bibfnamefont {S.}~\bibnamefont {Sarkar}},\ }\bibfield  {title} {\bibinfo {title} {A challenge to the standard cosmological model},\ }\href {https://doi.org/10.3847/2041-8213/ac88c0} {\bibfield  {journal} {\bibinfo  {journal} {The Astrophysical Journal Letters}\ }\textbf {\bibinfo {volume} {937}},\ \bibinfo {pages} {L31} (\bibinfo {year} {2022})}\BibitemShut {NoStop}%
\bibitem [{\citenamefont {Wagenveld}\ \emph {et~al.}(2025)\citenamefont {Wagenveld}, \citenamefont {von Hausegger}, \citenamefont {Klöckner},\ and\ \citenamefont {Schwarz}}]{Wagenveld_2025}%
  \BibitemOpen
  \bibfield  {author} {\bibinfo {author} {\bibfnamefont {J.~D.}\ \bibnamefont {Wagenveld}}, \bibinfo {author} {\bibfnamefont {S.}~\bibnamefont {von Hausegger}}, \bibinfo {author} {\bibfnamefont {H.-R.}\ \bibnamefont {Klöckner}},\ and\ \bibinfo {author} {\bibfnamefont {D.~J.}\ \bibnamefont {Schwarz}},\ }\bibfield  {title} {\bibinfo {title} {The kinematic contribution to the cosmic number count dipole},\ }\href {https://doi.org/10.1051/0004-6361/202453397} {\bibfield  {journal} {\bibinfo  {journal} {Astronomy \& Astrophysics}\ }\textbf {\bibinfo {volume} {697}},\ \bibinfo {pages} {A112} (\bibinfo {year} {2025})}\BibitemShut {NoStop}%
\bibitem [{\citenamefont {{Ellis}}\ and\ \citenamefont {{Baldwin}}(1984)}]{EllisBaldwin1984}%
  \BibitemOpen
  \bibfield  {author} {\bibinfo {author} {\bibfnamefont {G.~F.~R.}\ \bibnamefont {{Ellis}}}\ and\ \bibinfo {author} {\bibfnamefont {J.~E.}\ \bibnamefont {{Baldwin}}},\ }\bibfield  {title} {\bibinfo {title} {{On the expected anisotropy of radio source counts}},\ }\href {https://doi.org/10.1093/mnras/206.2.377} {\bibfield  {journal} {\bibinfo  {journal} {Monthly Notices of the Royal Astronomical Society}\ }\textbf {\bibinfo {volume} {206}},\ \bibinfo {pages} {377} (\bibinfo {year} {1984})}\BibitemShut {NoStop}%
\bibitem [{\citenamefont {Secrest}\ \emph {et~al.}(2021)\citenamefont {Secrest}, \citenamefont {Hausegger}, \citenamefont {Rameez}, \citenamefont {Mohayaee}, \citenamefont {Sarkar},\ and\ \citenamefont {Colin}}]{Secrest_2021}%
  \BibitemOpen
  \bibfield  {author} {\bibinfo {author} {\bibfnamefont {N.~J.}\ \bibnamefont {Secrest}}, \bibinfo {author} {\bibfnamefont {S.~v.}\ \bibnamefont {Hausegger}}, \bibinfo {author} {\bibfnamefont {M.}~\bibnamefont {Rameez}}, \bibinfo {author} {\bibfnamefont {R.}~\bibnamefont {Mohayaee}}, \bibinfo {author} {\bibfnamefont {S.}~\bibnamefont {Sarkar}},\ and\ \bibinfo {author} {\bibfnamefont {J.}~\bibnamefont {Colin}},\ }\bibfield  {title} {\bibinfo {title} {A test of the cosmological principle with quasars},\ }\href {https://doi.org/10.3847/2041-8213/abdd40} {\bibfield  {journal} {\bibinfo  {journal} {The Astrophysical Journal Letters}\ }\textbf {\bibinfo {volume} {908}},\ \bibinfo {pages} {L51} (\bibinfo {year} {2021})}\BibitemShut {NoStop}%
\bibitem [{\citenamefont {{Macaulay}}\ \emph {et~al.}(2011)\citenamefont {{Macaulay}}, \citenamefont {{Feldman}}, \citenamefont {{Ferreira}}, \citenamefont {{Hudson}},\ and\ \citenamefont {{Watkins}}}]{Macaulay_2011}%
  \BibitemOpen
  \bibfield  {author} {\bibinfo {author} {\bibfnamefont {E.}~\bibnamefont {{Macaulay}}}, \bibinfo {author} {\bibfnamefont {H.}~\bibnamefont {{Feldman}}}, \bibinfo {author} {\bibfnamefont {P.~G.}\ \bibnamefont {{Ferreira}}}, \bibinfo {author} {\bibfnamefont {M.~J.}\ \bibnamefont {{Hudson}}},\ and\ \bibinfo {author} {\bibfnamefont {R.}~\bibnamefont {{Watkins}}},\ }\bibfield  {title} {\bibinfo {title} {{A slight excess of large-scale power from moments of the peculiar velocity field}},\ }\href {https://doi.org/10.1111/j.1365-2966.2011.18426.x} {\bibfield  {journal} {\bibinfo  {journal} {\mnras}\ }\textbf {\bibinfo {volume} {414}},\ \bibinfo {pages} {621} (\bibinfo {year} {2011})},\ \Eprint {https://arxiv.org/abs/1010.2651} {arXiv:1010.2651 [astro-ph.CO]} \BibitemShut {NoStop}%
\bibitem [{\citenamefont {Doré}\ \emph {et~al.}(2015)\citenamefont {Doré}, \citenamefont {Bock}, \citenamefont {Ashby},\ and\ \citenamefont {Capak}}]{dore2015_spherex}%
  \BibitemOpen
  \bibfield  {author} {\bibinfo {author} {\bibfnamefont {O.}~\bibnamefont {Doré}}, \bibinfo {author} {\bibfnamefont {J.}~\bibnamefont {Bock}}, \bibinfo {author} {\bibfnamefont {M.}~\bibnamefont {Ashby}},\ and\ \bibinfo {author} {\bibfnamefont {P.}~\bibnamefont {Capak}},\ }\href {https://arxiv.org/abs/1412.4872} {\bibinfo {title} {Cosmology with the spherex all-sky spectral survey}} (\bibinfo {year} {2015}),\ \Eprint {https://arxiv.org/abs/1412.4872} {arXiv:1412.4872 [astro-ph.CO]} \BibitemShut {NoStop}%
\bibitem [{\citenamefont {Bock}\ \emph {et~al.}(2026)\citenamefont {Bock}, \citenamefont {Aboobaker},\ and\ \citenamefont {Adamo}}]{Bock_2026_SPHEREx}%
  \BibitemOpen
  \bibfield  {author} {\bibinfo {author} {\bibfnamefont {J.~J.}\ \bibnamefont {Bock}}, \bibinfo {author} {\bibfnamefont {A.~M.}\ \bibnamefont {Aboobaker}},\ and\ \bibinfo {author} {\bibfnamefont {J.}~\bibnamefont {Adamo}},\ }\bibfield  {title} {\bibinfo {title} {The spherex satellite mission},\ }\href {https://doi.org/10.3847/1538-4357/ae2be2} {\bibfield  {journal} {\bibinfo  {journal} {The Astrophysical Journal}\ }\textbf {\bibinfo {volume} {999}},\ \bibinfo {pages} {139} (\bibinfo {year} {2026})}\BibitemShut {NoStop}%
\bibitem [{\citenamefont {Laureijs}\ \emph {et~al.}(2011)\citenamefont {Laureijs}, \citenamefont {Amiaux}, \citenamefont {Arduini}, \citenamefont {Auguères}, \citenamefont {Brinchmann},\ and\ \citenamefont {Cole}}]{laureijs2011_euclid}%
  \BibitemOpen
  \bibfield  {author} {\bibinfo {author} {\bibfnamefont {R.}~\bibnamefont {Laureijs}}, \bibinfo {author} {\bibfnamefont {J.}~\bibnamefont {Amiaux}}, \bibinfo {author} {\bibfnamefont {S.}~\bibnamefont {Arduini}}, \bibinfo {author} {\bibfnamefont {J.~L.}\ \bibnamefont {Auguères}}, \bibinfo {author} {\bibfnamefont {J.}~\bibnamefont {Brinchmann}},\ and\ \bibinfo {author} {\bibfnamefont {R.}~\bibnamefont {Cole}},\ }\href {https://arxiv.org/abs/1110.3193} {\bibinfo {title} {Euclid definition study report}} (\bibinfo {year} {2011}),\ \Eprint {https://arxiv.org/abs/1110.3193} {arXiv:1110.3193 [astro-ph.CO]} \BibitemShut {NoStop}%
\bibitem [{\citenamefont {{Amendola}}\ \emph {et~al.}(2018)\citenamefont {{Amendola}}, \citenamefont {{Appleby}}, \citenamefont {{Avgoustidis}}, \citenamefont {{Bacon}}, \citenamefont {{Baker}}, \citenamefont {{Baldi}}, \citenamefont {{Bartolo}}, \citenamefont {{Blanchard}}, \citenamefont {{Bonvin}}, \citenamefont {{Borgani}}, \citenamefont {{Branchini}}, \citenamefont {{Burrage}}, \citenamefont {{Camera}}, \citenamefont {{Carbone}}, \citenamefont {{Casarini}}, \citenamefont {{Cropper}}, \citenamefont {{de Rham}}, \citenamefont {{Dietrich}}, \citenamefont {{Di Porto}}, \citenamefont {{Durrer}}, \citenamefont {{Ealet}}, \citenamefont {{Ferreira}}, \citenamefont {{Finelli}}, \citenamefont {{Garc{\'\i}a-Bellido}}, \citenamefont {{Giannantonio}}, \citenamefont {{Guzzo}}, \citenamefont {{Heavens}}, \citenamefont {{Heisenberg}}, \citenamefont {{Heymans}}, \citenamefont {{Hoekstra}}, \citenamefont {{Hollenstein}}, \citenamefont {{Holmes}}, \citenamefont {{Hwang}}, \citenamefont {{Jahnke}}, \citenamefont
  {{Kitching}}, \citenamefont {{Koivisto}}, \citenamefont {{Kunz}}, \citenamefont {{La Vacca}}, \citenamefont {{Linder}}, \citenamefont {{March}}, \citenamefont {{Marra}}, \citenamefont {{Martins}}, \citenamefont {{Majerotto}}, \citenamefont {{Markovic}}, \citenamefont {{Marsh}}, \citenamefont {{Marulli}}, \citenamefont {{Massey}}, \citenamefont {{Mellier}}, \citenamefont {{Montanari}}, \citenamefont {{Mota}}, \citenamefont {{Nunes}}, \citenamefont {{Percival}}, \citenamefont {{Pettorino}}, \citenamefont {{Porciani}}, \citenamefont {{Quercellini}}, \citenamefont {{Read}}, \citenamefont {{Rinaldi}}, \citenamefont {{Sapone}}, \citenamefont {{Sawicki}}, \citenamefont {{Scaramella}}, \citenamefont {{Skordis}}, \citenamefont {{Simpson}}, \citenamefont {{Taylor}}, \citenamefont {{Thomas}}, \citenamefont {{Trotta}}, \citenamefont {{Verde}}, \citenamefont {{Vernizzi}}, \citenamefont {{Vollmer}}, \citenamefont {{Wang}}, \citenamefont {{Weller}}, \citenamefont {{Zlosnik}},\ and\ \citenamefont {{The Euclid Theory
  Working Group}}}]{Amendola_2018_euclid}%
  \BibitemOpen
  \bibfield  {author} {\bibinfo {author} {\bibfnamefont {L.}~\bibnamefont {{Amendola}}}, \bibinfo {author} {\bibfnamefont {S.}~\bibnamefont {{Appleby}}}, \bibinfo {author} {\bibfnamefont {A.}~\bibnamefont {{Avgoustidis}}}, \bibinfo {author} {\bibfnamefont {D.}~\bibnamefont {{Bacon}}}, \bibinfo {author} {\bibfnamefont {T.}~\bibnamefont {{Baker}}}, \bibinfo {author} {\bibfnamefont {M.}~\bibnamefont {{Baldi}}}, \bibinfo {author} {\bibfnamefont {N.}~\bibnamefont {{Bartolo}}}, \bibinfo {author} {\bibfnamefont {A.}~\bibnamefont {{Blanchard}}}, \bibinfo {author} {\bibfnamefont {C.}~\bibnamefont {{Bonvin}}}, \bibinfo {author} {\bibfnamefont {S.}~\bibnamefont {{Borgani}}}, \bibinfo {author} {\bibfnamefont {E.}~\bibnamefont {{Branchini}}}, \bibinfo {author} {\bibfnamefont {C.}~\bibnamefont {{Burrage}}}, \bibinfo {author} {\bibfnamefont {S.}~\bibnamefont {{Camera}}}, \bibinfo {author} {\bibfnamefont {C.}~\bibnamefont {{Carbone}}}, \bibinfo {author} {\bibfnamefont {L.}~\bibnamefont {{Casarini}}}, \bibinfo {author}
  {\bibfnamefont {M.}~\bibnamefont {{Cropper}}}, \bibinfo {author} {\bibfnamefont {C.}~\bibnamefont {{de Rham}}}, \bibinfo {author} {\bibfnamefont {J.~P.}\ \bibnamefont {{Dietrich}}}, \bibinfo {author} {\bibfnamefont {C.}~\bibnamefont {{Di Porto}}}, \bibinfo {author} {\bibfnamefont {R.}~\bibnamefont {{Durrer}}}, \bibinfo {author} {\bibfnamefont {A.}~\bibnamefont {{Ealet}}}, \bibinfo {author} {\bibfnamefont {P.~G.}\ \bibnamefont {{Ferreira}}}, \bibinfo {author} {\bibfnamefont {F.}~\bibnamefont {{Finelli}}}, \bibinfo {author} {\bibfnamefont {J.}~\bibnamefont {{Garc{\'\i}a-Bellido}}}, \bibinfo {author} {\bibfnamefont {T.}~\bibnamefont {{Giannantonio}}}, \bibinfo {author} {\bibfnamefont {L.}~\bibnamefont {{Guzzo}}}, \bibinfo {author} {\bibfnamefont {A.}~\bibnamefont {{Heavens}}}, \bibinfo {author} {\bibfnamefont {L.}~\bibnamefont {{Heisenberg}}}, \bibinfo {author} {\bibfnamefont {C.}~\bibnamefont {{Heymans}}}, \bibinfo {author} {\bibfnamefont {H.}~\bibnamefont {{Hoekstra}}}, \bibinfo {author} {\bibfnamefont
  {L.}~\bibnamefont {{Hollenstein}}}, \bibinfo {author} {\bibfnamefont {R.}~\bibnamefont {{Holmes}}}, \bibinfo {author} {\bibfnamefont {Z.}~\bibnamefont {{Hwang}}}, \bibinfo {author} {\bibfnamefont {K.}~\bibnamefont {{Jahnke}}}, \bibinfo {author} {\bibfnamefont {T.~D.}\ \bibnamefont {{Kitching}}}, \bibinfo {author} {\bibfnamefont {T.}~\bibnamefont {{Koivisto}}}, \bibinfo {author} {\bibfnamefont {M.}~\bibnamefont {{Kunz}}}, \bibinfo {author} {\bibfnamefont {G.}~\bibnamefont {{La Vacca}}}, \bibinfo {author} {\bibfnamefont {E.}~\bibnamefont {{Linder}}}, \bibinfo {author} {\bibfnamefont {M.}~\bibnamefont {{March}}}, \bibinfo {author} {\bibfnamefont {V.}~\bibnamefont {{Marra}}}, \bibinfo {author} {\bibfnamefont {C.}~\bibnamefont {{Martins}}}, \bibinfo {author} {\bibfnamefont {E.}~\bibnamefont {{Majerotto}}}, \bibinfo {author} {\bibfnamefont {D.}~\bibnamefont {{Markovic}}}, \bibinfo {author} {\bibfnamefont {D.}~\bibnamefont {{Marsh}}}, \bibinfo {author} {\bibfnamefont {F.}~\bibnamefont {{Marulli}}}, \bibinfo
  {author} {\bibfnamefont {R.}~\bibnamefont {{Massey}}}, \bibinfo {author} {\bibfnamefont {Y.}~\bibnamefont {{Mellier}}}, \bibinfo {author} {\bibfnamefont {F.}~\bibnamefont {{Montanari}}}, \bibinfo {author} {\bibfnamefont {D.~F.}\ \bibnamefont {{Mota}}}, \bibinfo {author} {\bibfnamefont {N.~J.}\ \bibnamefont {{Nunes}}}, \bibinfo {author} {\bibfnamefont {W.}~\bibnamefont {{Percival}}}, \bibinfo {author} {\bibfnamefont {V.}~\bibnamefont {{Pettorino}}}, \bibinfo {author} {\bibfnamefont {C.}~\bibnamefont {{Porciani}}}, \bibinfo {author} {\bibfnamefont {C.}~\bibnamefont {{Quercellini}}}, \bibinfo {author} {\bibfnamefont {J.}~\bibnamefont {{Read}}}, \bibinfo {author} {\bibfnamefont {M.}~\bibnamefont {{Rinaldi}}}, \bibinfo {author} {\bibfnamefont {D.}~\bibnamefont {{Sapone}}}, \bibinfo {author} {\bibfnamefont {I.}~\bibnamefont {{Sawicki}}}, \bibinfo {author} {\bibfnamefont {R.}~\bibnamefont {{Scaramella}}}, \bibinfo {author} {\bibfnamefont {C.}~\bibnamefont {{Skordis}}}, \bibinfo {author} {\bibfnamefont
  {F.}~\bibnamefont {{Simpson}}}, \bibinfo {author} {\bibfnamefont {A.}~\bibnamefont {{Taylor}}}, \bibinfo {author} {\bibfnamefont {S.}~\bibnamefont {{Thomas}}}, \bibinfo {author} {\bibfnamefont {R.}~\bibnamefont {{Trotta}}}, \bibinfo {author} {\bibfnamefont {L.}~\bibnamefont {{Verde}}}, \bibinfo {author} {\bibfnamefont {F.}~\bibnamefont {{Vernizzi}}}, \bibinfo {author} {\bibfnamefont {A.}~\bibnamefont {{Vollmer}}}, \bibinfo {author} {\bibfnamefont {Y.}~\bibnamefont {{Wang}}}, \bibinfo {author} {\bibfnamefont {J.}~\bibnamefont {{Weller}}}, \bibinfo {author} {\bibfnamefont {T.}~\bibnamefont {{Zlosnik}}},\ and\ \bibinfo {author} {\bibnamefont {{The Euclid Theory Working Group}}},\ }\bibfield  {title} {\bibinfo {title} {{Cosmology and fundamental physics with the Euclid satellite}},\ }\href {https://doi.org/10.1007/s41114-017-0010-3} {\bibfield  {journal} {\bibinfo  {journal} {Living Reviews in Relativity}\ }\textbf {\bibinfo {volume} {21}},\ \bibinfo {eid} {2} (\bibinfo {year} {2018})},\ \Eprint
  {https://arxiv.org/abs/1606.00180} {arXiv:1606.00180 [astro-ph.CO]} \BibitemShut {NoStop}%
\bibitem [{\citenamefont {{Ivezi{\'c}}}\ \emph {et~al.}(2019)\citenamefont {{Ivezi{\'c}}}, \citenamefont {{Kahn}}, \citenamefont {{Tyson}}, \citenamefont {{Abel}},\ and\ \citenamefont {{Acosta}}}]{Ivezic_2019_lsst}%
  \BibitemOpen
  \bibfield  {author} {\bibinfo {author} {\bibfnamefont {{\v{Z}}.}~\bibnamefont {{Ivezi{\'c}}}}, \bibinfo {author} {\bibfnamefont {S.~M.}\ \bibnamefont {{Kahn}}}, \bibinfo {author} {\bibfnamefont {J.~A.}\ \bibnamefont {{Tyson}}}, \bibinfo {author} {\bibfnamefont {B.}~\bibnamefont {{Abel}}},\ and\ \bibinfo {author} {\bibfnamefont {E.}~\bibnamefont {{Acosta}}},\ }\bibfield  {title} {\bibinfo {title} {{LSST: From Science Drivers to Reference Design and Anticipated Data Products}},\ }\href {https://doi.org/10.3847/1538-4357/ab042c} {\bibfield  {journal} {\bibinfo  {journal} {\apj}\ }\textbf {\bibinfo {volume} {873}},\ \bibinfo {eid} {111} (\bibinfo {year} {2019})},\ \Eprint {https://arxiv.org/abs/0805.2366} {arXiv:0805.2366 [astro-ph]} \BibitemShut {NoStop}%
\bibitem [{\citenamefont {Ade}\ \emph {et~al.}(2019)\citenamefont {Ade}, \citenamefont {Aguirre}, \citenamefont {Ahmed}, \citenamefont {Aiola}, \citenamefont {Ali}, \citenamefont {Alonso},\ and\ \citenamefont {Alvarez}}]{Ade_2019_SO}%
  \BibitemOpen
  \bibfield  {author} {\bibinfo {author} {\bibfnamefont {P.}~\bibnamefont {Ade}}, \bibinfo {author} {\bibfnamefont {J.}~\bibnamefont {Aguirre}}, \bibinfo {author} {\bibfnamefont {Z.}~\bibnamefont {Ahmed}}, \bibinfo {author} {\bibfnamefont {S.}~\bibnamefont {Aiola}}, \bibinfo {author} {\bibfnamefont {A.}~\bibnamefont {Ali}}, \bibinfo {author} {\bibfnamefont {D.}~\bibnamefont {Alonso}},\ and\ \bibinfo {author} {\bibfnamefont {M.~A.}\ \bibnamefont {Alvarez}},\ }\bibfield  {title} {\bibinfo {title} {The simons observatory: science goals and forecasts},\ }\href {https://doi.org/10.1088/1475-7516/2019/02/056} {\bibfield  {journal} {\bibinfo  {journal} {Journal of Cosmology and Astroparticle Physics}\ }\textbf {\bibinfo {volume} {2019}}\bibinfo  {number} { (02)},\ \bibinfo {pages} {056–056}}\BibitemShut {NoStop}%
\bibitem [{\citenamefont {Sehgal}\ \emph {et~al.}(2020)\citenamefont {Sehgal}, \citenamefont {Aiola}, \citenamefont {Akrami}, \citenamefont {moni Basu}, \citenamefont {Boylan-Kolchin}, \citenamefont {Bryan},\ and\ \citenamefont {Casey}}]{sehgal2020_cmb-hd}%
  \BibitemOpen
\bibfield  {number} {  }\bibfield  {author} {\bibinfo {author} {\bibfnamefont {N.}~\bibnamefont {Sehgal}}, \bibinfo {author} {\bibfnamefont {S.}~\bibnamefont {Aiola}}, \bibinfo {author} {\bibfnamefont {Y.}~\bibnamefont {Akrami}}, \bibinfo {author} {\bibfnamefont {K.}~\bibnamefont {moni Basu}}, \bibinfo {author} {\bibfnamefont {M.}~\bibnamefont {Boylan-Kolchin}}, \bibinfo {author} {\bibfnamefont {S.}~\bibnamefont {Bryan}},\ and\ \bibinfo {author} {\bibfnamefont {C.~M.}\ \bibnamefont {Casey}},\ }\href {https://arxiv.org/abs/2002.12714} {\bibinfo {title} {Cmb-hd: Astro2020 rfi response}} (\bibinfo {year} {2020}),\ \Eprint {https://arxiv.org/abs/2002.12714} {arXiv:2002.12714 [astro-ph.CO]} \BibitemShut {NoStop}%
\bibitem [{\citenamefont {Harris}\ \emph {et~al.}(2020)\citenamefont {Harris}, \citenamefont {Millman}, \citenamefont {van~der Walt} \emph {et~al.}}]{numpy}%
  \BibitemOpen
  \bibfield  {author} {\bibinfo {author} {\bibfnamefont {C.~R.}\ \bibnamefont {Harris}}, \bibinfo {author} {\bibfnamefont {K.~J.}\ \bibnamefont {Millman}}, \bibinfo {author} {\bibfnamefont {S.~J.}\ \bibnamefont {van~der Walt}}, \emph {et~al.},\ }\bibfield  {title} {\bibinfo {title} {Array programming with {NumPy}},\ }\href {https://doi.org/10.1038/s41586-020-2649-2} {\bibfield  {journal} {\bibinfo  {journal} {Nature}\ }\textbf {\bibinfo {volume} {585}},\ \bibinfo {pages} {357} (\bibinfo {year} {2020})}\BibitemShut {NoStop}%
\bibitem [{\citenamefont {Virtanen}\ \emph {et~al.}(2020)\citenamefont {Virtanen}, \citenamefont {Gommers}, \citenamefont {Oliphant} \emph {et~al.}}]{scipy}%
  \BibitemOpen
  \bibfield  {author} {\bibinfo {author} {\bibfnamefont {P.}~\bibnamefont {Virtanen}}, \bibinfo {author} {\bibfnamefont {R.}~\bibnamefont {Gommers}}, \bibinfo {author} {\bibfnamefont {T.~E.}\ \bibnamefont {Oliphant}}, \emph {et~al.},\ }\bibfield  {title} {\bibinfo {title} {{{SciPy} 1.0: fundamental algorithms for scientific computing in Python}},\ }\href {https://doi.org/10.1038/s41592-019-0686-2} {\bibfield  {journal} {\bibinfo  {journal} {Nature Methods}\ }\textbf {\bibinfo {volume} {17}},\ \bibinfo {pages} {261} (\bibinfo {year} {2020})}\BibitemShut {NoStop}%
\bibitem [{\citenamefont {Hunter}(2007)}]{matplotlib}%
  \BibitemOpen
  \bibfield  {author} {\bibinfo {author} {\bibfnamefont {J.~D.}\ \bibnamefont {Hunter}},\ }\bibfield  {title} {\bibinfo {title} {Matplotlib: A 2d graphics environment},\ }\href {https://doi.org/10.1109/MCSE.2007.55} {\bibfield  {journal} {\bibinfo  {journal} {Computing in Science \& Engineering}\ }\textbf {\bibinfo {volume} {9}},\ \bibinfo {pages} {90} (\bibinfo {year} {2007})}\BibitemShut {NoStop}%
\bibitem [{\citenamefont {Chiang}\ \emph {et~al.}(2020)\citenamefont {Chiang}, \citenamefont {M{\'e}nard},\ and\ \citenamefont {Taghizadeh-Popp}}]{tomographer_website}%
  \BibitemOpen
  \bibfield  {author} {\bibinfo {author} {\bibfnamefont {Y.-K.}\ \bibnamefont {Chiang}}, \bibinfo {author} {\bibfnamefont {B.}~\bibnamefont {M{\'e}nard}},\ and\ \bibinfo {author} {\bibfnamefont {M.}~\bibnamefont {Taghizadeh-Popp}},\ }\href@noop {} {\bibinfo {title} {{Tomographer: Estimating Redshift Distributions}}},\ \bibinfo {howpublished} {\url{http://tomographer.org/}} (\bibinfo {year} {2020}),\ \bibinfo {note} {accessed: 2026-03-01}\BibitemShut {NoStop}%
\bibitem [{\citenamefont {{M{\'e}nard}}\ \emph {et~al.}(2013)\citenamefont {{M{\'e}nard}}, \citenamefont {{Scranton}}, \citenamefont {{Schmidt}}, \citenamefont {{Morrison}}, \citenamefont {{Jeong}}, \citenamefont {{Budavari}},\ and\ \citenamefont {{Rahman}}}]{menard_2013_tomog}%
  \BibitemOpen
  \bibfield  {author} {\bibinfo {author} {\bibfnamefont {B.}~\bibnamefont {{M{\'e}nard}}}, \bibinfo {author} {\bibfnamefont {R.}~\bibnamefont {{Scranton}}}, \bibinfo {author} {\bibfnamefont {S.}~\bibnamefont {{Schmidt}}}, \bibinfo {author} {\bibfnamefont {C.}~\bibnamefont {{Morrison}}}, \bibinfo {author} {\bibfnamefont {D.}~\bibnamefont {{Jeong}}}, \bibinfo {author} {\bibfnamefont {T.}~\bibnamefont {{Budavari}}},\ and\ \bibinfo {author} {\bibfnamefont {M.}~\bibnamefont {{Rahman}}},\ }\bibfield  {title} {\bibinfo {title} {{Clustering-based redshift estimation: method and application to data}},\ }\href {https://doi.org/10.48550/arXiv.1303.4722} {\bibfield  {journal} {\bibinfo  {journal} {arXiv e-prints}\ ,\ \bibinfo {eid} {arXiv:1303.4722}} (\bibinfo {year} {2013})},\ \Eprint {https://arxiv.org/abs/1303.4722} {arXiv:1303.4722 [astro-ph.CO]} \BibitemShut {NoStop}%
\bibitem [{\citenamefont {{Chiang}}\ \emph {et~al.}(2019)\citenamefont {{Chiang}}, \citenamefont {{M{\'e}nard}},\ and\ \citenamefont {{Schiminovich}}}]{chiang_2019_tomog}%
  \BibitemOpen
  \bibfield  {author} {\bibinfo {author} {\bibfnamefont {Y.-K.}\ \bibnamefont {{Chiang}}}, \bibinfo {author} {\bibfnamefont {B.}~\bibnamefont {{M{\'e}nard}}},\ and\ \bibinfo {author} {\bibfnamefont {D.}~\bibnamefont {{Schiminovich}}},\ }\bibfield  {title} {\bibinfo {title} {{Broadband Intensity Tomography: Spectral Tagging of the Cosmic UV Background}},\ }\href {https://doi.org/10.3847/1538-4357/ab1b35} {\bibfield  {journal} {\bibinfo  {journal} {Astrophys. J.}\ }\textbf {\bibinfo {volume} {877}},\ \bibinfo {eid} {150} (\bibinfo {year} {2019})},\ \Eprint {https://arxiv.org/abs/1810.00885} {arXiv:1810.00885 [astro-ph.CO]} \BibitemShut {NoStop}%
\bibitem [{\citenamefont {Kvasiuk}\ and\ \citenamefont {M{\"u}nchmeyer}(2024)}]{Kvasiuk:2023nje}%
  \BibitemOpen
  \bibfield  {author} {\bibinfo {author} {\bibfnamefont {Y.}~\bibnamefont {Kvasiuk}}\ and\ \bibinfo {author} {\bibfnamefont {M.}~\bibnamefont {M{\"u}nchmeyer}},\ }\bibfield  {title} {\bibinfo {title} {{Autodifferentiable likelihood pipeline for the cross-correlation of CMB and large-scale structure due to the kinetic Sunyaev-Zeldovich effect}},\ }\href {https://doi.org/10.1103/PhysRevD.109.083515} {\bibfield  {journal} {\bibinfo  {journal} {Phys. Rev. D}\ }\textbf {\bibinfo {volume} {109}},\ \bibinfo {pages} {083515} (\bibinfo {year} {2024})},\ \Eprint {https://arxiv.org/abs/2305.08903} {arXiv:2305.08903 [astro-ph.CO]} \BibitemShut {NoStop}%
\bibitem [{\citenamefont {Giri}\ and\ \citenamefont {Smith}(2022)}]{Giri:2020pkk}%
  \BibitemOpen
  \bibfield  {author} {\bibinfo {author} {\bibfnamefont {U.}~\bibnamefont {Giri}}\ and\ \bibinfo {author} {\bibfnamefont {K.~M.}\ \bibnamefont {Smith}},\ }\bibfield  {title} {\bibinfo {title} {{Exploring KSZ velocity reconstruction with N-body simulations and the halo~model}},\ }\href {https://doi.org/10.1088/1475-7516/2022/09/028} {\bibfield  {journal} {\bibinfo  {journal} {JCAP}\ }\textbf {\bibinfo {volume} {09}},\ \bibinfo {pages} {028}},\ \Eprint {https://arxiv.org/abs/2010.07193} {arXiv:2010.07193 [astro-ph.CO]} \BibitemShut {NoStop}%
\bibitem [{\citenamefont {Takahashi}\ \emph {et~al.}(2020)\citenamefont {Takahashi}, \citenamefont {Ioka}, \citenamefont {Mori},\ and\ \citenamefont {Funahashi}}]{Takahashi2020}%
  \BibitemOpen
  \bibfield  {author} {\bibinfo {author} {\bibfnamefont {R.}~\bibnamefont {Takahashi}}, \bibinfo {author} {\bibfnamefont {K.}~\bibnamefont {Ioka}}, \bibinfo {author} {\bibfnamefont {A.}~\bibnamefont {Mori}},\ and\ \bibinfo {author} {\bibfnamefont {K.}~\bibnamefont {Funahashi}},\ }\bibfield  {title} {\bibinfo {title} {{Statistical modelling of the cosmological dispersion measure}},\ }\href {https://doi.org/10.1093/mnras/stab170} {\bibfield  {journal} {\bibinfo  {journal} {Monthly Notices of the Royal Astronomical Society}\ }\textbf {\bibinfo {volume} {502}},\ \bibinfo {pages} {2615} (\bibinfo {year} {2020})},\ \Eprint {https://arxiv.org/abs/2010.01560} {arXiv:2010.01560} \BibitemShut {NoStop}%
\bibitem [{\citenamefont {Coulton}\ \emph {et~al.}(2024)\citenamefont {Coulton} \emph {et~al.}}]{ACT:2023wcq}%
  \BibitemOpen
  \bibfield  {author} {\bibinfo {author} {\bibfnamefont {W.}~\bibnamefont {Coulton}} \emph {et~al.} (\bibinfo {collaboration} {ACT}),\ }\bibfield  {title} {\bibinfo {title} {{Atacama Cosmology Telescope: High-resolution component-separated maps across one third of the sky}},\ }\href {https://doi.org/10.1103/PhysRevD.109.063530} {\bibfield  {journal} {\bibinfo  {journal} {Phys. Rev. D}\ }\textbf {\bibinfo {volume} {109}},\ \bibinfo {pages} {063530} (\bibinfo {year} {2024})},\ \Eprint {https://arxiv.org/abs/2307.01258} {arXiv:2307.01258 [astro-ph.CO]} \BibitemShut {NoStop}%
\end{thebibliography}%
\end{document}